\documentclass[twocolumn,tighten,twocolappendix]{aastex63}

\usepackage{verbatim} 
\usepackage{natbib} 
\usepackage{amsmath} 
\usepackage{amsbsy}
\usepackage{amssymb}
\usepackage{mathrsfs} 
\usepackage{graphicx}
\usepackage{epstopdf}
\usepackage{ctable} 
\usepackage{fixltx2e} 
\usepackage{url}
\usepackage{newtxtext,newtxmath}
\usepackage[T1]{fontenc} 
\usepackage{aecompl}
\usepackage{soul}
\usepackage{comment}
\usepackage{commath}
\usepackage{enumitem}

\newcommand{\Msun}{M$_{\odot}$}
\newcommand{\Msunyr}{M$_{\odot}$yr$^{-1}$}
\newcommand{\kms}{km\,s$^{-1}$}
\newcommand{\Mkpc}{M$_{\odot}$kpc$^{-2}$}
\newcommand{\Myrkpc}{M$_{\odot}$yr$^{-1}$kpc$^{-2}$}

\newcommand{\hmpc}{h^{-1}{\rm Mpc}}

\newcommand{\eqso}{\texttt{pre-QSO}}
\newcommand{\qso}{\texttt{full-QSO}}
\newcommand{\agn}{\texttt{late-AGN}}
\newcommand{\pathL}{.}
\newcommand{\vsf}{-0.5cm}

\shorttitle{Cosmological hyper-refinement simulations of quasar fueling}
\shortauthors{Daniel Angl{\'e}s-Alc{\'a}zar et al.}

\hypersetup{linkcolor=blue,citecolor=blue,filecolor=blue,urlcolor=blue}

\begin{document}

\title{\vspace{-0.2in}Cosmological simulations of quasar fueling to sub-parsec scales \\ using Lagrangian hyper-refinement}

\correspondingauthor{Daniel Angl{\'e}s-Alc{\'a}zar}
\email{angles-alcazar@uconn.edu}

\author{Daniel Angl{\'e}s-Alc{\'a}zar}
\affiliation{Department of Physics, University of Connecticut, 196 Auditorium Road, U-3046, Storrs, CT 06269-3046, USA}
\affiliation{Center for Computational Astrophysics, Flatiron Institute, 162 Fifth Avenue, New York, NY 10010, USA}

\author{Eliot Quataert}
\affiliation{Department of Astronomy and Theoretical Astrophysics Center, University of California Berkeley, Berkeley, CA 94720, USA}
\affiliation{Department of Astrophysical Sciences, Princeton University, Princeton, NJ 08544, USA }

\author{Philip F.~Hopkins}
\affiliation{TAPIR, Mailcode 350-17, California Institute of Technology, Pasadena, CA 91125, USA}

\author{Rachel S.~Somerville}
\affiliation{Center for Computational Astrophysics, Flatiron Institute, 162 Fifth Avenue, New York, NY 10010, USA}
\affiliation{Department of Physics and Astronomy, Rutgers University, 136 Frelinghuysen Rd, Piscataway, NJ 08854, USA}

\author{Christopher C. Hayward}
\affiliation{Center for Computational Astrophysics, Flatiron Institute, 162 Fifth Avenue, New York, NY 10010, USA}

\author{Claude-Andr{\'e} Faucher-Gigu{\`e}re}
\affiliation{CIERA and Department of Physics and Astronomy, Northwestern University, 2145 Sheridan Road, Evanston, IL 60208, USA}

\author{Greg L. Bryan}
\affiliation{Department of Astronomy, Columbia University, 550 West 120th Street, New York, NY 10027, USA}
\affiliation{Center for Computational Astrophysics, Flatiron Institute, 162 Fifth Avenue, New York, NY 10010, USA}

\author{Du\v{s}an Kere\v{s}}
\affiliation{Department of Physics, CASS, University of California at San Diego, 9500 Gilman Drive, La Jolla, CA 92093, USA}

\author{Lars Hernquist}
\affiliation{Harvard-Smithsonian Center for Astrophysics, 60 Garden Street, Cambridge, MA 02138, USA}

\author{James M. Stone}
\affiliation{School of Natural Sciences, Institute for Advanced Study, 1 Einstein Drive, Princeton, NJ 08540}

\begin{abstract}

We present cosmological hydrodynamic simulations of a quasar-mass halo ($M_{\rm halo} \approx 10^{12.5}\,{\rm M}_{\odot}~{\rm at}~z=2$) that for the first time resolve gas transport down to the inner 0.1\,pc surrounding the central massive black hole.  We model a multi-phase interstellar medium including stellar feedback by supernovae, stellar winds, and radiation, and a hyper-Lagrangian refinement technique increasing the resolution dynamically approaching the black hole.  We do not include black hole feedback.
We show that the sub-pc inflow rate (1) can reach $\sim$6\,\Msunyr~roughly in steady state during the epoch of peak nuclear gas density ($z\sim 2$), sufficient to power a luminous quasar, (2) is highly time variable in the pre-quasar phase, spanning 0.001--10\,\Msunyr~on Myr timescales, and (3) is limited to short ($\sim$2\,Myr) active phases (0.01--0.1\,\Msunyr) followed by longer periods of inactivity at lower nuclear gas density and late times ($z\sim1$), owing to the formation of a hot central cavity.  
Inflowing gas is primarily cool, rotational support dominates over turbulence and thermal pressure, and star formation can consume as much gas as provided by inflows across 1\,pc--10\,kpc.  Gravitational torques from multi-scale stellar non-axisymmetries dominate angular momentum transport over gas self-torquing and pressure gradients, with accretion weakly dependent on black hole mass.  
Sub-pc inflow rates correlate with nuclear (but decouple from global) star formation and can exceed the Eddington rate by $\times10$.  
The black hole can move $\sim$10\,pc from the galaxy center on $\sim$0.1\,Myr. 
Accreting gas forms pc-scale, rotationally supported, obscuring structures often misaligned with the galaxy-scale disk.  
These simulations open a new avenue to investigate black hole--galaxy co-evolution.\\ \\

\end{abstract}

\section{Introduction}

The inflow of gas from large scales down to galactic nuclei plays a key role in galaxy formation, driving the growth of central massive black holes and a variety of related phenomena: from bright quasars (QSOs) that outshine their host galaxies \citep[with bolometric luminosities reaching $L_{\rm bol} \sim 10^{46}$--$10^{48}$\,erg\,s$^{-1}$, e.g.][]{Fan2001,Mortlock2011,Trakhtenbrot2011,Banados2018,Zakamska2019} to active galactic nuclei (AGN) ``feedback'' in the form of fast nuclear outflows \citep[e.g.][]{Tombesi2013,Nardini2015}, galaxy-scale winds \citep[e.g.][]{Rupke2011,Greene2012_QSOoutflow,Liu2013,Cicone2014,Harrison2014,Zakamska2014,RamosAlmeida2017,Wylezalek2020}
and radio-emitting jets \citep{Fabian2012,Hlavacek-Larrondo2012} that may have a significant impact on galaxy evolution \citep[e.g.][]{Silk1998,DiMatteo2005,Murray2005,Faucher-Giguere2012_WindModel,Richings2018_MolecularOutflow}.    
The scaling relations between central black hole mass and properties of their host galaxies \citep[e.g.][]{Haring2004,Hopkins2007_BHplaneObs,Bentz2009,Bennert2011,Kormendy2013,McConnell2013,Reines2015,Graham2016} and the similarity between the global cosmic histories of star formation and black hole accretion \citep{Silverman2008,Aird2010,Rodighiero2010,Heckman2014,Madau2014} further suggest some form of black hole--galaxy co-evolution over cosmological timescales.  
It is thus crucial to understand the mechanisms driving gas inflows from galactic scales down to the black hole accretion disk in a full cosmological context, which remains a major challenge.

Modern large volume cosmological hydrodynamic simulations such as Magneticum \citep{Hirschmann2014}, Horizon-AGN \citep{Dubois2014,Volonteri2016}, Eagle \citep{Schaye2015,Rosas-Guevara2016}, Illustris \citep{Genel2014,Vogelsberger2014,Vogelsberger2014_Nature,Sijacki2015}, BlueTides \citep{Feng2016}, Romulus \citep{Tremmel2017_Romulus,Sharma2020}, IllustrisTNG \citep{Weinberger2017,Pillepich2018,Habouzit2019}, and SIMBA \citep{Dave2019_Simba,Thomas2019,Borrow2020} have implemented sub-grid models of black hole growth and feedback with increasing success at reproducing global galaxy properties and black hole observables \citep[see][for a recent comparison]{Habouzit2021_Mbh}.  
However, despite much recent progress, the typical $\sim$kpc scale resolution available 
(orders of magnitude from resolving accretion disk feeding on sub-pc scales)
makes sub-grid black hole modeling very schematic by necessity and a major source of uncertainty.  With strong degeneracies between accretion and feedback, the successes of most current models rely on carefully tuning sub-grid parameters that significantly limit their predictive power \citep{SomervilleDave2015}.
For example, simulations implementing Bondi accretion invariably predict that AGN feedback self-regulation drives the black hole--galaxy scaling relations \citep[e.g.][]{DiMatteo2008,Sijacki2015}, while simulations implementing gravitational torque-driven accretion emphasize the role of a common gas supply for star formation and black hole growth \citep{Angles-Alcazar2013,Angles-Alcazar2015,Angles-Alcazar2017_BHfeedback,Dave2019_Simba}. 
The inferred impact of AGN feedback depends not only on the assumed efficiency and accretion model, but also on the sub-grid treatment of the interstellar medium (ISM), star formation, and stellar feedback.

Cosmological ``zoom-in'' simulations provide a higher resolution framework to study massive black holes in galaxies \citep[e.g.][]{Bellovary2013,Costa2014,Choi2015_CosmoSim,Choi2018,Blank2019}.  The Feedback In Realistic Environments (FIRE) project\footnote{\url{http://fire.northwestern.edu}.} \citep{Hopkins2014_FIRE,Hopkins2018_FIRE2methods} implements stellar feedback on the scale of star-forming regions, injecting energy, momentum, mass, and metals from supernovae (SNe), stellar winds, and radiation directly based on the predictions of stellar population synthesis models.  
This yields a realistic ISM with multi-phase structure and self-consistent generation of galactic winds \citep{Muratov2015,Angles-Alcazar2017_BaryonCycle,Pandya2021} while reproducing a variety of galaxy properties \citep[e.g.][]{Ma2016_Metallicity,Ma2017_StellarDisk,Wetzel2016,Feldmann2017_MassiveFIRE,Sparre2017_SFbursts,Garrison-Kimmel2018,Cochrane2019,Liang2019_MassiveFIRE} and circumgalactic medium (CGM) observables \citep[e.g.][]{Faucher-Giguere2015,Faucher-Giguere2016,Hafen2017}.  
The increased resolution and detailed stellar feedback have important implications for black hole scaling relations, AGN demographics, and galaxy quenching:  stellar feedback can significantly suppress early black hole growth by efficiently evacuating gas from galactic nuclei \citep{Angles-Alcazar2017_BHsOnFIRE}.  
Qualitatively similar effects are seen in simulations implementing Bondi accretion \citep{Dubois2015_SNa,Bower2017,Habouzit2017,Prieto2017,McAlpine2018,Trebitsch2018,Lupi2019} but the details crucially depend on resolution, ISM physics, and black hole parameterization.

Idealized (non-cosmological) simulations offer the possibility to study gas transport in galaxies in a more controlled setup and at much higher resolution than typically available in cosmological simulations.  Early models showed that large-scale tidal torques induced by galaxy mergers or bar/spiral wave instabilities in self-gravitating disks can lead to angular momentum transport and rapid inflow of gas to the central sub-kpc of galaxies \citep{Hernquist1989,Barnes1991,Hernquist1995,Barnes1996}, confirmed by more recent simulations \citep[e.g.][]{Springel2005_BHmodel,Younger2008,Hopkins2009}.  Gas inflow to $<$100\,pc scales requires additional processes, as large-scale torques become less efficient \citep{Jogee2006}.  Subsequent secondary instabilities occurring at smaller scales \citep[``bars-within-bars'';][]{Shlosman1989,Shlosman1990} are a promising mechanism to transport gas down to the black hole accretion disk, where turbulent MHD processes and/or MHD winds are expected to dominate \citep{BlandfordPayne1982,BalbusHawley1998}. Idealized simulations with sub-pc resolution have shown that such cascades of instabilities do occur under certain conditions \citep{Escala2007,Mayer2010,Hopkins2010_MultiScale,Hopkins2011_Analytic,Emsellem2015,Mayer2015,Hopkins2016_NuclearSims}.

These idealized studies have greatly informed our understanding of black hole fueling, providing the basis for more physically motivated accretion prescriptions that can be implemented in cosmological simulations \citep[e.g.][]{Hopkins2011_Analytic,Angles-Alcazar2017_BHfeedback,Dave2019_Simba}.
However, idealized simulations are limited by artificial initial conditions that do not reflect the complexity seen in cosmological simulations, where the more realistic larger scale environment includes cold filamentary accretion \citep[e.g.][]{Keres2005,Brooks2009,Dekel2009}, pressure-supported gas in hot halos \citep[e.g.][]{Correa2018,Stern2020}, fountain flows \citep{Oppenheimer2010,Christensen2016,Angles-Alcazar2017_BaryonCycle}, and global mass transport in disks \citep[e.g.][]{El-Badry2016}. 
Idealized models are also more likely to experience artificial transients (in e.g. morphological and star formation properties), cannot reach steady state owing to gas consumption/evacuation and the lack of larger scale inflow, and cannot self-consistently determine the origin of gravitational torques (and if e.g. instabilities are initiated by mergers or internal processes) owing to their limited dynamic range of scales.
In addition, many previous idealized simulations have been limited by the use of simplified sub-grid ISM models inherited from simulations with $\sim$kpc resolution, missing the small-scale multi-phase structure of the ISM in the nuclear region.

In this work, we aim to model black hole growth in a full cosmological galaxy formation context while reaching sub-pc resolution, bridging the crucial gap between idealized models and cosmological hydrodynamic simulations.
Our goal is to model explicitly the inflow of gas down to the scales $\sim$0.1\,pc below which the accretion disk is expected to form around the central black hole in a massive galaxy \citep[e.g.][]{Goodman2003}, 
including multi-scale gravitational torques between the gas and stellar components, gas consumption by star formation, turbulence and gas ejection driven by stellar feedback, and non trivial black hole dynamics in the nuclear potential.  We achieve this by implementing multi-phase ISM physics from the FIRE project, a novel hyper-Lagrangian refinement technique that increases the resolution dynamically closer to the black hole, and a sub-pc scale treatment of black hole growth that reduces many of the uncertainties introduced by parameterized sub-grid models on 0.1--1\,kpc scales.

As a first application of this methodology, we focus on understanding the mechanisms responsible for gas transport across spatial scales and black hole growth during qualitatively distinct phases of a massive galaxy ($M_{\star} \sim 6$--20\,$\times 10^{10}$\,\Msun) before, during, and after its peak of nuclear gas density at $z\sim 2.5\rightarrow1$.  
This allows us to investigate the physical conditions conducive to a broad range of black hole accretion properties, from large QSO-like inflow rates down to extended periods of inactivity, which represents a unique challenge for current models.  Given the complex multi-physics involved, we choose to neglect the effects of black hole feedback in an attempt to reduce model uncertainties and avoid making prior assumptions about the efficiency of AGN feedback.
The results presented here provide the baseline for future simulations including AGN feedback.

We describe our methodology in \S\ref{sec:methods}, focusing on the galaxy formation model (\S\ref{sec:fire}) and the use of pre-existing FIRE cosmological zoom-in simulations as the initial conditions (\S\ref{sec:ics}) to implement our new hyper-Lagrangian refinement technique (\S\ref{sec:split}). 
We present an overview of our simulations in \S\ref{sec:overview}, illustrating the radical increase in resolution relative to previous models (\S\ref{sec:resmaps}) and their unique dynamic range (\S\ref{sec:maps}).  
We then quantify the galaxy radial structure and thermodynamic state of gas over five orders of magnitude in spatial scales (\S\ref{sec:profiles}), which provides significant insight into the physics driving black hole growth and the connection between accretion and galaxy properties on larger scales.
Our main results are presented in \S\ref{sec:multi}, where we measure for the first time resolved accretion rates at 0.1\,pc under different conditions at the center of a massive galaxy, reaching peak inflow rates sufficient to power a luminous quasar (\S\ref{sec:mdot}).  We study the connection between star formation and gas inflow rate across the full range of spatial scales 0.1\,pc--10\,kpc (\S\ref{sec:SfrMdot}), with important implications for black hole fueling and the star formation-AGN connection.
We quantify the angular momentum properties of gas across scales (\S\ref{sec:angmom}) and evaluate the relative contributions of gravitational torques and pressure gradients to angular momentum transport (\S\ref{sec:torque}), a crucial step toward identifying the dominant mechanisms driving gas inflows.
We further study the intrinsic dependence of accretion rate on black hole mass (\S\ref{sec:bhdep}), with important implications for sub-grid black hole accretion prescriptions.
We discuss our results and conclude in \S\ref{sec:discussion}.
Additional details about the robustness of our methodology are presented in \ref{sec:appendix:tests}, including systematic tests of resolution convergence (\ref{sec:appendix:res}) and other numerics (\ref{sec:appendix:gravtest}).

This work has been developed with the support of the Simulating Multiscale Astrophysics to Understand Galaxies (SMAUG) consortium\footnote{\url{https://www.simonsfoundation.org/flatiron/center-for-computational-astrophysics/galaxy-formation/smaug/}}, which attempts to reduce the uncertainties associated with sub-grid parameterizations of key baryonic processes in galaxy formation to improve the accuracy of fundamental cosmological measurements. 
Other work in SMAUG has centered on star formation and stellar feedback \citep{Kim2020,Li2019_smaug,Li2020_smaug,Motwani2020}, the physics of the CGM \citep{Fielding2020}, and the connection between semi-analytic models and cosmological hydrodynamic simulations \citep{Pandya2020}.  Our work complements these by providing a key step towards developing better sub-grid models for black hole accretion in galaxy formation, which play a key role in the evolution of massive galaxies and large-scale structure.

\section{Methods}\label{sec:methods}

\subsection{Cosmological galaxy formation model}\label{sec:fire}

We use the FIRE-2 galaxy formation model presented in \citet{Hopkins2018_FIRE2methods}, which is an updated numerical implementation of the original FIRE simulations \citep{Hopkins2014_FIRE} with more accurate hydrodynamics, gravitational softening, and supernova coupling algorithms.  FIRE-2 employs the GIZMO\footnote{\url{http://www.tapir.caltech.edu/~phopkins/Site/GIZMO.html}} $N$-body+hydrodynamics code \citep{Hopkins2015_Gizmo} in its ``meshless finite mass" (MFM) mode.  MFM is a mesh-free, Lagrangian finite-volume Godunov method that combines the advantages of particle-based and grid-based methods: MFM provides automatic adaptivity in resolution while minimizing advection and angular momentum conservation errors relative to AMR codes and exhibits superior performance in shock-capturing and fluid-mixing problems while avoiding the low-order errors inherent to smooth particle hydrodynamics (SPH).
As shown in \citet{Hopkins2015_Gizmo}, MFM performs well with different particle masses unlike many SPH methods where the errors increase when having unequal-mass particles interacting, which is crucial for our hyper-refinement simulations.

We adopt a ``standard" flat $\Lambda$CDM cosmology with parameters $H_0 = 69.7$\,km\,s$^{-1}$Mpc$^{-1}$, $\Omega_{\rm M} = 1-\Omega_{\rm \Lambda} = 0.2821$, $\Omega_{\rm b} = 0.0461$, $\sigma_8 = 0.817$, and $n_s = 0.9646$ \citep{Hinshaw2013}. 
Gravitational forces are computed using an improved version of the tree-particle-mesh gravity solver of the GADGET-3 code \citep{Springel2005_Gadget}, including adaptive gravitational softenings for the gas component.  Following \citet{Price2007}, we assume that the mass distribution corresponding to each element has the same functional form as the interaction kernel in MFM.  The time integration is fully adaptive with a power-of-two hierarchy for assigning individual time-steps \citep{Springel2005_Gadget}, including a time-step limiter to handle strong feedback events \citep{Durier2012}.

\subsubsection{Cooling, star formation, and stellar feedback}

We incorporate radiative cooling and heating processes from $T = 10$--$10^{10}$\,K as detailed in \citet{Hopkins2018_FIRE2methods}, including free-free, photoionization/recombination, Compton, photo-electric, metal-line, molecular, fine-structure, dust collisional, and cosmic ray\footnote{We include approximate cosmic ray heating in the dense ISM but do not include full cosmic ray transport and feedback as in e.g. \citet{Chan2019_CRs} and \citet{Hopkins2020_FIREmultiphys}.} processes. 
Star formation occurs only in gas which is (1) locally self-gravitating following the criterion of \citet{Hopkins2013_StarFormationLaw}, (2) molecular with self-shielding fraction given by the local Sobolev approximation \citep{Krumholz2011}, (3) Jeans unstable with gas element mass larger than the thermal Jeans mass, and (4) above a minimum hydrogen number density $n_{\rm H} \geq 1000$\,cm$^{-3}$. 
Eligible gas elements are converted into collisionless star particles with a probability set by their SFR density $\dot{\rho}_{\star}$ integrated over a timestep, where 
$\dot{\rho}_{\star} = \rho_{\rm mol}  / t_{\rm ff}$ and $t_{\rm ff}$ is the local free-fall time.  This is consistent with the rate at which small, locally self-gravitating clumps fragment in higher-resolution simulations of turbulent clouds \citep[e.g.][]{Padoan2011} and analytic models based on turbulent fragmentation \citep[e.g.][]{Guszejnov2015,Guszejnov2016}, which may only be a small fraction of the dense gas mass.  The global efficiency of star formation is self-regulated by stellar feedback at $\sim$1--10\% per free-fall time and is insensitive to the details of the star formation model \citep[e.g.,][]{Faucher-Giguere2013,Hopkins2014_FIRE,Orr2018}.

Each star particle represents a single stellar population with known mass, age, and metallicity.  
Stellar feedback is implemented at the scale of star-forming regions following explicitly the time dependence of several different mechanisms, including energy, momentum, mass, and metal injection from (1) Type Ia and Type II Supernovae (SNe), (2) continuous stellar mass-loss from OB and AGB winds, (3) photoionization and photoelectric heating, and (4) local and long-range momentum flux from radiation pressure. 
All feedback quantities and their time dependence are tabulated directly from the stellar population synthesis model {\sc starburst99} \citep{Leitherer1999} assuming a \citet{Kroupa2001} initial mass function (IMF) and non-rotating, non-binary stellar populations.

The radiative feedback implementation is described in detail in \citet{Hopkins2020_RadiativeFeedback}.
The age and metallicity-dependent IMF-averaged spectrum of each star particle is locally extinguished by the surrounding gas using a Sobolev approximation with frequency and metallicity-dependent opacities.  The emerging luminosities (including re-radiated dust emission) are propagated through a tree structure in the optically thin limit, yielding a long-range incident flux which is then corrected for local extinction at the location of the gas element.  We also include a uniform, redshift-dependent photo-ionizing background \citep{Faucher-Giguere2009} with self-shielding.  The resulting incident flux is used to compute the ionization state of the gas, radiative heating/cooling rates, and radiation pressure from photon absorption including UV/optical single-scattering and re-radiated infrared photons. 

The SNe feedback algorithm, fully described in \citet{Hopkins2018_SNeFeedback}, is constructed to ensure conservation of mass, energy, and momentum as well as isotropic injection in the rest frame of the star particle regardless of the geometry and dynamics of the surrounding gas distribution (non-trivial in Lagrangian codes).  
Each individual SN is treated independently, with IMF-averaged ejecta mass of 1.4\,\Msun~and 10.5\,\Msun~per SNe Ia and SNe II explosion, respectively, each with $10^{51}$\,erg of ejecta energy.
The correct relative amount of momentum and thermal energy is injected depending on the resolved coupling distance relative to the SN cooling radius.  This algorithm reproduces converged solutions in both energy and momentum independent of resolution, which is crucial for the multi-mass resolution simulations presented here.  The same feedback implementation has been successfully used in cosmological zoom-in simulations with mass resolution ranging from $m_{\rm b} \sim 20$\,\Msun~in dwarfs \citep{Wheeler2019} to $m_{\rm b} \sim 3\times10^4$\,\Msun~in massive galaxies \citep{Angles-Alcazar2017_BHsOnFIRE}.   
At the highest resolution reached here ($m_{\rm b} \sim 15$\,\Msun), individual star particles are still massive enough to accommodate a full SN II mass ejection, but the same algorithm performs well down to significantly lower particle masses, conserving mass in a time-average sense \citep{Grudic2020_STARFORGE}.
Stellar winds follow a similar implementation but with continuous rather than impulsive injection of mass, momentum, energy, and metals into surrounding gas particles.
The discrete nature of individual SNe is crucial in generating a multi-phase ISM and driving galactic winds, but IMF sampling on radiative feedback and stellar winds has weak effects on galaxy-scale properties \citep{Su2018_IMFsampling}.

\subsubsection{Black hole physics}

We follow the growth of a massive black hole located at the center of the main halo in the high resolution region of a pre-existing cosmological zoom-in simulation from \citet{Angles-Alcazar2017_BHsOnFIRE}, which implemented a sub-grid accretion model based on gravitational torques \citep{Hopkins2011_Analytic,Angles-Alcazar2017_BHfeedback} evaluated on $\sim100$\,pc scales.  The radical increase in resolution achieved by the hyper-Lagrangian refinement technique introduced here allows us to study black hole growth without the need for a sub-grid parameterization for the inflow rate from larger scales down to the accretion disk.

The black hole is represented by a collisionless particle with mass $M_{\rm BH}$ which is 
many orders of magnitude larger than the mass of high resolution gas and star particles that dominate the potential in the central $\sim$100\,pc.
This allows us to track the black hole dynamics explicitly owing to resolved gravitational forces without the need for a sub-grid dynamic friction prescription \citep[e.g.][]{Tremmel2015_DynFriction,Pfister2019,MaLinhao2021} or the artificial drag forces and re-positioning algorithms required in large volume cosmological simulations \citep[e.g.][]{DiMatteo2008,Dave2019_Simba}.   
The initial black hole mass in the hyper-refinement simulations is chosen to be $M_{\rm BH} = 10^8$\,\Msun~(see \S\ref{sec:bhdep} for simulations with different initial $M_{\rm BH}$), which is thus generally different from the evolving black hole in the pre-existing cosmological zoom-in simulation (\S\ref{sec:split}).

We model accretion by direct gravitational capture of gas by the central black hole. 
Gas particles located within the accretion radius $R_{\rm acc}$ are captured if (1) their velocity relative to the black hole is lower than the escape velocity and (2) the apocentric radius of the particle relative to the black hole is also within $R_{\rm acc}$.  We choose the accretion radius $R_{\rm acc} \equiv 0.1$\,pc so that it roughly corresponds to the scale below which the accretion disk is expected to form, $R_{\rm acc} \sim 10^4\,R_{\rm s}$ \citep[e.g.][]{Goodman2003}, where $R_{\rm s} = 2\,G\,M_{\rm BH}/c^2 \approx 10^{-5}$\,pc is the Schwarzschild radius for $M_{\rm BH} = 10^8$\,\Msun.  We do not allow the black hole to accrete star particles.
A similar explicit gravitational capture approach was used in \citet{Hopkins2011_Analytic,Hopkins2010_MultiScale} and \citet{Hopkins2016_NuclearSims} to study black hole growth in idealized nuclear scale simulations.  
For simplicity, we refer to the measured gas inflow rate through $R_{\rm acc}$ as ``black hole accretion rate" or ``$\dot{M}_{\rm BH}$'' but it should be interpreted as the instantaneous feeding rate onto the black hole accretion disk.  
This represents an upper limit to the actual black hole growth owing to e.g. mass loss in accretion disk winds \citep[e.g.][]{Proga2008,Yuan2012_AccDiskSim,Jiang2014}.
We neglect the effects of black hole feedback in an attempt to understand the mechanisms responsible for mass transport without making any assumptions about the efficiency of AGN feedback.

\subsection{Zoom-in simulations as initial conditions}\label{sec:ics}

We use the FIRE-2 zoom-in simulations of massive halos ($M_{\rm halo} \sim 10^{12.5-13}$\,\Msun~at z=1) presented in \citet{Angles-Alcazar2017_BHsOnFIRE} to identify interesting redshift snapshots for re-simulation at ultra-high resolution.  The initial conditions correspond to a subset of the {\bf A} series of MassiveFIRE halos that were originally simulated with the FIRE-1 model \citep{Feldmann2016,Feldmann2017_MassiveFIRE}.  For this work, we focus on halo {\bf A4} (studied also in \citealt{Cochrane2019} and \citealt{Wellons2020}), which was evolved from early times down to $z=1$ including an on-the-fly treatment of black hole growth  but not black hole feedback.  The mass resolution employed was $m_{\rm b} = 3.3 \times 10^4$\,\Msun~and $m_{\rm DM} = 1.7 \times 10^5$\,\Msun~for the baryonic and dark matter components.  The force softenings were $\epsilon_{\rm gas}^{\rm min} = 0.7$\,pc, $\epsilon_{\rm \star} = \epsilon_{\rm BH} = 7$\,pc, and $\epsilon_{\rm DM} = 57$\,pc, where $\epsilon_{\rm gas}^{\rm min}$ is the minimum adaptive force softening for gas (with $\epsilon_{\rm gas}$ identical to the kernel smoothing scale) and the softenings for the stellar ($\epsilon_{\rm \star}$), black hole ($\epsilon_{\rm BH}$), and dark matter ($\epsilon_{\rm DM}$) components are fixed in physical units at $z<9$.

\begin{figure}
\begin{center}
\includegraphics[width=0.48\textwidth]{\pathL/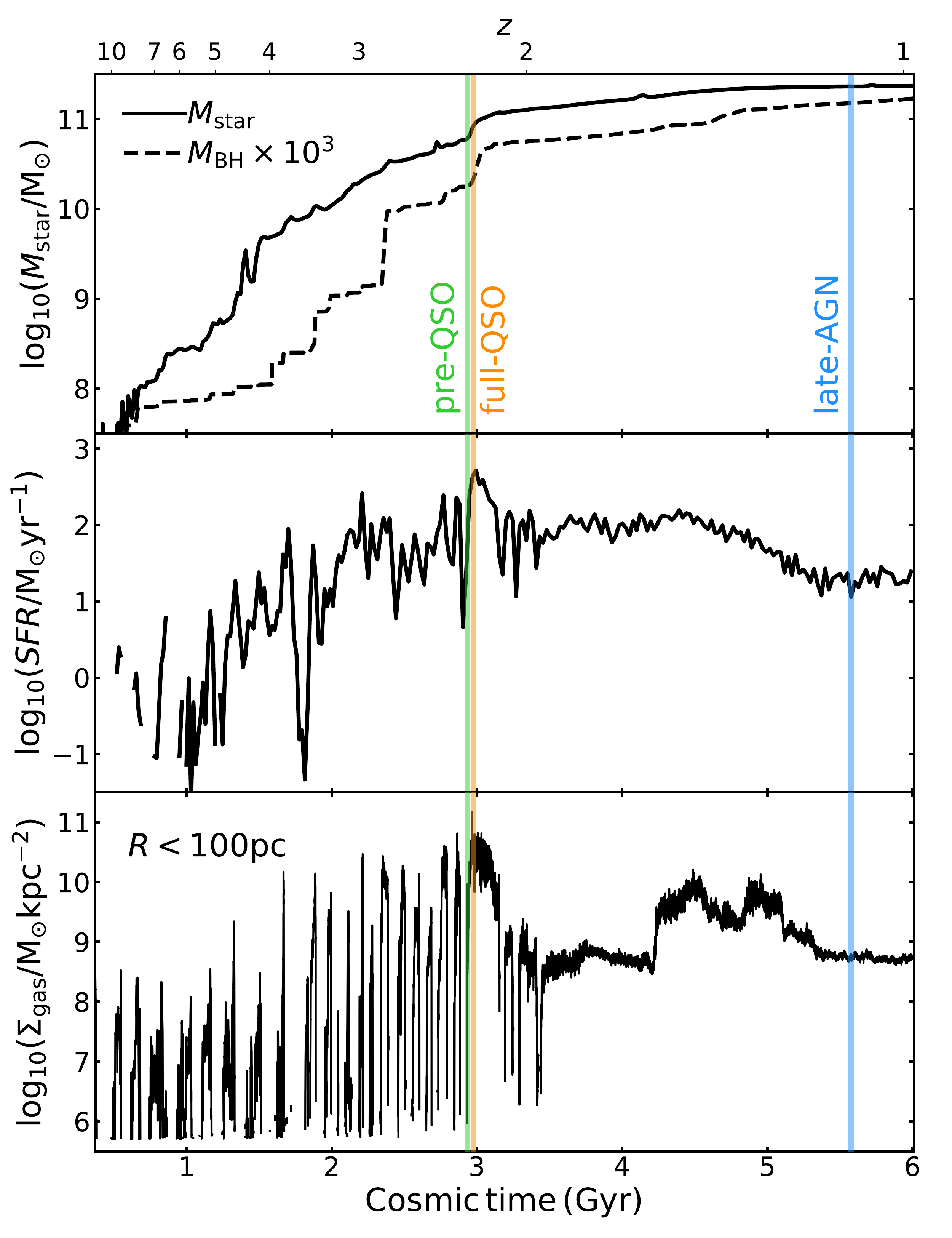}
\end{center}
\vspace{\vsf}
\caption{Stellar mass (top), SFR (middle), and nuclear gas surface density (within $R_0 \sim 10$--100\,pc; bottom) as a function of redshift for the central galaxy in halo {\bf A4} from \citet{Angles-Alcazar2017_BHsOnFIRE}, with $M_{\rm vir} \sim 5\times10^{12}$\,\Msun~at $z=1$.  
The dashed line in the top panel shows the mass of the central black hole multiplied by $\times$1000.
Vertical lines indicate the redshifts chosen for re-simulation with dynamic hyper-refinement of the gas component.  We select:
(1) $z = 2.25$ (orange) as the most optimistic conditions to trigger a luminous ``\qso'' phase, with the highest nuclear $\Sigma_{\rm gas}$ and near the peak of global SFR; 
(2) $z = 2.28$ (green) as the less extreme conditions prevailing $\sim$40\,Myr before the \qso~phase, representative of ``\eqso''~conditions; and
(3) $z = 1.10$ (blue) as the typical conditions with lower SFR and $\Sigma_{\rm gas}$ prevalent at late times, more representative of lower luminosity ``\agn''~conditions.
}
\vspace{-0.3cm}
\label{fig:ini}
\end{figure}

Figure~\ref{fig:ini} shows the global stellar mass ($M_{\star}$; top) and SFR (middle) as a function of redshift for the central galaxy of halo {\bf A4}.  Throughout this paper, global galaxy properties refer to integrated quantities within $0.1\,R_{\rm vir}$ of the host halo.  We identify halos using the Amiga Halo Finder \citep[AHF;][]{Gill2004_AHF,Knollmann2009_AHF}, modified such that the virial radius $R_{\rm vir}$ corresponds to the evolving overdensity definition of \citet{Bryan1998}.  
We also show the nuclear gas surface density ($\Sigma_{\rm gas}$; bottom) measured within the black hole kernel, which is defined to contain 256 gas resolution elements and has typical values in the range $R_0 \sim 10$--100\,pc.
Global galaxy properties are computed from the snapshots available at $\sim20$--25\,Myr intervals, but we take advantage of black hole-specific data outputs that recorded the physical conditions within $R_0$ (including $\Sigma_{\rm gas}$) for every timestep of the simulation, reaching time resolution $dt\sim 100$\,yr.

The evolution of the central galaxy in halo {\bf A4} is qualitatively similar to that of halo {\bf A2} described in detail in \citet{Angles-Alcazar2017_BHsOnFIRE}. 
During the first $\sim 3$\,Gyr, the stellar mass grows to $M_{\star} \sim 5\times10^{10}$\,\Msun~while undergoing strong bursts of star formation reaching $SFR \sim 10$--300\,\Msunyr.  Correlated SNe drive large scale winds that can evacuate a large fraction of ISM gas and temporarily shut down star formation, while recycling of wind material provides fuel for subsequent bursts of star formation \citep{Muratov2015,Angles-Alcazar2017_BaryonCycle}.  At later times, star formation becomes less bursty and galactic winds less efficient \citep{Stern2020}, with the galaxy growing more steadily to $M_{\star} \gtrsim 2\times10^{11}$\,\Msun~down to $z=1$.
This transition from bursty to steady star formation has a significant impact on the nuclear gas reservoir: $\Sigma_{\rm gas}$ fluctuates by orders of magnitude at early times but there is a more steady nuclear gas reservoir at late times.  \citet{Angles-Alcazar2017_BHsOnFIRE} showed that this has important implications for black hole growth, which can be significantly suppressed at early times relative to the host galaxy growth (see also \citealt{Catmabacak2020}; \citealt{Hopkins2021_WindAccretionCorrection}; Byrne et al. in prep.).

We focus on three qualitatively distinct host galaxy conditions to study black hole feeding at ultra-high resolution, indicated in Figure~\ref{fig:ini} by vertical lines of different colors: 
\begin{itemize}[itemsep=-5pt]
\item We identify $z = 2.25$ (orange) as the time at which galaxy {\bf A4} experienced the highest nuclear gas surface density, with $\Sigma_{\rm gas}(<100\,{\rm pc}) > 10^{11}$\,\Mkpc~and global SFR $\sim$200\,\Msunyr.  These extreme conditions are the most likely to trigger gas inflow rates sufficient to power a luminous quasar, which we denote as ``\qso'' phase.\\

\item We investigate the conditions leading to the \qso~phase by going $\sim$40\,Myr back in time to $z = 2.28$ (green), with lower nuclear gas density $\Sigma_{\rm gas}(<100\,{\rm pc}) \sim 10^{10}$\,\Mkpc~but highly variable conditions reaching higher global SFR $\sim 500$\,\Msunyr~(the peak SFR is missed in the parent cosmological simulation owing to limited output time resolution).  We denote these as the ``\eqso" initial conditions. \\

\item We compare the $z>2$ conditions near the peak of star formation with the late time conditions at $z = 1.10$ (blue) with significantly lower activity, where SFR $\sim$10\,\Msunyr~and $\Sigma_{\rm gas}(<100\,{\rm pc}) < 10^{9}$\,\Mkpc.  These conditions are presumably representative of more common, low-luminosity AGN, which we denote as ``\agn".
\end{itemize}
As we will show throughout the paper, the \qso~phase reaches a quasi-steady state over the whole hyper-refinement simulation time while accretion in the \eqso~and \agn~phases is highly time variable.

\begin{deluxetable*}{lccccccccc}\label{tbl:sims}
\tablenum{1}
\tablecaption{Simulation parameters (units are physical): 
{\bf (1)} Name: simulation designation. 
{\bf (2)} $z$: initial redshift of hyper-refinement.
{\bf (3)} $M_{\rm vir}$: halo virial mass.
{\bf (4)} $M_{\star}$: galaxy stellar mass.
{\bf (5)} $M_{\rm gas}$: galaxy gas mass.
{\bf (6)} SFR: galaxy star formation rate.
{\bf (7)} $\Delta t$: total duration (physical run-time) of the simulation at the full hyper-refinement level.
{\bf (8)} $dt_{\rm min}$: minimum timestep of black hole particle.
{\bf (9)} $m_{\rm g}$: minimum mass of gas resolution element.
{\bf (10)} $h_{\rm sml}^{\rm min}$: minimum gas smoothing length.}
\tablewidth{0pt}
\tablehead{
\multicolumn{1}{l}{Name} &
\multicolumn{1}{c}{$z$} &
\multicolumn{1}{c}{$M_{\rm vir}$ [\Msun]} &
\multicolumn{1}{c}{$M_{\star}$ [\Msun]} &
\multicolumn{1}{c}{$M_{\rm gas}$ [\Msun]} &
\multicolumn{1}{c}{SFR [\Msunyr]} &
\multicolumn{1}{c}{$\Delta t$ [Myr]} &
\multicolumn{1}{c}{$dt_{\rm min}$ [yr]} &
\multicolumn{1}{c}{$m_{\rm g}$ [\Msun]} &
\multicolumn{1}{c}{$h_{\rm sml}^{\rm min}$ [pc]}
}
\startdata
\eqso  &  2.28  & 2.31e12  &  5.90e10  &  2.15e10  &  523   &  3   &   0.15  & 15  & 0.08 \\
\qso   &  2.25  &  2.37e12  &  8.21e10  &  1.87e10  &  222 &   4  &  0.42  &  30 &  0.08 \\
\agn   &  1.10  &  4.44e12  &  2.31e11  &  9.63e9    &  13   & 20  &  2.55  &   15 &  0.15 
\enddata
\end{deluxetable*}

\subsection{Hyper-Lagrangian refinement}\label{sec:split}

\begin{figure}
\begin{center}
\includegraphics[width=0.45\textwidth]{\pathL/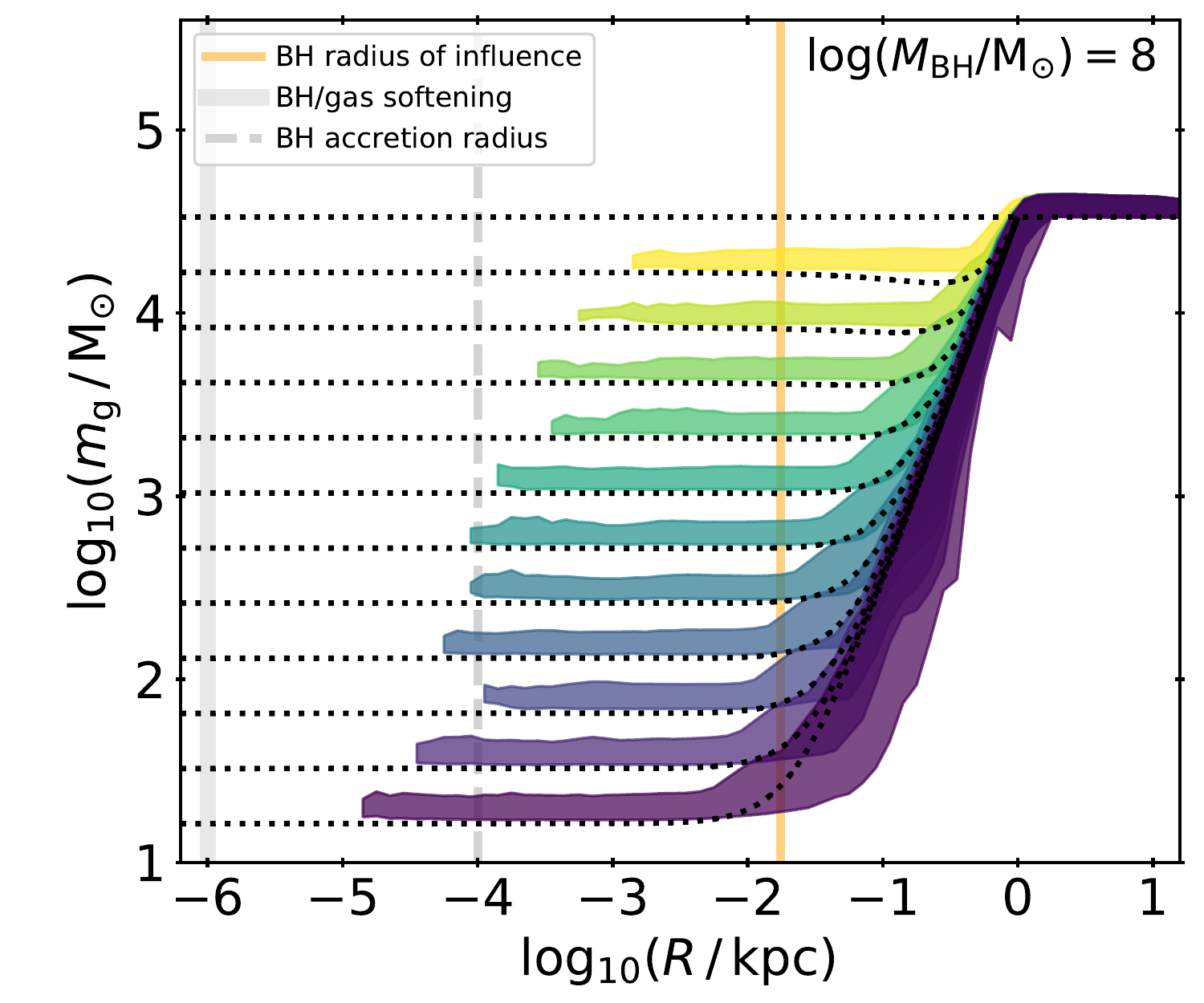}
\end{center}
\vspace{\vsf}
\caption{Gas mass resolution as a function of radial distance from the black hole for separate \eqso~simulations with increasing hyper-refinement level, from $\chi_{\rm ref}^{\rm min} = 1/2$ (yellow) to $\chi_{\rm ref}^{\rm min} = 1/2048$ (purple).  For each radial bin, shaded regions correspond to the 10\%--90\% percentiles of the distribution. 
Black dotted lines show the refinement condition $m_{\rm ref} \propto R^{2}$ for different target mass resolution near the black hole, normalized to match the original mass resolution of the cosmological zoom-in simulation at $R = 1$\,kpc.  
The vertical gray solid line indicates the softening length of the black hole particle ($\epsilon_{\rm BH}=0.001$\,pc), which is also the minimum (adaptive) force softening length for gas elements, while the vertical gray dashed line indicates the maximum size of the black hole kernel ($R_{\rm acc} = 0.1$\,pc). Gas elements at $R>R_{\rm acc}$ cannot be accreted by the black hole. The vertical orange line indicates the black hole radius of influence for $M_{\rm BH} = 10^8$\,\Msun~and \eqso~conditions.
Dynamic splitting improves the mass resolution from $m_{\rm g} = 3\times10^4$\,\Msun$\rightarrow 15$\,\Msun~in our highest resolution simulations.
}
\label{fig:ref}
\end{figure}

Lagrangian hydrodynamic methods are naturally adaptive, populating high density regions with a larger number of gas resolution elements.  In order to achieve super-Lagrangian refinement in MFM, we split gas resolution elements progressively as they approach the central black hole.
The distance $R$ from each gas particle to the black hole is computed while traversing the gravity tree 
\citep{Garrison-Kimmel2017}, which allows us to define a mass refinement factor $\chi_{\rm ref}(R)$ as an analytic function of $R$.
Gas resolution elements split into two new elements if $m_{\rm g} \geq \chi_{\rm ref}(R)\times 2\, m_{\rm b}$ for their current mass $m_{\rm g}$ and distance $R$, where $m_{\rm b}$ is the baryonic mass resolution in the original zoom-in simulation.  The mass of each new split element is $0.5\,m_{\rm g}$ and they are located at a distance $dr = {\rm min}(0.25\,h_{\rm sml}, 0.35\,d_{\rm ngb})$ from the position of the parent resolution element, in opposite sides along a random direction.  Here, $h_{\rm sml}$ is the smoothing length, $d_{\rm ngb}$ is the distance to the nearest neighbor, and the numerical factors are chosen to minimize perturbations to hydrodynamic quantities and avoid overlapping of fluid elements.    
The velocity of split elements is identical to that of the parent gas element, conserving total energy, momentum, and angular momentum in the splitting process.  We have also experimented with a first-order correction according to the existing velocity gradient, providing a more accurate reconstruction of the velocity field while introducing second order errors in the energy and angular momentum; we find no significant differences.  All other fluid quantities such as density and internal energy per unit mass remain unchanged, but are recomputed immediately after particle splitting following the standard formulation in MFM hydrodynamics for the new configuration of gas elements \citep{Hopkins2015_Gizmo}.
Split elements are allowed to merge onto higher mass elements when they are significantly under-massive relative to the resolution requirement, if $m_{\rm g} < \chi_{\rm ref}(R)\times m_{\rm b} / 200$, which is infrequent for the simulations presented here.
Our fiducial simulations only refine the gas component but we have also implemented splitting/merging of collisionless particles (see Appendix \ref{sec:appendix:gravtest} for details). 
In either case, newly formed stars have the same mass as the progenitor gas element regardless of the level of resolution.

Figure~\ref{fig:ref} shows the mass of gas resolution elements as a function of $R$ obtained for separate simulations using different levels of refinement for our fiducial $\chi_{\rm ref}(R)$ functional form:
\begin{equation}
\chi_{\rm ref} = 
\begin{cases}
 \chi_{\rm ref}^{\rm min} \times (1 - R/R_{\rm ref}) + (R/R_{\rm ref})^2  &  \text{if $R \leq R_{\rm ref}$} \\ 
 1 &  \text{if $R > R_{\rm ref}$}
\end{cases} 
\end{equation}
where $ \chi_{\rm ref}^{\rm min}$ is the desired refinement factor near the black hole, which we vary from $\chi_{\rm ref}^{\rm min} = 1/2$ (yellow) to $\chi_{\rm ref}^{\rm min} = 1/2048$ (purple), and $R_{\rm ref} = 1$\,kpc is the distance at which super-Lagrangian refinement begins to operate.  
The original baryonic mass resolution $m_{\rm b} = 3.3 \times 10^4$\,\Msun~is retained at $R>1$\,kpc, decreasing rapidly down to $m_{\rm g} \approx \chi_{\rm ref}\times m_{\rm b}$ closer to the black hole.  In our highest resolution simulations we reach $m_{\rm g} \approx 15$\,\Msun, which allows us to follow the inflowing gas well within the black hole radius of influence and down to the accretion aperture $R_{\rm acc} = 0.1$\,pc. 
Despite the strong gradient in resolution at $R =0.1$--1\,kpc, we do not see any significant artifacts owing to dynamical splitting and merging of gas elements, except for a short initial transient that we discard in the analysis.  We have experimented extensively with increasing the mass resolution slowly as the simulation evolves and also refining directly down to the highest resolution level; while the former minimizes the perturbations introduced, both methods yield similar long term behavior.  
We have also experimented with different radial functional forms for $\chi_{\rm ref}(R)$, but the flexibility of this particle splitting technique allows for hyper-refinement with any arbitrary geometry (see \citealt{Su2021_Jets} for hyper-refinement of magnetohydrodynamic jets propagating in a cluster environment).
 We find that $\chi_{\rm ref} \propto R^2$ provides a convenient mapping between standard resolution at 1\,kpc and $\chi_{\rm ref}^{\rm min}$ such that we reach uniform resolution at $\lesssim$100\,pc.

Table~\ref{tbl:sims} summarizes the main properties of our fiducial simulations.  The initial conditions are taken at each of the redshift snapshots corresponding to the \eqso, \qso, and \agn~conditions, where we remove all black hole particles other than the one located at the center of galaxy {\bf A4}.  The actual black hole mass is set to $M_{\rm BH} = 10^8$\,\Msun~such that we can compare the inflow rate under different conditions for the same $M_{\rm BH}$ (we test different $M_{\rm BH}$ values in additional simulations; see \S\ref{sec:bhdep}).  
We use the same gravitational softenings for dark matter, $\epsilon_{\rm DM} = 57$\,pc, but use fully-adaptive softenings for gas (matching the hydrodynamic solver) with an arbitrarily small minimum allowed (but in practice minimum softenings reached are given in Table~\ref{tbl:sims}), and treat the black hole correctly as a Keplerian point-like particle (using a much smaller softening, $\epsilon_{\rm BH}=0.001\,$pc, intentionally smaller than the ``accretion radius''). We adopt $\epsilon_{\star} = R_{\rm acc} = 0.1\,$pc for stars, motivated roughly by the softening of gas at the densities where stars form in the densest hyper-refined regions, but in Appendix \ref{sec:appendix:gravtest} we show that our results are robust relative to changes in $\epsilon_{\rm \star}$, including adaptive and variable softenings.
Hyper-refinement operates within $R_{\rm ref} = 1$\,kpc in all simulations presented here and we use fiducial maximum refinement factors $\chi_{\rm ref}^{\rm min} = 1/1024$ (\qso~phase) and $\chi_{\rm ref}^{\rm min} = 1/2048$ (\eqso~and \agn~phases).  Additional simulations with lower refinement levels are employed for resolution convergence tests (\ref{sec:appendix:res}).  
We save data snapshots for every 0.01\,Myr of evolution in all simulations.
Except for the radical increase in resolution and the treatment of black hole growth, our cosmological hyper-refinement simulations implement identical physics as the original FIRE-2 simulation.

We reach minimum timesteps $\sim$0.1\,yr and evolve our hyper-refinement simulations for $\Delta t \sim 3$--20\,Myr (Table~\ref{tbl:sims}).  
The initial conditions considered have characteristic dynamical times $t_{\rm dyn} \equiv R/v_{\rm c} \sim 1$--$2\,{\rm Myr} \rightarrow 0.1$--$0.2$\,Myr given the circular velocity $v_{\rm c} $ at $R\sim 1\,{\rm kpc} \rightarrow 100$\,pc scales.  We are thus evolving the simulations for $\sim$2--20 dynamical times at the scale where hyper-refinement starts (1\,kpc) and $\sim$20--200 dynamical times in the nuclear region ($\lesssim$100\,pc) within which the maximum resolution is achieved, with the inner 1--10\,pc evolving for hundreds to thousands of dynamical times.
Our simulations are thus reaching a reasonable statistical steady state in which the high resolution region matches the lower resolution conditions on larger scales in a physically meaningful way.  In addition, we find characteristic flow times $t_{\rm flow}\equiv M_{\rm gas} / \dot{M}_{\rm in} \sim 6$--22\,Myr given the enclosed gas mass and inflow rate at 1\,kpc (\S\ref{sec:profiles}), roughly similar to our total evolution times, which emphasizes the importance of including larger scale gas flows when evolving the nuclear region over many dynamical times.


\section{Overview of simulations}\label{sec:overview}

\subsection{From zoom-in initial conditions to hyper-refined galactic nuclei}\label{sec:resmaps}

\begin{figure}
\begin{center}
\includegraphics[width=0.45\textwidth]{\pathL/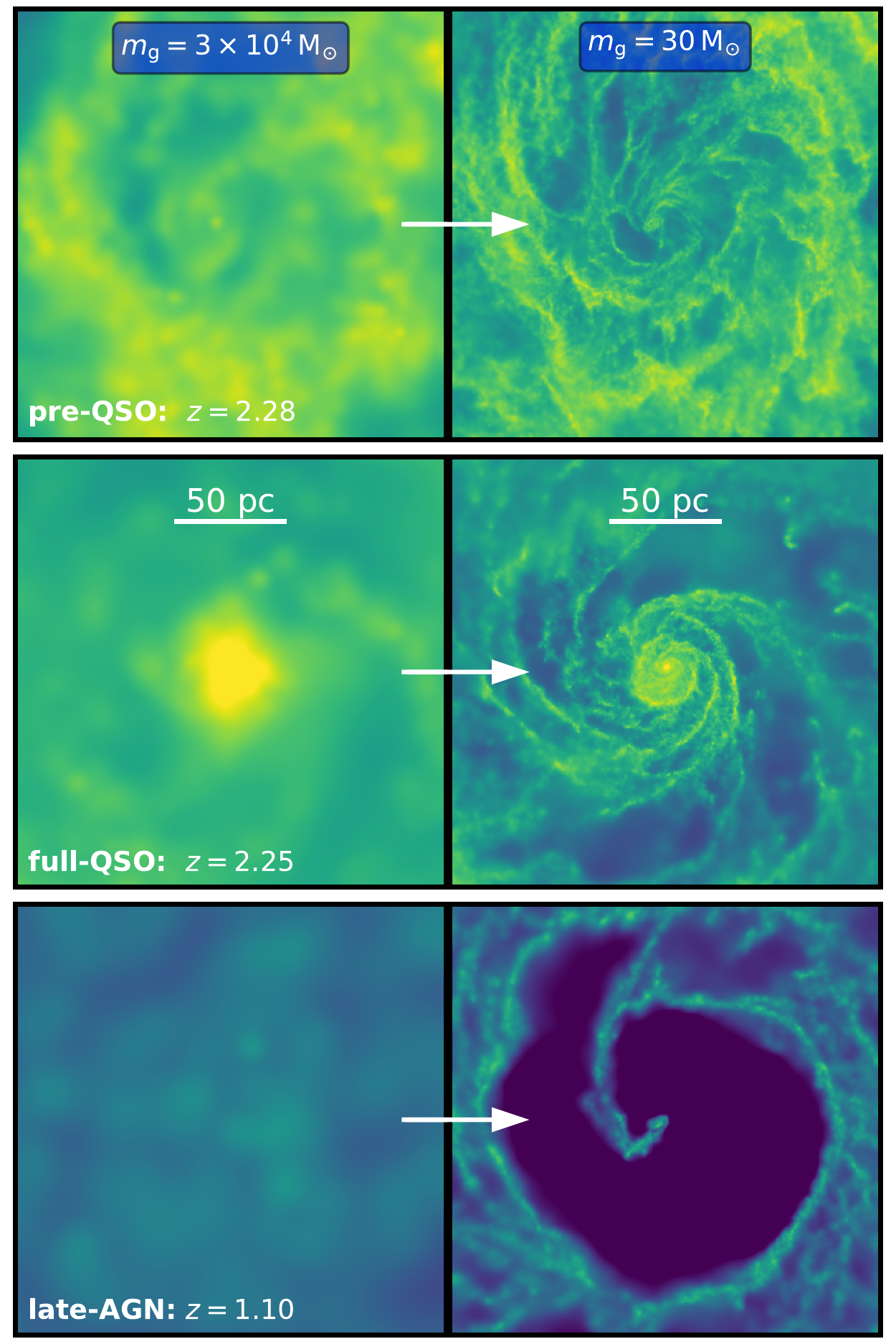}
\end{center}
\vspace{\vsf}
\caption{Projected gas mass surface density for a cubic volume of side 200\,pc around the central black hole in the \eqso~phase (top), \qso~phase (middle), and \agn~(bottom) simulations.  Left panels correspond to the original FIRE-2 simulation chosen for re-simulation, with mass resolution $m_{\rm b} = 3\times10^4$\,\Msun~\citep{Angles-Alcazar2017_BHsOnFIRE}.  
Right panels show the gas distributions obtained in the hyper-refinement simulations with $m_{\rm b} \sim 30$\,\Msun~at the same physical times (corresponding to $\sim [0.6,3.8,6.1]$\,Myr after the start of each hyper-refinement simulation, from top to bottom).
The color scale is logarithmic from $\Sigma_{\rm gas} = 10^{7.5}$\,\Mkpc~(purple) to $10^{10.5}$\,\Mkpc~(yellow).
}
\label{fig:resmap}
\end{figure}

Figure~\ref{fig:resmap} shows the gas distribution in the central 100\,pc for the three initial conditions considered, comparing the original resolution from the parent cosmological zoom-in simulation (left) and that obtained in the cosmological hyper-refinement simulations (right).  
Each pair of simulations is shown at the same evolution time, corresponding to $\sim$0.6\,Myr (\eqso), $\sim$3.8\,Myr (\qso), and $\sim$6.1\,Myr (\agn) after reaching mass resolution $m_{\rm b} \sim 30$\,\Msun~in the hyper-refinement simulation (we reach up to one resolution level higher in the \eqso~and \agn~conditions; Table~\ref{tbl:sims}).
The top panels correspond to the \eqso~conditions at $z=2.28$, with average $\Sigma_{\rm gas} \sim 10^{10}$\,\Mkpc~in the inner 100\,pc.  The initial conditions (top left) already capture a remarkable level of sub-structure for a massive galaxy simulated in a full cosmological context, with a clumpy, dense, circumnuclear ring and a hint of filamentary structures that appear to connect with a central gas concentration.  The hyper-refinement simulation (top right) confirms the presence of the ring-like structure in the nuclear region, which is now resolved by $N_{\rm gas} \sim 2\times 10^6$ resolution elements 
into a complex distribution of clumps, filaments, and spiral structures.

The middle panels of Figure~\ref{fig:resmap} show the nuclear gas distribution for the \qso~conditions at $z=2.25$, characterized by an extreme gas concentration reaching $\Sigma_{\rm gas} > 10^{11}$\,\Mkpc~in the central 10\,pc.  The parent zoom-in simulation suggests the presence of up to four nuclear spiral arms while the $\times1000$ increase in mass resolution yields a variety of morphological features with more than three orders of magnitude in density contrast, including an ultra-dense eccentric disc within $\sim 10$\,pc that breaks up into multiple spiral arms.
Finally, the bottom panels correspond to the \agn~conditions at $z=1.1$, where the parent zoom-in simulation shows a nearly homogeneous gas distribution with average $\Sigma_{\rm gas} \lesssim 10^{9}$\,\Mkpc~in the inner 100\,pc.  The cosmological hyper-refinement simulation yields a remarkably different gas distribution, with a collection of dense clumps and filaments embedded in a low density medium and the formation of a central cavity extending a few tens of pc.

\subsection{Mapping gas from Mpc to pc scales}\label{sec:maps}

\begin{figure*}
\begin{center}
\includegraphics[width=1\textwidth]{\pathL/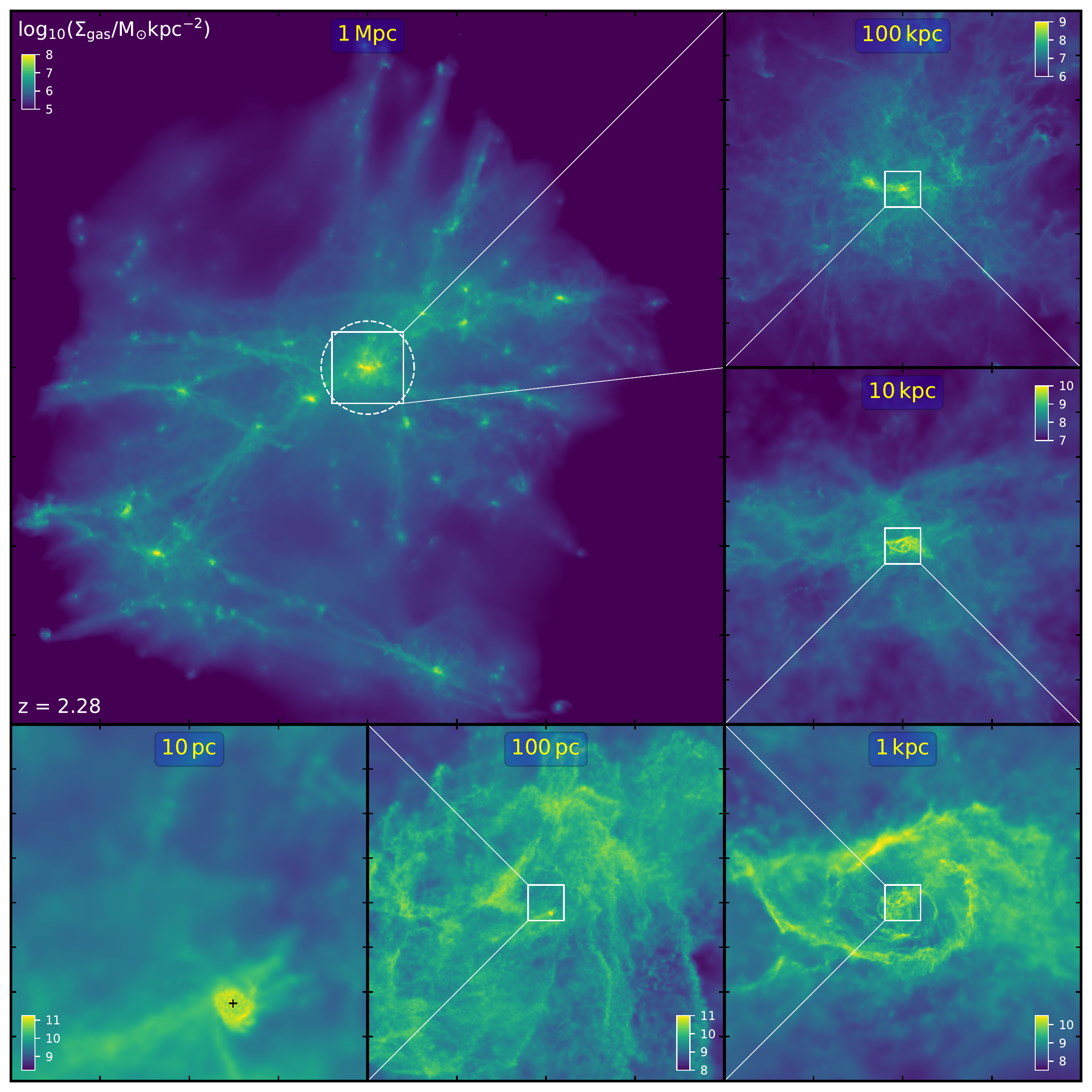}
\end{center}
\vspace{\vsf}
\caption{
Multi-scale gas mass surface density distribution for the \eqso~conditions at $z=2.28$, corresponding to $\sim$1.5\,Myr after the start of the hyper-refinement simulation.
The top left panel shows the central 1\,Mpc, with the white dashed line indicating $R_{\rm vir} \sim 130$\,kpc of the central halo.  Subsequent panels progressively zoom into the central 10\,pc of the main galaxy, with dynamic hyper-refinement occuring at $<$1\,kpc.  
All panels are centered on the center of mass of the stellar component within the inner 1\,kpc of the main halo.
The location of the central massive black hole is indicated by a + symbol in the lower left panel.  The simulation captures $>6$ orders of magnitude variation in gas surface density ($\Sigma_{\rm gas} \sim 10^{5-11}$\,\Mkpc) and a dynamic range spanning over 6 orders of magnitude in spatial scale.
}
\label{fig:zoom_eqso} 
\end{figure*}

\begin{figure*}
\begin{center}
\includegraphics[width=1\textwidth]{\pathL/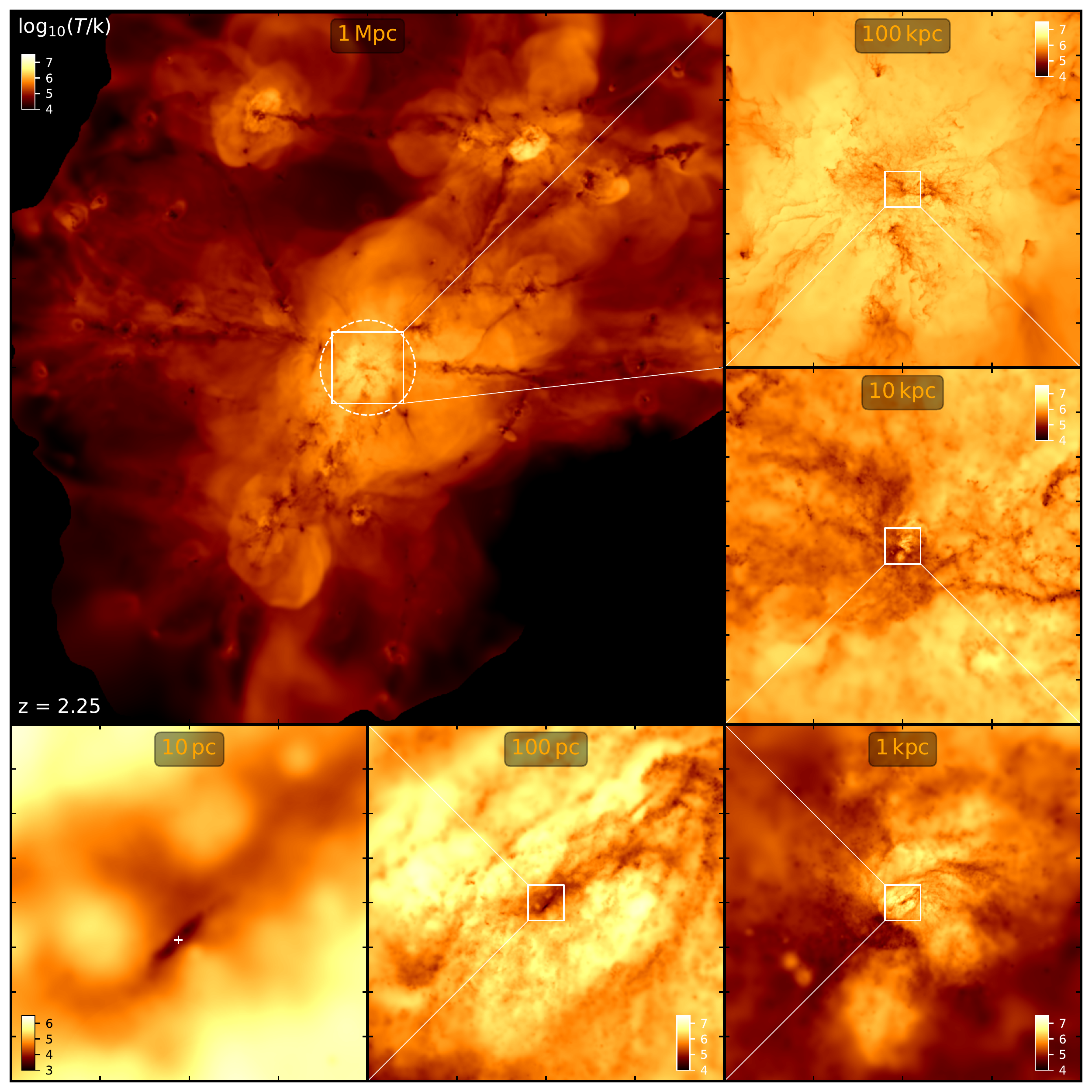}
\end{center}
\vspace{\vsf}
\caption{
Multi-scale mass-weighted temperature distribution for the \qso~conditions at $z=2.25$, corresponding to $\sim$3.6\,Myr after the start of the hyper-refinement simulation.
The top left panel shows the central 1\,Mpc, with the white dashed line indicating the virial radius of the central halo.  Subsequent panels progressively zoom into the central 10\,pc, where the location of the massive black hole is indicated by a + symbol.
We model simultaneously large scale filaments of cool gas penetrating the virial shock of dark matter halos all the way down to multi-phase ISM gas in the central 10\,pc, where the black hole accretes from a cold, pc-scale gas disk embedded in a hot medium.
See Figure~\ref{fig:qsomap} for the corresponding multi-scale gas mass surface density distribution in the \qso~phase.
}
\label{fig:zoom_qso} 
\end{figure*}

\begin{figure*}
\begin{center}
\includegraphics[width=1\textwidth]{\pathL/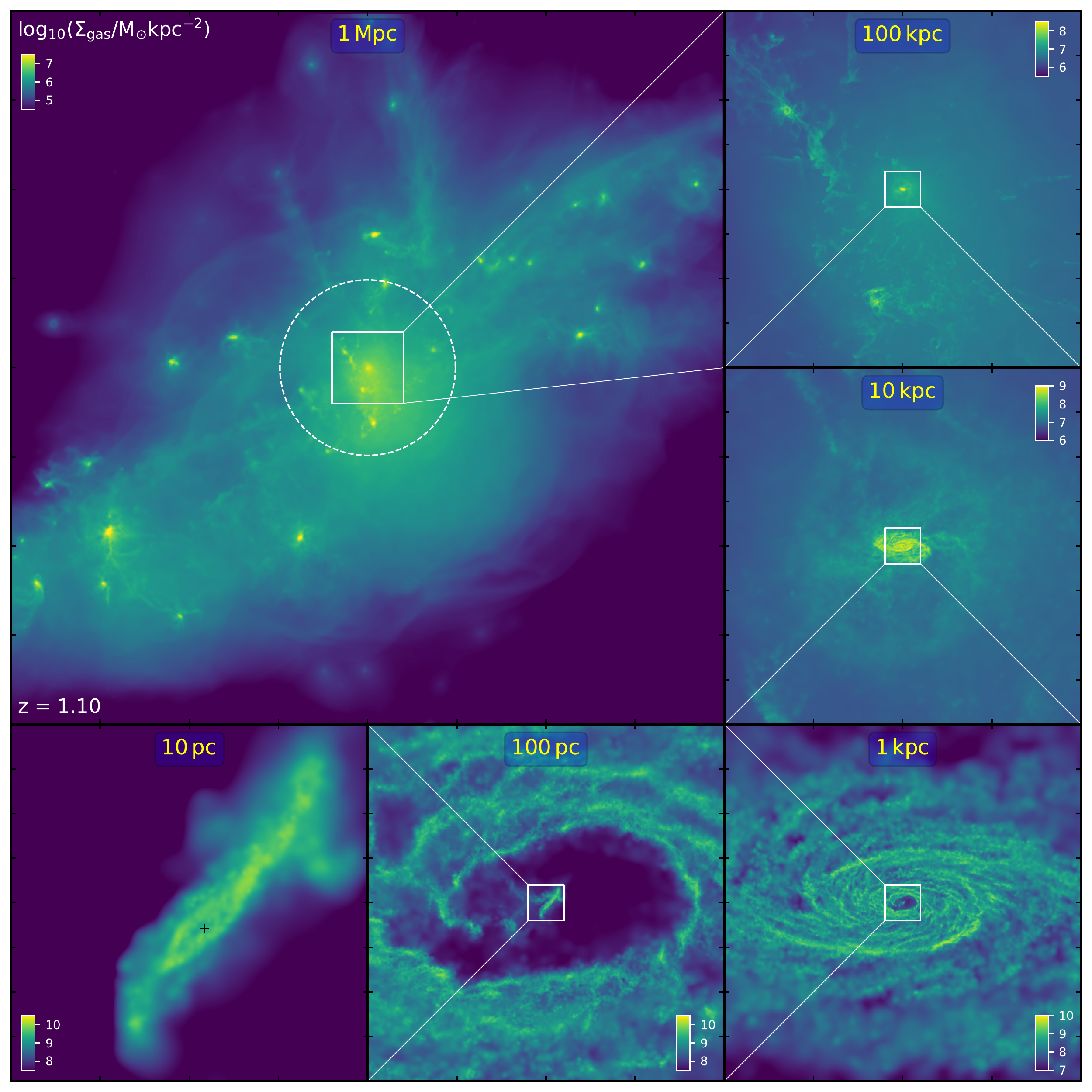}
\end{center}
\vspace{\vsf}
\caption{
Multi-scale gas mass surface density distribution for the \agn~conditions at $z=1.1$, corresponding to $\sim$6.2\,Myr after the start of the hyper-refinement simulation.
The top left panel shows the central 1\,Mpc with the white dashed line indicating $R_{\rm vir} \sim 250$\,kpc of the central halo. Subsequent panels progressively zoom into the central 10\,pc of the main galaxy.  The location of the central black hole is indicated by a + symbol in the lower left panel.  A low density cavity forms in the central 100\,pc of the thin, rotationally supported gas disk, suppressing black hole growth.  Gravitational instabilities in the inner edge of the circumnuclear disk drive gas clumps and filaments toward the central 10\,pc, occasionally feeding the black hole.   
}
\label{fig:zoom_agn} 
\end{figure*}

Figure~\ref{fig:zoom_eqso} illustrates the full dynamic range of our simulations by showing the gas distribution for the \eqso~conditions on different scales, from 1\,Mpc to the central 10\,pc.
The top left panel shows the full extent of the zoom-in region, which is populated with high resolution dark matter and gas elements and embedded in a low-resolution volume of $[100\,\hmpc]^3$ that allows us to capture tidal torques from large scale structures.  High density peaks correspond to the location of galaxies, tracing large scale filaments.  The white dashed circle in the center indicates the virial radius of the main halo, with $R_{\rm vir} = 130$\,kpc and  $M_{\rm vir} = 2.3\times10^{12}$\,\Msun~at $z=2.28$.  Zooming into the central 100\,kpc, the CGM shows a complex gas distribution owing to the interplay between filamentary accretion from the IGM, cooling of hot halo gas, and galactic winds from the central galaxy and satellites \citep[e.g.][]{Hafen2019_CGMorigins,Hafen2020_CGMfates,Fielding2020}.  A merging satellite galaxy ($M_{\star} \sim 10^{10} $\,\Msun) is approaching the central ($M_{\star} = 5.9\times10^{10} $\,\Msun), currently on its second passage at a distance of $\sim 20$\,kpc and playing a minor role in driving instantaneous black hole growth (see \S\ref{sec:tordis}).

On 10\,kpc scales, the ISM shows a very irregular morphology, with gas surface densities in the range $\Sigma_{\rm gas} \sim 10^{7-10}$\,\Mkpc.
The highest densities are reached within the inner 1\,kpc in a clumpy ring-like structure of radius $\sim400$\,pc, with a secondary gas concentration forming in the inner 100\,pc.  The black hole responds dynamically to the turbulent clumpy gas distribution by orbiting up to 10\,pc away from the center of mass of the stellar component.  Gravitational capture of infalling clumps and filaments yields pc-scale gas disks around the black hole with changing orientation depending on the angular momentum of the infalling material.

Figure~\ref{fig:zoom_qso} shows projected mass-weighted temperature distributions from 1\,Mpc to 10\,pc scales for the \qso~conditions at $z=2.25$.  In this case, the Mpc scale view in the top left panel highlights the thin, large scale filaments of cool gas ($T<10^5$\,K) feeding the central halo and the expanding hot gas ($T>10^6$\,K) heated by accretion shocks and galactic winds, reaching well beyond  $R_{\rm vir} \gtrsim 130$\,kpc.  On 100\,kpc scales, the CGM shows a prominent multi-phase structure with cold clumps and filaments embedded in a hot medium of virialized gas ($T>10^{6.5}$\,K).  The projected temperature distribution in the central 10\,kpc is highly irregular, with a large covering fraction of cold, star forming gas.  Dust continuum radiative transfer calculations show that these conditions produce a submm-bright phase \citep{Cochrane2019}.
Zooming into the inner 1\,kpc, cold gas with low angular momentum falls toward the inner region while nuclear spiral arms form out of the turbulent ISM.  In the central 100\,pc, the simulation captures $\sim 4$ orders of magnitude variation in projected mass-weighted temperature, with prominent spiral structures of $T\sim10^{4-5}$\,K gas embedded in $T > 10^7$\,K gas heated by SNe.  The bottom left panel highlights the temperature contrast in the inner 10\,pc, where a cold ($T\sim 1000$\,K), pc-scale, rotationally supported gas disk forms around the black hole.  
See Figure~\ref{fig:qsomap} for the corresponding gas mass surface density distributions in the \qso~phase.

Figure~\ref{fig:zoom_agn} shows again projected gas surface density distributions on multiple scales, as in Figure~\ref{fig:zoom_eqso}, but in this case for the \agn~conditions at $z=1.1$.  The main halo has doubled in mass since $z\sim 2.3$, with $M_{\rm vir} = 4.4\times10^{12}$\,\Msun~and $R_{\rm vir} \sim 250$\,kpc, and the central galaxy has developed a massive stellar component ($M_{\star} = 2.3\times10^{11}$\,\Msun).  The CGM on 100\,kpc scales shows a smoother gas distribution compared to earlier times, though some dense gas structures are still prominent.  With the gas fraction within 0.1\,$R_{\rm vir}$ falling below $f_{\rm gas} < 5$\%~and $SFR\sim 10$\,\Msunyr, the galaxy forms a thin, rotationally supported disk in the inner $\sim 2$\,kpc.  A low-density contrast, bipolar structure appears to extend a few kpc above and below the plane of the disk, likely driven by an earlier episode of galactic winds.  Zooming into the central 1\,kpc, the disk structure is dominated by star-forming regions with $\Sigma_{\rm gas} > 10^{9}$\,\Mkpc~embedded in a low density medium, with a compact spiral structure superimposed.
In the central 100\,pc, gas consumption by star formation and ejection by stellar feedback form a cavity that extends a few tens of pc and starves the black hole of fuel \citep[see also][]{Angles-Alcazar2017_BHsOnFIRE}.  Gravitational torques from the stellar component drive gas clumps and filaments from the inner boundary of the circumnuclear disk down to the central 10\,pc, occasionally feeding the black hole (see \S\ref{sec:torque}).

\subsection{Galaxy radial structure and kinematics}\label{sec:profiles}

\begin{figure}
\begin{center}
\includegraphics[width=0.49\textwidth]{\pathL/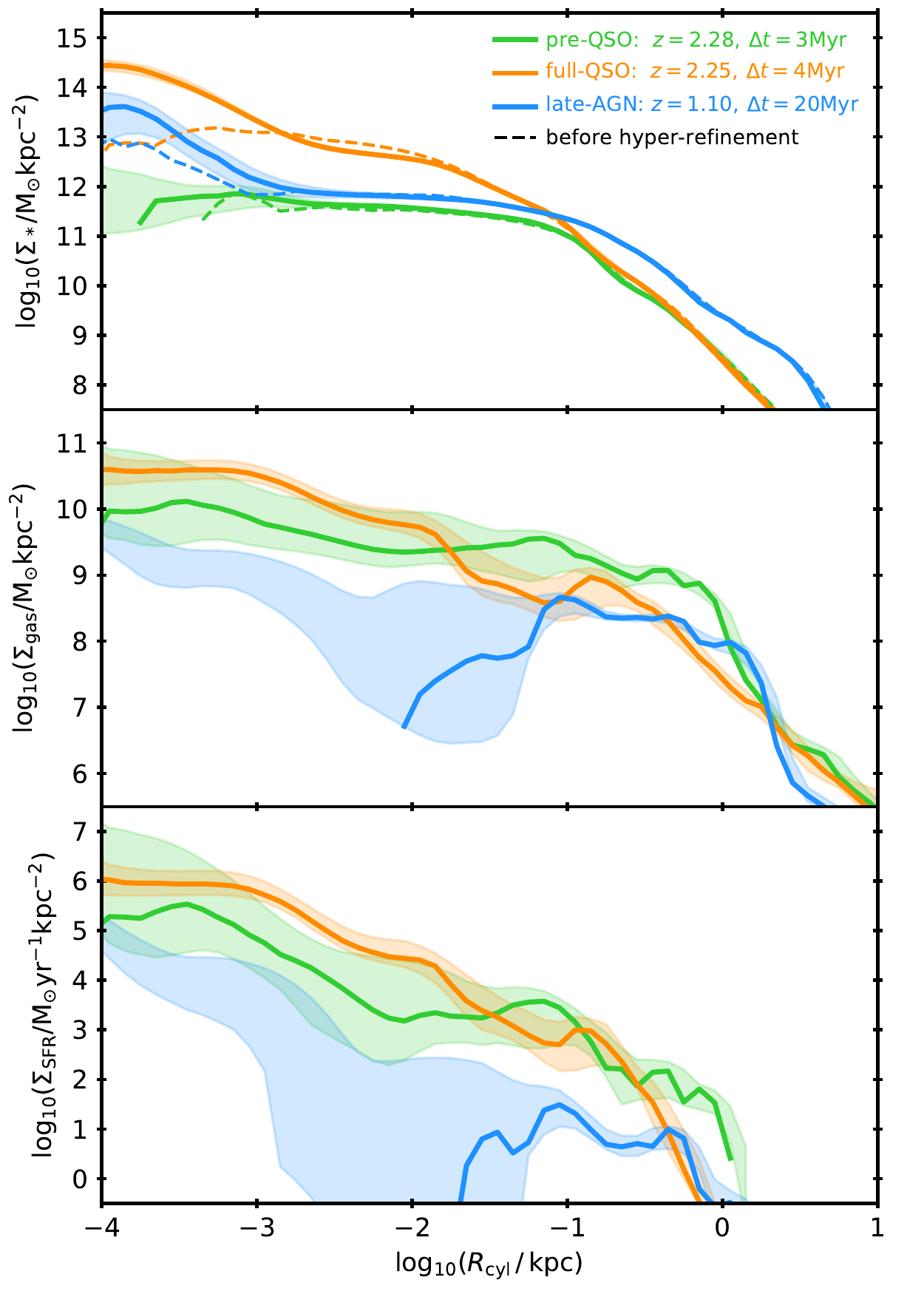}
\end{center}
\vspace{\vsf}
\caption{Stellar mass surface density (top), gas mass surface density (middle), and SFR surface density (bottom) as a function of cylindrical distance $R_{\rm cyl}$ from the black hole (defined with respect to the angular momentum axis of gas/stars at $R<R_{\rm cyl}$; see text) for the \eqso~(green), \qso~(orange), and  \agn~(blue) conditions.  Solid lines show median values in each radial bin considering the full evolution time $\Delta t$ at hyper-refined resolution, 
while shaded regions indicate the 10--90\% percentile range including only timesteps with non-zero values in each radial bin.  Dashed lines in the top panel correspond to the initial $\Sigma_{\star}$ at the start of the hyper-refinement simulations.
}
\label{fig:struct_vs_r}
\end{figure}

Figure~\ref{fig:struct_vs_r} shows the stellar, gas, and star formation rate surface densities as a function of cylindrical radial distance $R_{\rm cyl}$ for the three conditions simulated.  We define a cylindrical coordinate system (centered on the black hole) independently for each time and radial bin, where the z-axis is aligned with the total angular momentum inside of the 3D radial distance $R = R_{\rm cyl}$ to the black hole, also performed independently for the gas and stellar components.  Surface densities $\Sigma_{\star}$, $\Sigma_{\rm gas}$, and $\Sigma_{\rm SFR}$ are then computed within these local cylindrical radial bins (logarithmically spaced) by integrating the mass/SFR within $\pm100$\,pc along the vertical direction relative to the local cylindrical plane.  This allows us to capture spatial and temporal variations in the structure and kinematics of the galaxy, but our results are not sensitive to the exact choice of local radial bins, their vertical extension relative to the local galaxy plane, or the use of spherical rather than cylindrical radial bins. 
Note that the black hole and therefore the adopted reference frame can move relative to the center of mass of the galaxy  (Figure~\ref{fig:zoom_eqso}), with subsequent analysis reflecting the instantaneous conditions driving black hole growth. 
We compute median surface density profiles considering the full time evolution of each simulation.  In addition, we extract the 10--90\% range of variation including only timesteps with non-zero values in each radial bin, which allows us to study infrequent events such as the transport of gas clumps across the central cavity that forms in the \agn~phase.

The \eqso~conditions at $z=2.28$ (green) show a significant increase in stellar mass surface density $\Sigma_{\star} \sim 10^8 \rightarrow 10^{11}$\,\Mkpc~in the range $R_{\rm cyl} = 1 \rightarrow 0.1$\,kpc, with a mild increase to $\Sigma_{\star} \sim 10^{11.7}$\,\Mkpc~at smaller scales.  The initial stellar distribution (dashed line) is very similar to the median $\Sigma_{\star}$ found during the $\sim 3$\,Myr time evolution in the hyper-refinement simulation (solid line), but it does not extend to within the inner 1\,pc.  
New stars formed out of ultra-high resolution gas populate the inner $\sim$1\,pc region, but the stellar distribution on larger scales does not change significantly over the short $\sim$3\,Myr simulation period.
The shaded region indicates the 10--90\% range for each radial bin, reflecting the build up of stellar mass within 1\,pc, reaching $\Sigma_{\star} \sim 10^{12.5}$\,\Mkpc, but also variability in $\Sigma_{\star} $ owing to black hole dynamics relative to the stellar distribution. 
The \qso~phase at $z=2.25$ (orange), $\sim$40\,Myr later, shows very similar $\Sigma_{\star}$ at $>100$\,pc but significantly higher values in smaller scales relative to the \eqso~conditions.  The initial \qso~conditions had already reached $\Sigma_{\star} \sim 10^{13}$\,\Mkpc~within $\sim$10\,pc, increasing to even more extreme densities as new ultra-high resolution stars form in the inner $\sim$1\,pc.
Moving forward to $z=1.1$, the \agn~conditions (blue) show a more extended stellar distribution owing to size growth and expansion of the inner region.  Within $\sim$100\,pc, $\Sigma_{\star}$ is roughly $\times10$ lower compared to the \qso~conditions but the original stellar distribution retained the central core with $\Sigma_{\star} \sim 10^{13}$\,\Mkpc, further increasing during the $\sim$20\,Myr hyper-refinement simulation period.
The extreme stellar densities reached in the \qso~and \agn~conditions are likely unphysically large owing to the lack of AGN feedback.

The middle and bottom panels of Figure~\ref{fig:struct_vs_r} show the radial gas and SFR distributions. 
The highest gas surface densities on galactic scales ($>$100\,pc) are reached in the \eqso~conditions, with $\Sigma_{\rm gas} \gtrsim 10^9$\,\Mkpc~in the central 1\,kpc and $\Sigma_{\rm SFR} \sim 10 \rightarrow1000$\,\Myrkpc~increasing inward.  Within $\sim$1\,pc around the black hole, the \eqso~phase maintains a median $\Sigma_{\rm gas} \gtrsim 10^{10}$\,\Mkpc~over the $\sim$3\,Myr simulation period, with variations of over two orders of magnitude, and $\Sigma_{\rm SFR} \sim 10^5$--$10^7$\,\Myrkpc. 
The \qso~phase shows lower $\Sigma_{\rm gas}$ on galactic scales but more steady and significantly larger gas and SFR surface densities ($\sim$0.5 dex) at $<$10\,pc, with median $\Sigma_{\rm gas} \sim 10^{10.5}$\,\Mkpc~and $\Sigma_{\rm SFR} \sim 10^6$\,\Myrkpc~in the central 1\,pc.
The \agn~conditions show median $\Sigma_{\rm gas} \gtrsim 10^8$\,\Mkpc on galactic scales but a rapid decline at $<$100\,pc owing to the formation of a low-density central cavity.  Nonetheless, gas transported down to $<$1\,pc due to instabilities in the outer edge of the cavity can reach $\Sigma_{\rm gas} > 10^9$\,\Mkpc, with star formation happening all the way to the black hole accretion radius.

\begin{figure}
\begin{center}
\includegraphics[width=0.49\textwidth]{\pathL/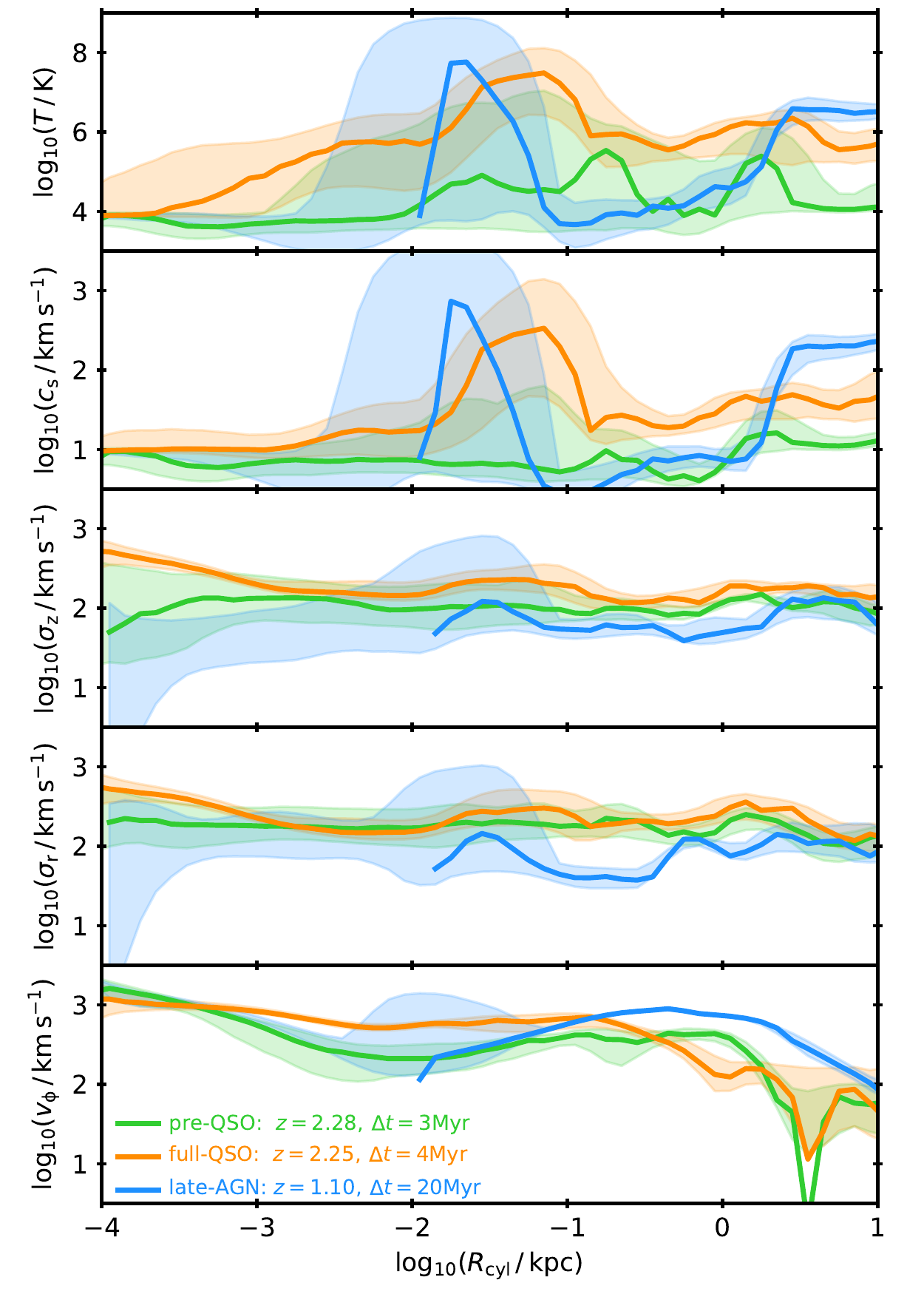}
\end{center}
\vspace{\vsf}
\caption{From top to bottom: average (mass-weighted) gas temperature, sound speed, vertical velocity dispersion, radial velocity dispersion, and azimuthal velocity as a function of cylindrical radial distance from the black hole for the \eqso~(green), \qso~(orange), and  \agn~(blue) conditions.  Solid lines show median values in each radial bin over the full evolution time $\Delta t$, while shaded regions indicate the 10--90\% percentile range including only timesteps with non-zero gas mass in each radial bin.  
}
\label{fig:kin_vs_r}
\end{figure}

Figure~\ref{fig:kin_vs_r} investigates the kinematic and thermal properties of gas from $R_{\rm cyl} = 0.1$\,pc--10\,kpc for our \eqso~(green), \qso~(orange), and  \agn~(blue) simulations.  The top two panels show the average (mass-weighted) temperature ($T$) and sound speed ($c_{\rm s}$) within each cylindrical radial bin, extending only $\pm10$\,pc along the vertical direction relative to the local plane.  As in Figure~\ref{fig:struct_vs_r}, solid lines show median values for the full hyper-refinement simulation time and shaded regions indicate the 10--90\% range variation in each radial bin. 
The \eqso~conditions are dominated by cool gas ($T\lesssim 10^5$\,K) in the entire range $R_{\rm cyl} = 0.1$\,pc--10\,kpc owing to the large amount of dense gas and thus rapid cooling times.  With average $c_{\rm s} \sim 10$\,\kms, thermal pressure support does not play a relevant dynamical role.  Nonetheless, heating by SNe following intense star formation can sometimes increase the temperature to $T > 10^6$\,K in the inner $\sim$\,100\,pc; note that the black hole can also move from cold to hot dominated regions as it orbits within the central 10\,pc.  The thermal properties in the \qso~conditions are rather different, with average $T\sim 10^6$\,K on galaxy scales and $T\gtrsim 10^7$\,K consistently in the central $\sim$10--100\,pc, where there is a stark contrast between the temperatures reached by the cold and hot ISM phases (see Figure~\ref{fig:zoom_qso}).  Despite this, gas within $\sim$1\,pc of the black hole is predominantly cold.  By the time we reach the \agn~conditions at $z=1.1$, the galaxy has already developed a hot halo and we find $T \sim 10^{6.5}$\,K at $\sim$10\,kpc scales, dropping to $T \sim 10^{4}$\,K within the $\sim$\,kpc scale gas disk.  At $<$100\,pc, the low-density central cavity seen in Figure~\ref{fig:zoom_agn} reaches $T \sim 10^{8}$\,K and $c_{\rm s}\sim 1000$\,\kms, while gas that penetrates the cavity down to the black hole accretion radius is predominantly cold.

\begin{figure*}
\begin{center}
\includegraphics[width=0.99\textwidth]{\pathL/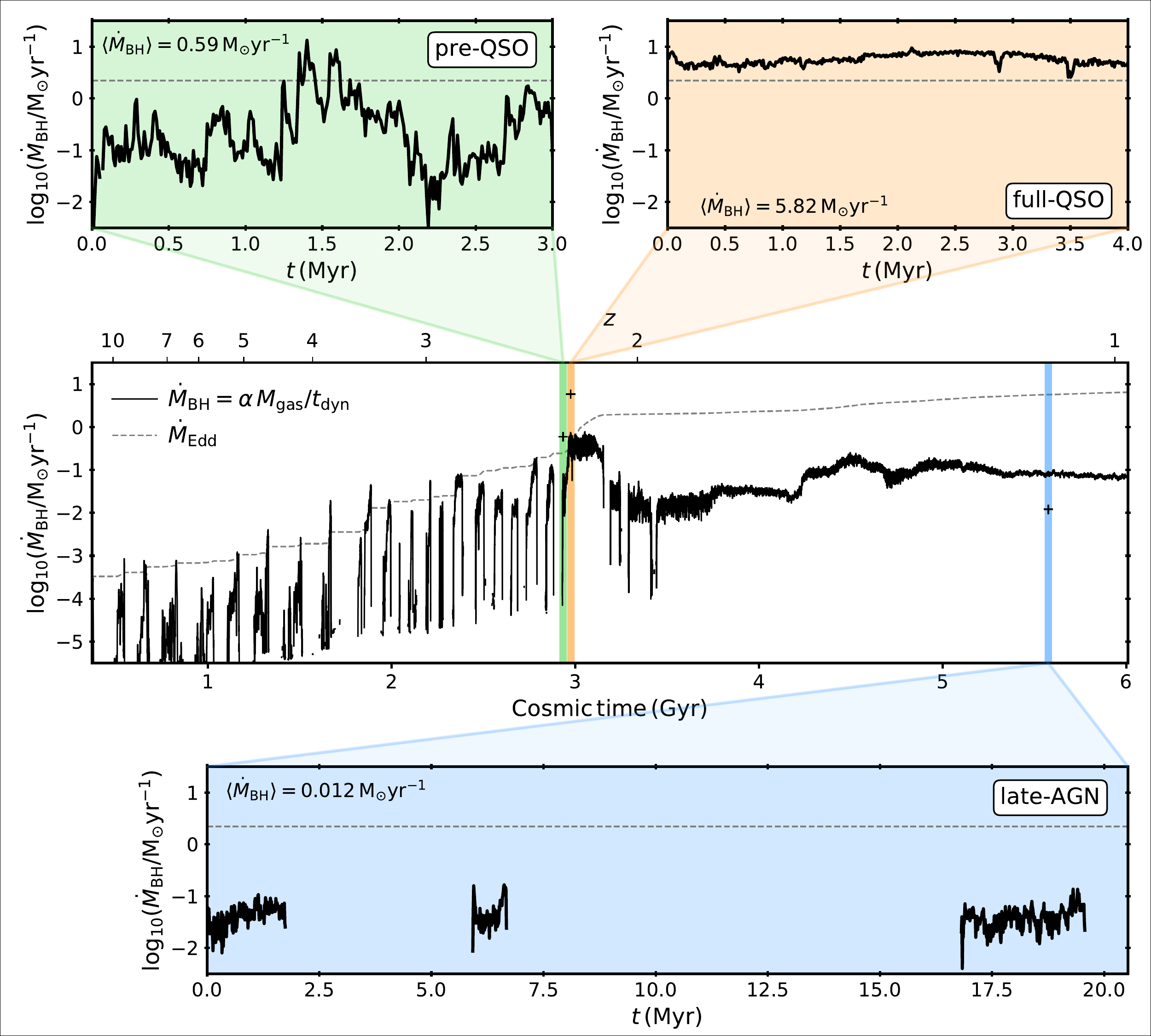}
\end{center}
\caption{
Black hole accretion history from $z = 10 \rightarrow 1$ inferred from sub-grid modeling in a full cosmological simulation (middle panel) compared to the explicit gas inflow rates measured at $R_{\rm acc}=0.1$\,pc in re-simulations with dynamic hyper-refinement (colored panels).
The black solid line in the middle panel shows the free-fall accretion estimator $\dot{M}_{\rm BH} = \alpha \, M_{\rm gas}/t_{\rm dyn}$ evaluated for the central black hole in halo {\bf A4} from \citet{Angles-Alcazar2017_BHsOnFIRE}, where $M_{\rm gas}$ is the gas mass within $R_0\sim100$\,pc, $t_{\rm dyn} = (R_{0}^{3}/GM_{\rm tot})^{1/2}$, $M_{\rm tot}$ is the total mass within $R_0$, and $\alpha \equiv 5\times10^{-4}$ is set such that $M_{\rm BH}/M_{\rm bulge} \sim 0.002$ at $z=1$.  The dashed gray line shows the Eddington rate ($\dot{M}_{\rm Edd}$) corresponding to the growing black hole.  Vertical lines indicate the redshifts chosen for re-simulation with dynamic hyper-refinement for the \eqso~($z = 2.28$; green), \qso~($z = 2.25$; orange), and \agn~($z = 1.10$; blue) conditions.  
Inset panels show the explicit accretion histories measured at $R_{\rm acc}=0.1$\,pc in each case for re-simulations considering an initial black hole mass $M_{\rm BH}^{\rm ini} = 10^8$\,\Msun, with horizontal gray dashed lines indicating $\dot{M}_{\rm Edd}(M_{\rm BH}^{\rm ini})$.  Average accretion rates for re-simulation periods are shown in each inset panel and indicated as + symbols in the middle panel.
Dynamic hyper-refinement simulations capture over four orders of magnitude variability in $\dot{M}_{\rm BH}$ and predict qualitatively distinct phases of black hole growth, with (1) strong variability in \eqso~conditions with $\dot{M}_{\rm BH} = 0.001$--10\,\Msunyr, (2) steady super-Eddington feeding in the \qso~phase (in the absence of AGN feedback), and (3) low luminosity in the \agn~conditions with $\sim$25\% duty cycle.
}
\label{fig:mdot_vs_z}
\end{figure*}

The three bottom panels of Figure~\ref{fig:kin_vs_r} show the radial dependence of the vertical velocity dispersion ($\sigma_{\rm z}$), the radial velocity dispersion ($\sigma_{\rm r}$), and the azimuthal velocity ($v_{\phi}$) in cylindrical coordinates for the gas component.   
Velocity dispersions are defined as $\sigma_{\rm z}^2 \equiv \langle v_{\rm z}^{2} - \langle v_{\rm z} \rangle^{2} \rangle$ and $\sigma_{\rm r}^2 \equiv \langle v_{\rm r}^{2} - \langle v_{\rm r} \rangle^{2} \rangle$.
The \eqso~conditions show remarkably flat velocity dispersion profiles, with $\sigma_{\rm z} \sim 100$\,\kms~in the entire range $R_{\rm cyl} = 0.1$\,pc--10\,kpc and slightly higher dispersion $\sigma_{\rm r}$ in the radial direction \citep[see also][]{Wellons2020}.  The galaxy-scale azimuthal velocity reaches $v_{\phi} \sim 400$\,\kms~at $\sim 100$\,pc--1\,kpc.  In the central 10\,pc, the azimuthal velocity increases to $>1000$\,\kms~with $v_{\phi} \propto R^{-1/2}$, as expected for Keplerian rotation within the black hole radius of influence.       
With increased stellar and gas mass surface densities, the \qso~conditions show even higher levels of turbulence.  In this case, both $\sigma_{\rm z}$ and $\sigma_{\rm r}$  increase from $\sim 200$\,\kms~on galaxy scales to $\sim 500$\,\kms~at $\sim 0.1$\,pc.  Nonetheless, rotational motion still dominates with $v_{\phi} > 600$\,\kms~at $R_{\rm cyl}<100$\,pc.
At later times, the rotational velocity in the \agn~conditions reaches $v_{\phi} > 800$\,\kms~on kpc scales owing to the build up of a massive stellar component.  The gas distribution shows a rotationally supported disk with $\sim 50$\,\kms~turbulent motions.  The cool gas that penetrates the inner hot cavity (blue shaded region within $\sim$10\,pc) follows a roughly Keplerian rotation curve with $v_{\phi} \propto R^{-1/2}$.

Overall, despite the high level of turbulent motions in the \eqso~and \qso~conditions and the role of thermal pressure in the \agn~central cavity, rotational support appears to dominate in the full range of scales analyzed, with 
$c_{\rm s} < \sigma_{\rm z} \sim \sigma_{\rm r} < v_{\phi}$.   
The \qso~conditions reach extreme stellar densities (Figure~\ref{fig:struct_vs_r}), higher than the maximum stellar surface density predicted by analytic models due to the failure of stellar feedback \citep{Hopkins2010_MaxStellarDen,Grudic2019}.
In addition, the large $v_{\phi}$ velocities reached in the \agn~conditions are comparable only to the most extreme compact galaxies observed \citep[e.g.][]{vanDokkum2015,Diamond-Stanic2021}, suggesting the need for AGN feedback or some other additional mechanism to suppress nuclear star formation \citep{Wellons2020,Parsotan2021}.
We investigate next if large inflow rates down to sub-pc scales can occur, a prerequisite for efficient ``QSO-mode'' black hole feedback.

\section{Multi-scale gas Inflows}\label{sec:multi}

\subsection{Explicit accretion at $<0.1$\,pc}\label{sec:mdot}

Figure~\ref{fig:mdot_vs_z} shows $\dot{M}_{\rm BH}$ as a function of time measured at 0.1\,pc in hyper-refinement simulations for the \eqso~($z = 2.28$; green), \qso~($z = 2.25$; orange), and \agn~($z = 1.10$; blue) conditions, considering in all cases a black hole with initial mass $M_{\rm BH}^{\rm ini} = 10^8$\,\Msun. 
For context, the middle panel shows the black hole accretion history from $z = 10 \rightarrow 1$ obtained by post-processing the original cosmological zoom-in simulation adopting the ``free-fall" sub-grid accretion estimator $\dot{M}_{\rm BH} = \alpha \, M_{\rm gas}/t_{\rm dyn}$, where $t_{\rm dyn} = (R_{0}^{3}/GM_{\rm tot})^{1/2}$ is the dynamical time within $R_0\sim100$\,pc and $M_{\rm gas}$ and $M_{\rm tot}$ are the gas mass and total mass within $R_0$.  We define $\alpha \equiv 5\times10^{-4}$ such that $M_{\rm BH}/M_{\rm bulge} \sim 0.002$ at $z=1$, which results in accretion rates roughly consistent with the gravitational torque accretion estimator used in \citet{Angles-Alcazar2017_BHsOnFIRE}, but the characteristic accretion history obtained is independent of this choice.
In the absence of AGN feedback, the sub-grid accretion model predicts short accretion episodes that can reach or even exceed the Eddington rate at early times ($z > 2.3$) and a transition to more steady accretion at late times, resembling the evolution in nuclear gas surface density shown in Figure~\ref{fig:ini}.

\begin{figure}
\begin{center}
\includegraphics[width=0.45\textwidth]{\pathL/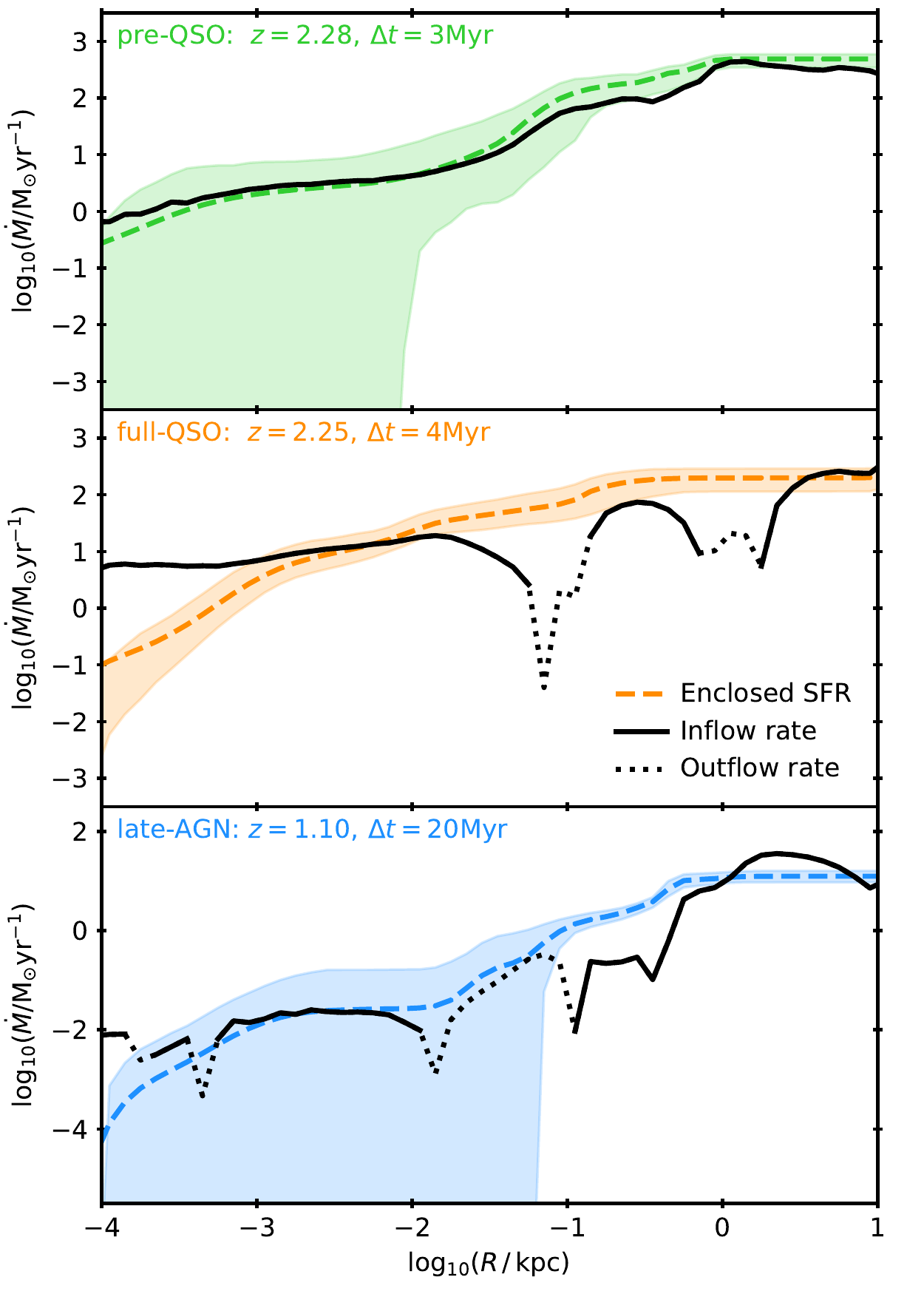}
\end{center}
\vspace{\vsf}
\caption{Time-averaged gas mass flow rate as a function of spherical radial distance $R$ from the central black hole for the \eqso~(top), \qso~(middle), and \agn~(bottom) conditions.  Black solid and dotted lines indicate radial bins with net inflow and outflow rate respectively. 
In each panel, the colored dashed line shows the time-averaged enclosed SFR and the shaded region indicates the 10--90\% percentile range achieved in each radial bin during the full evolution time $\Delta t$.
The enclosed SFR generally traces the radial flow rate profile, with some indication for comparatively lower star formation at $<1$\,pc.
}
\label{fig:MdotSFR}
\end{figure}

The \eqso~conditions at $z = 2.28$ target the epoch when the bursty-to-steady transition is happening.  The top left panel of Figure~\ref{fig:mdot_vs_z} (green) shows high level of variability in accretion when measured at 0.1\,pc, with $\dot{M}_{\rm BH} = 0.001$--10\,\Msunyr~and average $\langle \dot{M}_{\rm BH} \rangle = 0.6$\,\Msunyr~during a period of $\sim$3\,Myr.  For $M_{\rm BH}^{\rm ini} = 10^8$\,\Msun, this implies an average Eddington ratio $\langle \lambda_{\rm Edd} \rangle \approx 0.3$, where $\lambda_{\rm Edd} \equiv \dot{M}_{\rm BH}/\dot{M}_{\rm Edd}$ and $\dot{M}_{\rm Edd} \approx 2.2$\,\Msunyr~is the Eddington rate. 
The \qso~phase at $z = 2.25$ (orange panel; top right) coincides with the peak of black hole accretion according to the free-fall sub-grid accretion model, occurring $\sim$40\,Myr after the \eqso~conditions.  In this case, the hyper-refinement simulation predicts $\langle \dot{M}_{\rm BH} \rangle \approx 6$\,\Msunyr~ with very little variability during a period of $\sim$4\,Myr, corresponding to $\langle \lambda_{\rm Edd} \rangle \approx 2.6$.
This implies that quasi-steady super-Eddington feeding rates onto the accretion disk are possible in the absence of AGN feedback.
The \agn~conditions at $z = 1.1$ (blue panel; bottom) show significantly lower accretion rate, with $\langle \dot{M}_{\rm BH} \rangle = 0.01$\,\Msunyr~during a period of $\sim$20\,Myr, corresponding to $\langle \lambda_{\rm Edd} \rangle \approx 0.005$.  The formation of a low-density central cavity filled with $T\sim 10^7$\,K gas (Figure~\ref{fig:kin_vs_r}) inhibits black hole growth during $\sim$75\%~of the time (with $\dot{M}_{\rm BH} = 0$), while the instantaneous accretion rate ranges from $\dot{M}_{\rm BH} = 0.001$--0.1\,\Msunyr~when cold gas clumps and filaments destabilize from the inner edge of the circumnuclear disk and fall toward the black hole (Figure~\ref{fig:zoom_agn}).  The finite mass of gas resolution elements ($m_{\rm g} \sim 15$\,\Msun)~implies an upper limit of roughly $\dot{M}_{\rm BH} < 10^{-6}$\,\Msunyr~during the inactive periods in the \agn~phase.

Overall, our hyper-refinement simulations predict qualitatively distinct accretion phases with properties that are not trivial to infer even from the $\sim$100\,pc physical conditions available in state-of-the-art cosmological zoom-in simulations.  Remarkably, we show that it is possible to feed the central 0.1\,pc at a rate sufficient to power a luminous QSO ($L_{\rm bol} \approx 3\times10^{46}$\,erg\,s$^{-1}$ for $\dot{M}_{\rm BH} = 6$\,\Msunyr~and $\epsilon_{\rm rad}=0.1$) during the epoch of peak nuclear gas density in the central galaxy of a massive halo ($M_{\rm vir}\sim10^{12.5}$\,\Msun) at $z\sim2$.  At the same time, our simulations predict strong variability in the conditions preceding the QSO phase and low-duty cycle and lower luminosity for the more common conditions prevalent at later times.

\subsection{Star formation--inflow connection}\label{sec:SfrMdot}

Figure~\ref{fig:MdotSFR} investigates the connection between gas inflow rate and SFR over five orders of magnitude in spatial scales.  For each hyper-refinement simulation, we compute the net mass flow rate across spherical shells and show the time-averaged radial profile as either solid or dotted lines for net inflow and outflow rate, respectively.  
Note that large-scale, coherent galactic outflow events driven by stellar feedback are common at higher redshift but absent in the conditions analyzed here, where inflows and outflows can coexist in the same radial domain.
Colored dashed lines show the time-averaged enclosed SFR as a function of radial distance, computed from the instantaneous SFR of individual gas elements, and shaded regions indicate the 10--90\% variation in each radial bin.

In the \eqso~conditions (top), inflows dominate over outflows in the entire radial range $R = 0.1$\,pc--10\,kpc, with the peak inflow rate reaching $\dot{M}_{\rm in} \sim 475$\,\Msunyr~at $\sim$1\,kpc and a monotonic decrease to $\dot{M}_{\rm in} \lesssim 1$\,\Msunyr~at $\sim$0.1\,pc.  The enclosed SFR shows significant variability occurring during only $\sim$3\,Myr, indicated by the green shaded region, with the global peak of star formation reaching $\sim$1000\,\Msunyr.  As expected, the relative variation in enclosed SFR increases at lower radii, owing to the shorter dynamical times and the non-trivial motion of the black hole relative to the surrounding gas distribution.
Despite this, the time-averaged enclosed SFR profile roughly matches the net inflow rate in the entire range 0.1\,pc--10\,kpc. 

The \qso~conditions (middle) display similar large scale inflow rates as the \eqso~simulation, but show net outflow in the intermediate range $\sim$50\,pc--1\,kpc (coincident with the maximum temperatures reached) and significantly higher net inflow rates in the inner region.  The enclosed SFR is again roughly similar to the net inflow rate on $\gtrsim$1\,kpc and $\sim$10\,pc scales, but decreases rapidly in the inner 1\,pc.   As expected, the global inflow rate and SFR are significantly lower in the \agn~conditions (bottom), but the inflow rate profile again roughly follows the enclosed SFR distribution.  In this case, outflows dominate in the range $\sim$10--100\,pc owing to the formation of a hot, low-density central cavity, but cold gas clumps driven inward by instabilities yield a net inflow rate of $\sim$0.01\,\Msunyr~at 0.1\,pc while feeding star formation at a rate $\sim$100 times lower.  
Overall, these results show that star formation can consume as much gas as provided by inflows at all scales, with some indication for comparatively lower star formation relative to the inflow rate in the central $\sim$1\,pc.

\begin{figure}
\begin{center}
\includegraphics[width=0.45\textwidth]{\pathL/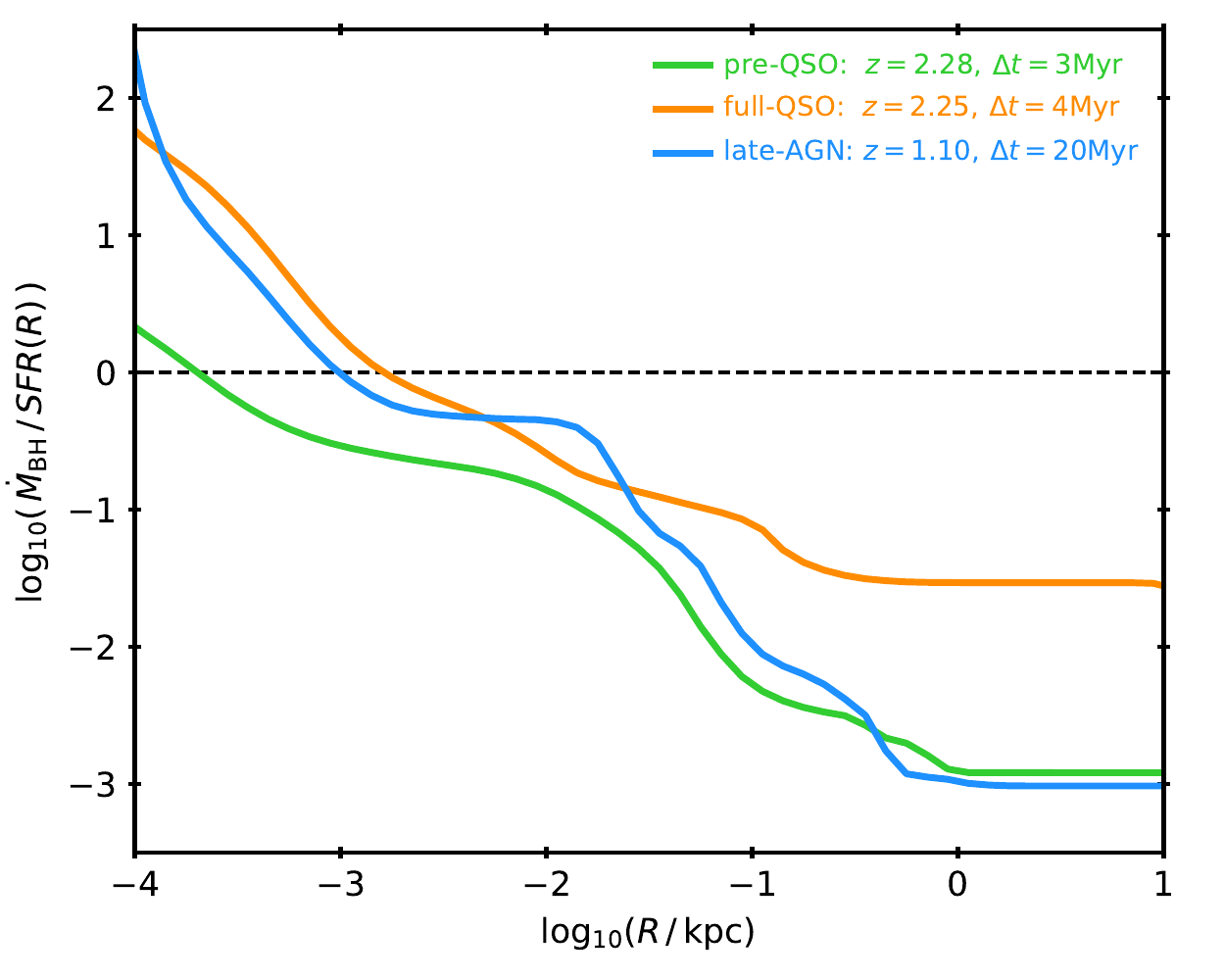}
\end{center}
\vspace{\vsf}
\caption{Ratio of the average black hole accretion rate, measured as explicit inflow rate at 0.1\,pc, to the average enclosed SFR as a function of radial distance.
On galaxy scales, $\dot{M}_{\rm BH} \sim$ SFR/50 for the \qso~phase, decreasing to $\dot{M}_{\rm BH} \sim$ SFR/1000 for the \eqso~(green) and \agn~(blue) conditions.
}
\label{fig:MdotSFRratio}
\end{figure}

Figure~\ref{fig:MdotSFRratio} shows more directly the relation between black hole accretion and radially integrated SFR on different scales, where we plot the time-averaged ratio $\dot{M}_{\rm BH}/$SFR as a function of radial distance.  $\dot{M}_{\rm BH}$ can greatly exceed the enclosed SFR for the smallest scales probed here, reaching $\dot{M}_{\rm BH}\sim 100\times$SFR at 0.1\,pc for the \qso~and \agn~conditions.  In all cases, the SFR within $\sim$10\,pc already exceeds the accretion rate, and $\dot{M}_{\rm BH}/$SFR decreases for the global ($\sim$1--10\,kpc) star formation rate down to 1/1000 in the \eqso~and \agn~conditions and 1/50 for the \qso~phase.  
These accretion to global SFR ratios have important implications for black hole--galaxy scaling relations (\S\ref{sec:sfragn}).

\subsection{Cool gas inflow and angular momentum}\label{sec:angmom}

\begin{figure}
\begin{center}
\includegraphics[width=0.45\textwidth]{\pathL/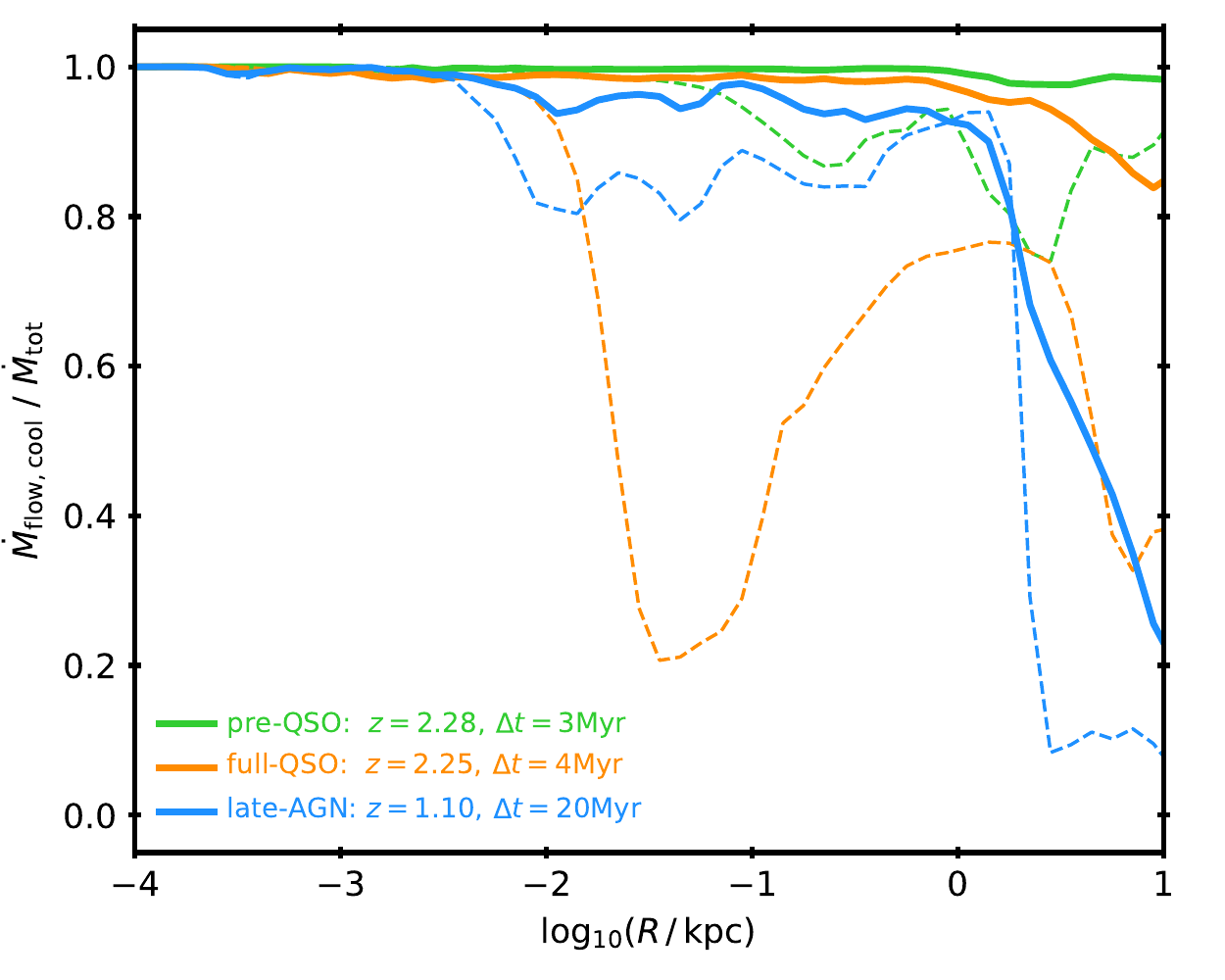}
\end{center}
\vspace{\vsf}
\caption{Fraction of gas inflow rate (solid) and outflow rate (dashed) in cool gas ($\dot{M}_{\rm flow,cool}$; $T < 10^{4.5}$\,K) relative to total ($\dot{M}_{\rm tot}$) as a function of radial distance from the central black hole for the \eqso~(green), \qso~(orange), and \agn~(blue) conditions.  
Cool gas dominates the inflow rate in all cases at $R<1$\,kpc.
\vspace{-0.2cm}
}
\label{fig:ColdIN}
\end{figure}

Figure~\ref{fig:ColdIN} shows the fractional contribution of cool gas ($T < 10^{4.5}$\,K) to the total inflow rate as a function of radial distance (solid lines).  In the \eqso~conditions, cool gas comprises almost the totality of inflowing gas within 10\,kpc.  For the \qso~conditions, cool gas contributes $>90$\%~of the inflow rate within 1\,kpc, dropping to $\sim$85\%~at 10\,kpc.  Similarly, the cool gas component represents $\sim90$\%~of the inflow rate within 1\,kpc for the \agn~conditions, while in this case hot gas clearly dominates the inflow rate at 10\,kpc.   
As expected, hot gas ($T > 10^{4.5}$\,K) contributes a higher fraction of the mass outflow rate (dashed lines), which clearly dominates over the cool component at $R>1$\,kpc for the \qso~and \agn~conditions and for the range 10\,pc--100\,pc in the \qso~phase. 
Thus, while the mass-weighted temperature can exceed $10^6$\,K within the central 100\,pc (top panel of Figure~\ref{fig:kin_vs_r}), fueling of the central black hole relies primarily on the supply of cool gas in the nuclear region for all three conditions analyzed, with thermal pressure support playing a minor role.

\begin{figure}
\begin{center}
\includegraphics[width=0.45\textwidth]{\pathL/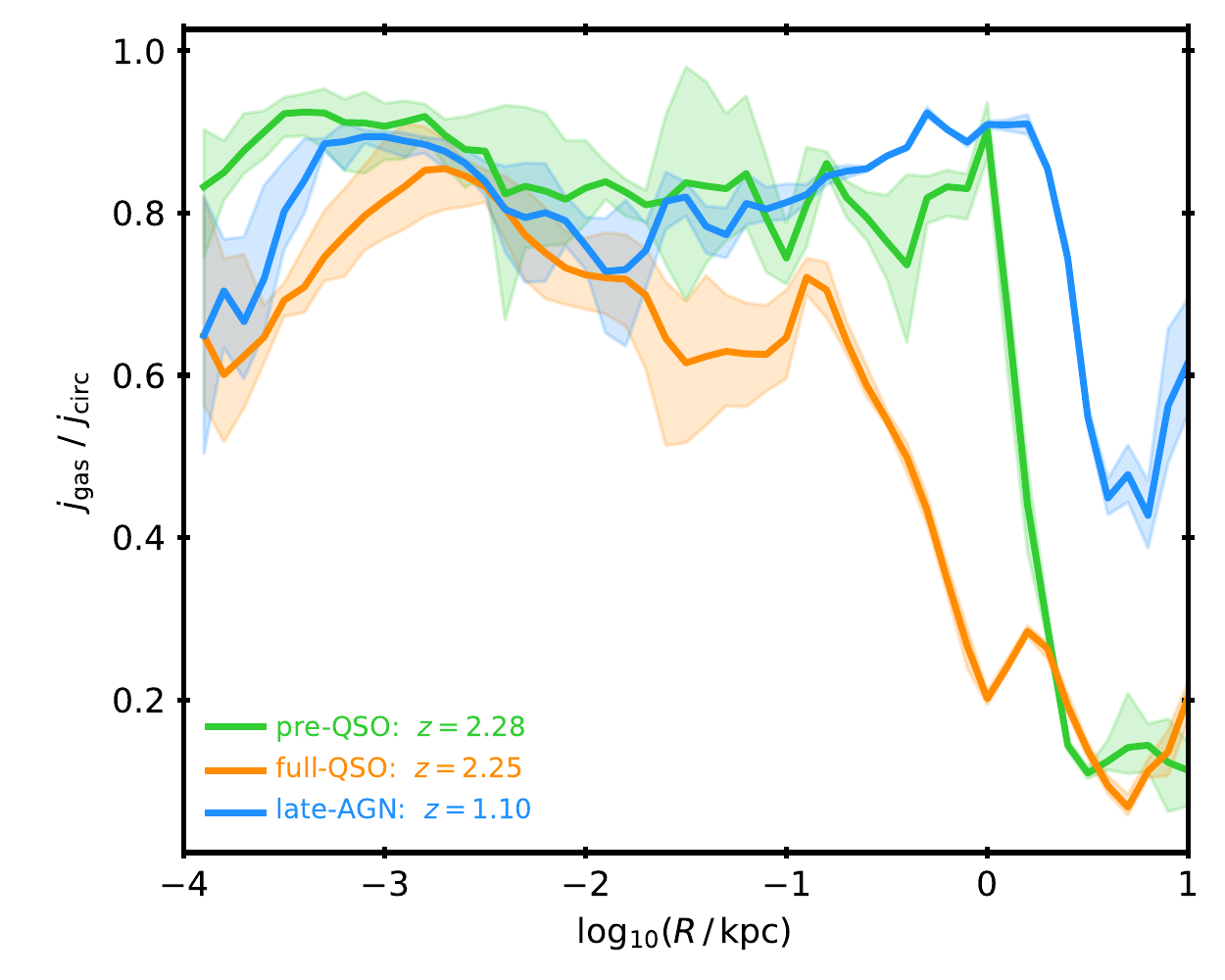}
\end{center}
\vspace{\vsf}
\caption{Ratio of angular momentum per unit mass of cool ($T < 10^{4.5}$\,K) gas ($j_{\rm gas}$) to that expected for circular motion assuming spherically symmetric enclosed mass distribution ($j_{\rm circ} = \sqrt{G\,M_{\rm enc}(R)\,R}$) as a function of distance from the black hole for the \eqso~(green), \qso~(orange), and \agn~(blue) conditions. 
Solid lines show time-averaged $j_{\rm gas}/j_{\rm circ}$ while shaded regions indicate the 10--90\% percentile range achieved in each radial bin.
The specific angular momentum of gas is typically only $\sim$10--40\% below that of fully-circular orbits at $R<100\,$pc; given that it is also cool (Figure~\ref{fig:ColdIN}), it is primarily supported by angular momentum.
}
\label{fig:Jr}
\end{figure}

Figure~\ref{fig:Jr} shows the radial profile of the specific angular momentum of cool gas, $j_{\rm gas}$, normalized by the specific angular momentum assuming circular motion, $j_{\rm circ} = \sqrt{G\,M_{\rm enc}(R)\,R}$, where $M_{\rm enc}(R)$ is the total enclosed mass within $R$.  We compute $j_{\rm gas}$ as the magnitude of the angular momentum of cool gas in each spherical radial shell divided by its cool gas mass content, averaging over time for each simulation.  In all cases, $j_{\rm gas}$ varies by $\sim$4 orders of magnitude from kpc to sub-pc scales, roughly tracing $j_{\rm circ}$.
The \agn~conditions show nearly circular motion within $\sim$1\,kpc, roughly the size of the thin gas disk shown in Figure~\ref{fig:zoom_agn}, but also show strong variations of $j_{\rm gas}$ at $<100$\,pc owing to the formation of a hot central cavity where cold gas clumps and filaments penetrate with $\sim$25\%~duty cycle. 
The \qso~conditions show the lowest $j_{\rm gas}$ relative to $j_{\rm circ}$ at 10\,pc--1\,kpc, but all cases show substantial angular momentum support, with $j_{\rm gas} \sim 0.8 j_{\rm circ}$ at $<10$\,pc, indicating the need for torques to remove angular momentum and drive gas inflows down to smaller scales.

\begin{figure}
\begin{center}
\includegraphics[width=0.45\textwidth]{\pathL/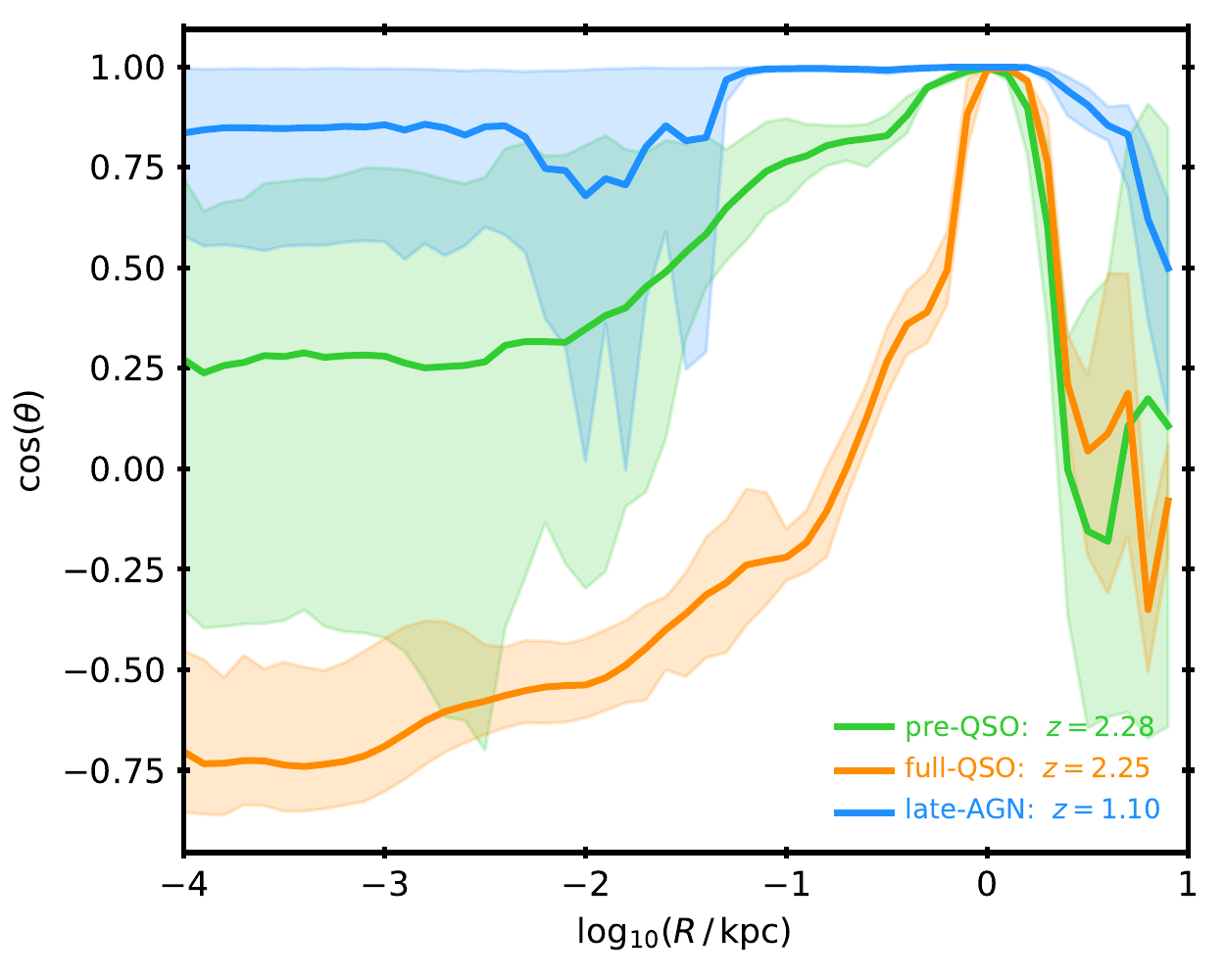}
\end{center}
\vspace{\vsf}
\caption{Direction of the angular momentum of cool ($T < 10^{4.5}$\,K) gas as a function of distance to the black hole for the \eqso~(green), \qso~(orange), and \agn~(blue) conditions.
Solid lines show time-averaged ${\rm cos}(\theta)$, where $\theta$ is the angle between the angular momentum vector in each radial shell and that of the angular momentum at $R=1$\,kpc.  Shaded regions indicate the 10--90\% variability range in each case.
The inner $\sim$10\,pc gas distribution can be significantly misaligned relative to the galaxy scale disk and the degree of misalignment shows strong time variability.
}
\label{fig:Jdir}
\end{figure}

Figure~\ref{fig:Jdir} shows the angle $\theta$ between the angular momentum vector of cool gas in spherical radial shells and that of the angular momentum of cool gas in a thin spherical radial shell centered at $R=1$\,kpc.  
The \agn~conditions show very good alignment across scales, with average $\theta \sim 0^{\circ}$ at 0.1--1\,kpc corresponding to a well defined, rotationally supported disk.  On smaller scales ($<100$\,pc), cold gas clumps penetrating the central cavity tend to maintain the larger scale angular momentum direction but can be occasionally torqued into other directions all the way to $\theta > 90^{\circ}$.   
The gas rich, turbulent conditions in the \eqso~phase show a very different situation, with a persistent misalignment of $\theta \sim 70^{\circ}$ between the galaxy-scale disk and the nuclear gas disk at $<100$\,pc.  The strong variability in the inner $\sim$10\,pc manifests as rapid changes in angular momentum direction, covering the full range $0^{\circ} < \theta < 180^{\circ}$.   
The \qso~conditions show less time variability but an increasing misalignment from kpc scales all the way to counter-rotation on pc-scales.
Thus, despite the similarity between the radial profiles of $j_{\rm gas}$ and $j_{\rm circ}$, the angular momentum direction can change significantly across scales.  The pc-scale gas disk ultimately responsible for feeding the black hole can form in any direction relative to the galaxy scale disk.

\subsection{Gravitational torques and pressure gradients}\label{sec:torque}

\begin{figure}
\begin{center}
\includegraphics[width=0.45\textwidth]{\pathL/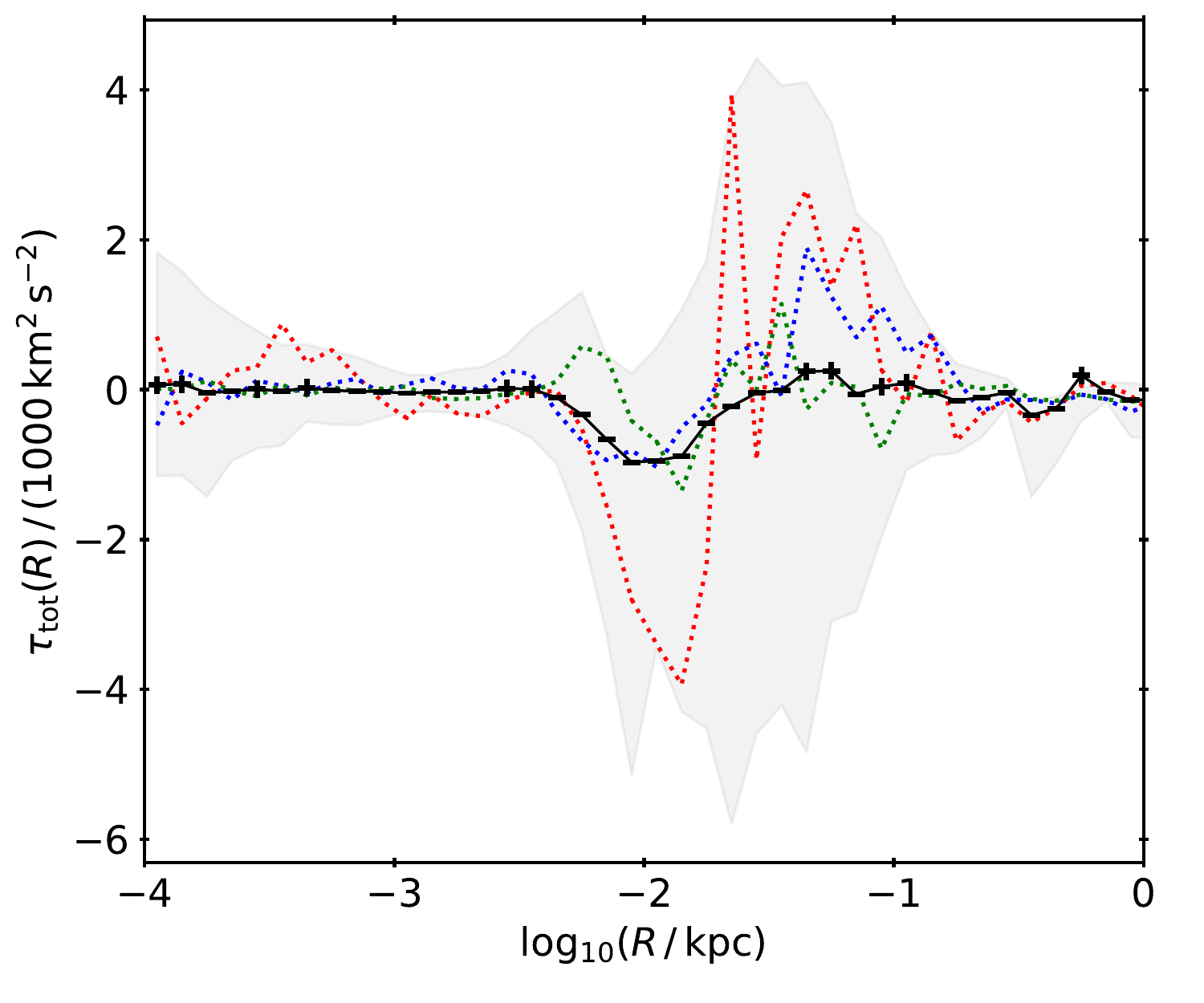}
\end{center}
\vspace{\vsf}
\caption{Total torque per unit mass acting on cool gas ($T < 10^{4.5}$\,K) as a function of radial distance from the black hole for the \qso~conditions, averaging over narrow radial annuli (negative torques decrease the gas angular momentum).  
The black solid line shows the time-averaged specific torque over $\sim$4\,Myr, where radial bins with negative and positive average torque are indicated by the $-$ and $+$ symbols, respectively.  The gray shaded region indicates the 10--90\% percentile range in each radial bin.  Dotted lines of different colors show the instantaneous radial profile of the specific torque at three randomly chosen times.
The torque radial profile changes with time reflecting the state of the system, with persistent sign changes also in a time-average sense.
}
\label{fig:trq}
\end{figure}

Figure~\ref{fig:trq} illustrates the magnitude and time variability of the total torque exerted on the cool gas component ($T < 10^{4.5}$\,K) for the \qso~conditions.  
We compute the average specific torque in each spherical radial bin as:
\begin{equation}\label{eq:trq}
{\pmb \tau}_{j} = \frac{1}{M_{\rm bin}} \, \sum_{i} m_{i}\, {\bf{r}}_{i} \times {\bf a}_{j}({\bf r}_{i}),
\end{equation} 
where $M_{\rm bin} = \sum m_{i}$, the sum over $i$ refers to the cool gas elements with mass $m_{i}$ inside the radial bin, and ${\bf a}_{j}({\bf r}_{i})$ is the acceleration exerted by different components on gas element $i$ located at ${\bf r}_{i}$.  We compute separately the torque due to gravitational forces from gas, stars, or dark matter, where ${\bf a}_{j} = \nabla \Phi_{j}({\bf r}_{i})$ and $\Phi_{j}$ is the gravitational potential for each component, and the torque from pressure forces, where ${\bf a}_{j} = \nabla P({\bf r}_{i}) /  \rho({\bf r}_{i})$ and $P$ and $\rho$ are the gas pressure and density, respectively\footnote{Torques from gravitational forces and pressure gradients are computed in post-processing using M. Grudi\'c's {\sc pykdgrav} and {\sc meshoid} Python repositories available at \url{https://github.com/mikegrudic}.}.  
We then compute the total torque per unit mass in the direction of the angular momentum ${\bf L}$ in each radial bin, $\tau_{\rm tot} = [ {\pmb \tau}_{\rm grav} + {\pmb \tau}_{\rm hydro} ]_{\hat{L}}$, where ${\pmb \tau}_{\rm grav} = {\pmb \tau}_{\rm gas} +  {\pmb \tau}_{\rm star} + {\pmb \tau}_{\rm DM}$.

The dotted lines of different colors in Figure~\ref{fig:trq} show the radial dependence of the specific torque $\tau_{\rm tot}$ at three randomly chosen times during the \qso~phase.  We find very strong instantaneous specific torques on the cool gas component that can exceed 1000\,km$^2$\,s$^{-2}$, with significant time variability and sign changes over the entire radial range 0.1\,pc--1\,kpc. The black solid line indicates the time-averaged specific torque $\tau_{\rm tot}$, which shows that there are also persistent sign changes for the radial profile in a time-average sense.  The transition between negative to positive average specific torque at $\sim 20$\,pc roughly corresponds to the transition seen in Figure~\ref{fig:MdotSFR} from inflow dominated dynamics to outflow dominated for the \qso~conditions.
Qualitatively similar results are found for the \eqso~and \agn~conditions.

\begin{figure}
\begin{center}
\includegraphics[width=0.45\textwidth]{\pathL/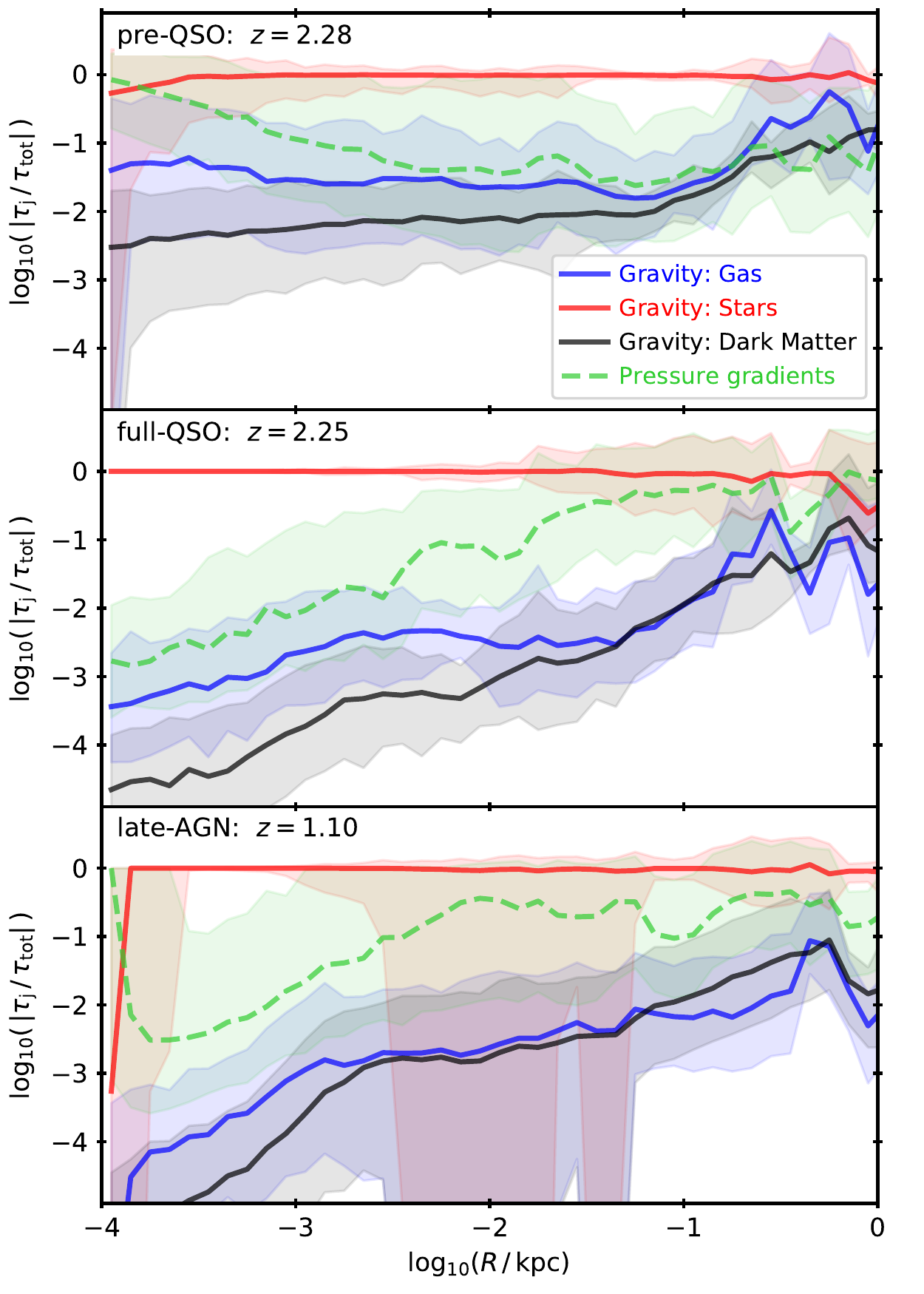}
\end{center}
\vspace{\vsf}
\caption{Radial profile of the specific torque on cool gas ($T < 10^{4.5}$\,K) from different components normalized by the total torque, shown for the \eqso~(top), \qso~(middle), and \agn~(bottom) simulations.  We compare gravitational torques from the gas itself (blue), stars (red), and dark matter (black), and hydrodynamic torques from pressure gradients (green dashed).  Lines show median values while shaded regions indicate the 10--90\% percentile range in each radial bin.  
The net torque on the gas is dominated by gravitational torques from the stars at all radii.
}
\label{fig:trqRat}
\end{figure}

Figure~\ref{fig:trqRat} compares the contribution of gravitational forces from other gas (blue), stars (red), and dark matter (black), as well as pressure forces (green) to the total specific torque on the cool gas component as a function of radial distance.  Note that fractional contributions can exceed unity ($|\tau_{\rm j} / \tau_{\rm tot}| > 1$) when different components contribute to $\tau_{\rm tot}$ with opposite sign.
The top panel shows that gravitational torques from the stars dominate the total torque for most radii in the \eqso~conditions.  Dark matter contributes the least to gravitational torques, followed by gas self-gravity which can become comparable to the stellar torques at galactic scales ($>500$\,pc).  Torques from pressure forces are typically 1--2 orders of magnitude lower than stellar torques but increase in the inner 1\,pc to the extent that pressure gradients can become the dominant mechanism of angular momentum transport at $\sim 0.1$\,pc.  However, local viscous stresses driven by MHD turbulence, not captured here, are believed to dominate angular momentum transport in the accretion disk around the central BH \citep[e.g.][]{BalbusHawley1998,Goodman2003}.  The \qso~conditions (middle panel) show qualitatively similar results, with stellar torques clearly dominating in the central 100\,pc and, in this case, pressure gradients becoming important at larger scales.  The \agn~conditions (bottom panel) show very large scatter in the fractional contribution of different torque components owing to the formation of a low density central cavity.  Nonetheless, stellar torques are again the dominant component.

\begin{figure}
\begin{center}
\includegraphics[width=0.45\textwidth]{\pathL/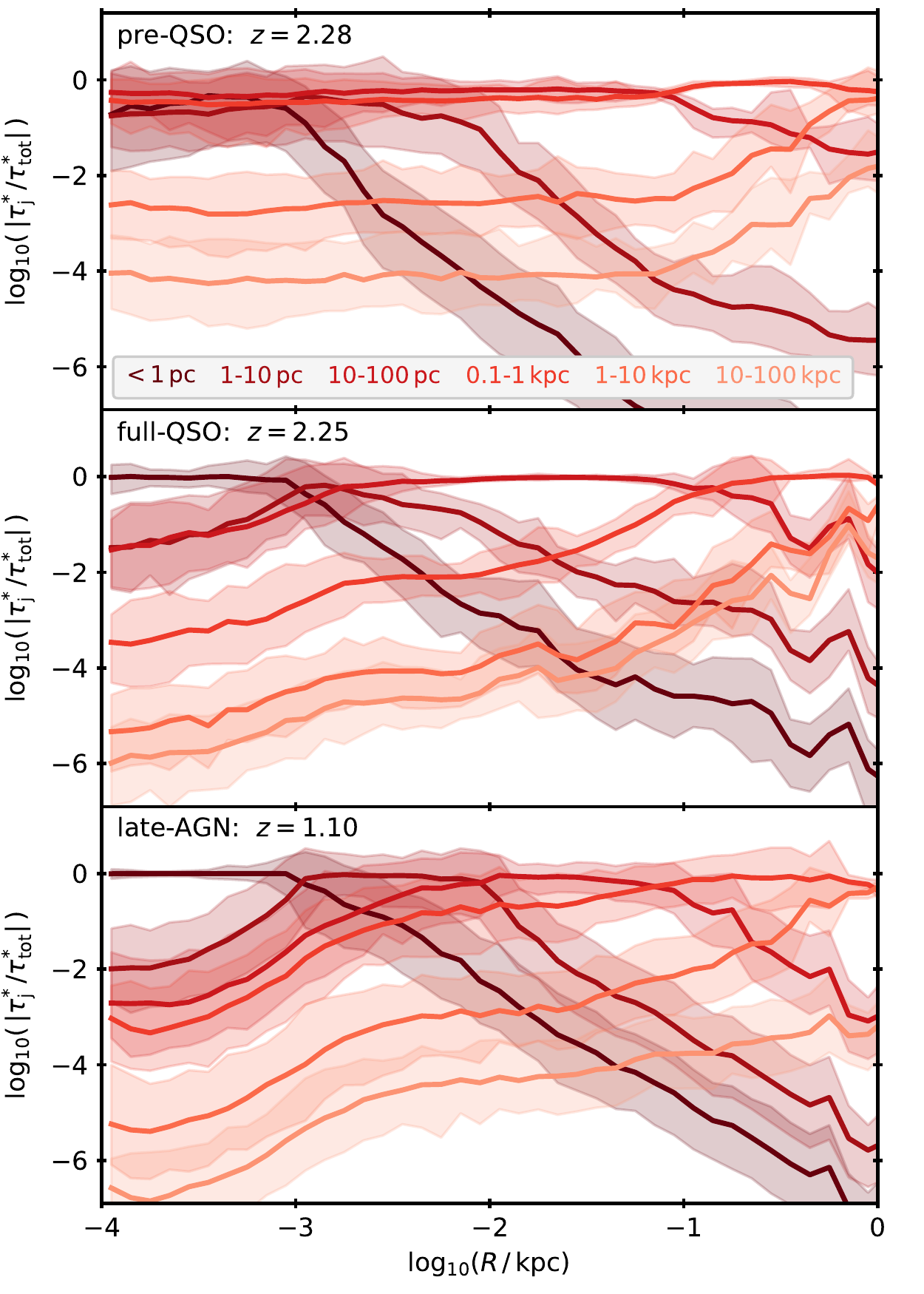}
\end{center}
\vspace{\vsf}
\caption{Radial profile of the fraction of (specific) stellar gravitational torque on cool gas ($T < 10^{4.5}$\,K) contributed by stars located in different radial annuli, from $R < 1$\,pc to 10\,kpc\,$< R < 100$\,kpc, as indicated by the color scale.  Solid lines show median values while shaded regions indicate the 10--90\,\% percentile range in each radial bin.  
The total gravitational torque is generally dominated by nearby stars in the same radial annuli as the gas, but non-local torques can be important and even dominate the central gas inflow in some cases.
}
\label{fig:trqStarRat}
\end{figure}

Figure~\ref{fig:trqStarRat} shows the fractional contribution of stars at different radii to the total stellar gravitational torque.  We compute the stellar torque on the gas located in each radial bin as in Equation~\ref{eq:trq} but considering only the gravitational potential produced by stars located in an independent set of radial annuli.  
The bottom panel shows a very clear connection between the radial distribution of gas and the location of the stars that dominate the gravitational torque for the \agn~conditions.  The torque in the central 1\,pc is dominated by stars within 1\,pc, the torque on gas at 1\,pc\,$< R < 10$\,pc is dominated by stars in the same radial bin, and similarly for radial annuli on larger scales.  Stars located at $R>100$\,pc contribute only $\sim 10$\% of the total torque at 10\,pc and $<1$\% at 1\,pc, suggesting that the instantaneous gas inflow rate at $<1$\,pc is decoupled from the larger scale stellar distribution.
The middle panel shows qualitatively similar results for the \qso~conditions, with the main difference being that the torque within 1\,pc\,$< R < 10$\,pc is dominated by stars at 10\,pc\,$< R < 100$\,pc.
The \eqso~conditions (top panel) show enhanced coupling of scales due to strong large scale non-axisymmetries and lower stellar surface densities in the central ($<$10\,pc) region.  While the overall trend is similar, the total torque at $R<1$\,pc contains significant contributions from stars in the entire radial range 0.1\,pc--1\,kpc, suggesting that the instantaneous accretion rate in this case is sensitive to the large scale stellar distribution.  This result would be difficult to capture without our hyper-refinement technique enabling us to cover a large dynamic range in scales and should be considered in the development of improved sub-grid accretion models.

\begin{figure}
\begin{center}
\includegraphics[width=0.48\textwidth]{\pathL/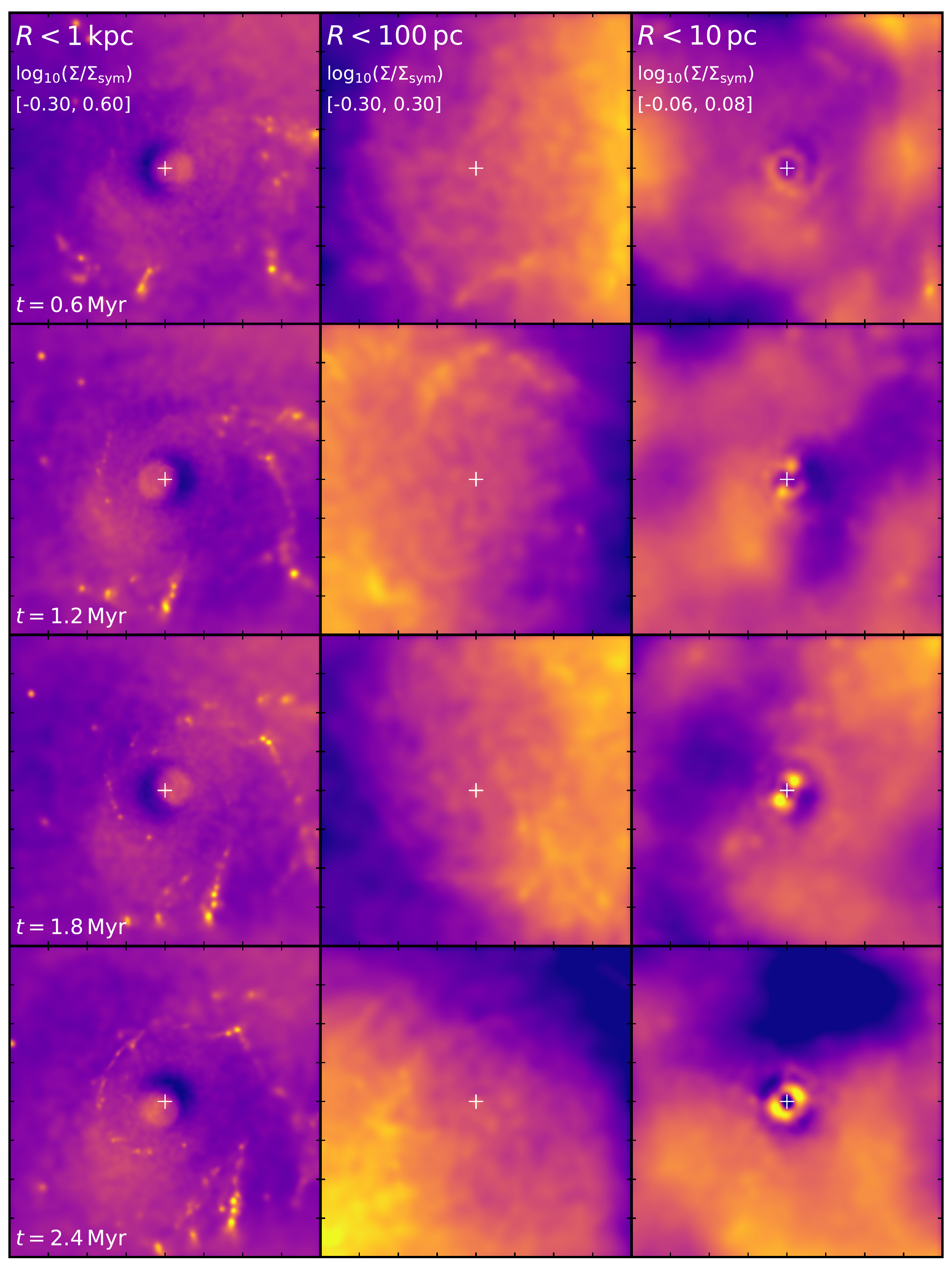}
\end{center}
\vspace{\vsf}
\caption{Ratio of projected stellar mass surface density ($\Sigma_{\star}$) to the azimuthally-averaged, axisymmetric component ($\Sigma_{\star,{\rm sym}}$) for the central 1\,kpc (left), 100\,pc (middle), and 10\,pc (right) in the \eqso~simulation.
The color scale is logarithmic and extends from the minimum (blue) to maximum (yellow) values of $\log_{10}(\Sigma_{\star} / \Sigma_{\star,{\rm sym}}) = \log_{10}(\Sigma_{\star}(R,\,\phi) / \langle \Sigma_{\star}(R) \rangle)$ indicated for each scale.
The black hole location is indicated by the + sign and time evolution is shown in 0.6\,Myr intervals from top to bottom.
Strong non-axisymmetries with a diversity of morphologies are present in the stellar component from $\sim$kpc scales down to the inner 10\,pc.
}
\label{fig:starmap}
\end{figure}

Figure~\ref{fig:starmap} illustrates the non-axisymmetries in the stellar component that yield the multi-scale gravitational torques quantified in Figure~\ref{fig:trqStarRat} (top panel) for the \eqso~conditions.  We make stellar mass surface density maps ($\Sigma_{\star}$) on different scales, oriented face-on relative to the angular momentum of gas within 100\,pc, and compute the corresponding axisymmetric component ($\Sigma_{\star,{\rm sym}}$) by azimuthally-averaging $\Sigma_{\star}$ in cylindrical radial bins around the black hole.  We show the ratio $\Sigma_{\star} / \Sigma_{\star,{\rm sym}}$ from 1\,kpc (left) to 10\,pc (right), which highlights the strong non-axisymmetries present in all scales.   Spiral arms with prominent stellar clumps are visible on sub-kpc scales along with a smooth asymmetric component that extends down to $\sim$100\,pc scales, where a dense, eccentric stellar disk dominates the gravitational potential.  The non-axisymmetric component reaches order unity, with $\Sigma_{\star} / \Sigma_{\star,{\rm sym}} \sim 0.5$--2 in the strong non-linear regime.   
On $<10$\,pc scales (comparable to the orbital motion of the black hole), the stellar distribution becomes less coherent and non-axisymmetries of order $\Sigma_{\star} / \Sigma_{\star,{\rm sym}} \sim 0.8$--1.2 appear and vary on short timescales, while a pc-scale stellar disk forms around the black hole with frequent changes of orientation. 
Remarkably, in the \eqso~conditions, the full range of stellar non-axisymmetries from $\sim$kpc to $\sim$pc scales contributes significantly to the total torque on sub-pc scale gas immediately responsible for black hole fueling.

\subsection{Black hole mass dependence}\label{sec:bhdep}

\begin{figure}
\begin{center}
\includegraphics[width=0.48\textwidth]{\pathL/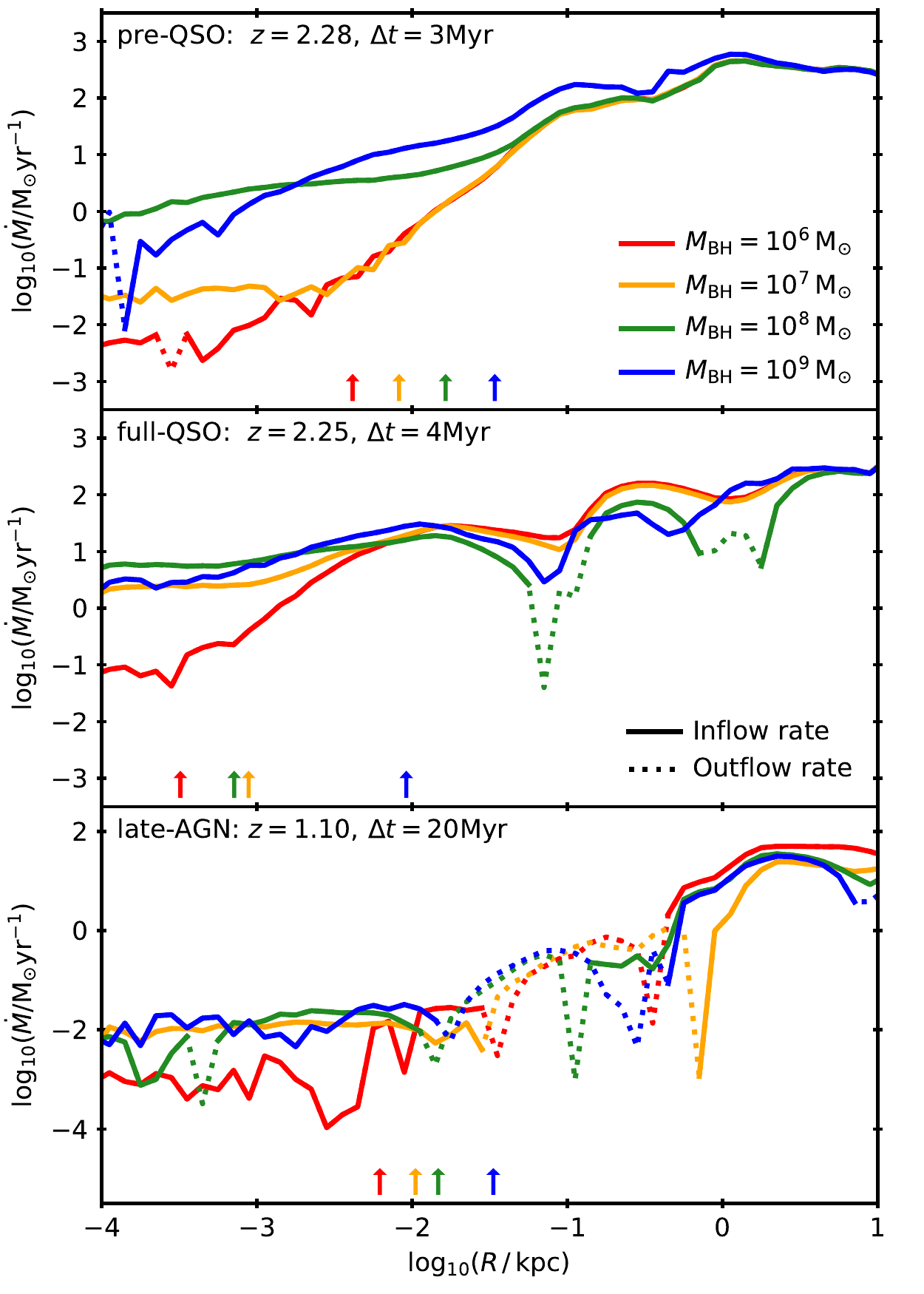}
\end{center}
\vspace{-0.6cm}
\caption{Time average gas mass flow rate as a function of radial distance from the black hole for the \eqso~(top), \qso~(middle), and \agn~(bottom) conditions.  We compare simulations with identical resolution and initial conditions but different initial black hole mass, with $M_{\rm BH} = 10^6$\,\Msun~(red),  $10^7$\,\Msun~(orange), $10^8$\,\Msun~(green), and $10^9$\,\Msun~(blue).  Solid and dotted lines indicate radial bins with net inflow and outflow, respectively. 
Vertical arrows show the black hole radius of influence $R_{\rm BH}$ at the end of the simulation in each case, defined as $M_{\star}(<R_{\rm BH}) = M_{\rm BH}$.
The inflow rate is independent of $M_{\rm BH}$ on large ($\gtrsim$100\,pc) scales and only weakly dependent on $M_{\rm BH}$ on scales comparable to $R_{\rm BH}$.
}
\label{fig:Mbh_test}
\end{figure}

The inflow rates presented throughout the paper correspond to the fiducial simulations for a black hole with initial mass $M_{\rm BH} = 10^8$\,\Msun~accreting under different conditions.  Figure~\ref{fig:Mbh_test} explores the accretion rate dependence on $M_{\rm BH}$ for the \eqso~(top), \qso~(middle), and \agn~(bottom) conditions.  In each case, we replicate the fiducial simulation with identical resolution and initial conditions but vary the initial black hole mass in the range $M_{\rm BH} = 10^6$--10$^9$\,\Msun.  
Changing $M_{\rm BH}$ has a minor impact in the gas and stellar distributions in the initial conditions prior to hyper-refinement. 
All simulations are evolved for the same physical time except for $M_{\rm BH} = 10^9$\,\Msun~in the \eqso~and \qso~conditions, which only follow $\sim$1\,Myr of evolution owing to their significantly higher computational expense as a result of shorter dynamical times.

The inflow/outflow rates at $>$100\,pc are roughly similar in all simulations regardless of $M_{\rm BH}$, as expected given that the dynamical influence of the black hole is negligible on galactic scales.  Nonetheless, we see variations in inflow rate due to stochastic effects, including the movement of the black hole (which we use as reference frame) relative to the center of mass of the galaxy.  The \eqso~conditions show very similar inflow rates at $<$10\,pc for $M_{\rm BH} = 10^{6-7}$\,\Msun~while black holes with $M_{\rm BH} = 10^{8-9}$\,\Msun~show significantly higher inflow rates.
The highest average inflow rates at $\sim$1--100\,pc occur for $M_{\rm BH} = 10^9$\,\Msun, but sub-pc accretion rates are higher for $M_{\rm BH} = 10^8$\,\Msun.
This suggests an overall trend for higher $\dot{M}_{\rm BH}$ with increasing $M_{\rm BH}$ but subject to significant stochastic effects.  
The \qso~and \agn~phases are consistent with no dependence of gas inflow rate on black hole mass at all scales except for $M_{\rm BH} = 10^6$\,\Msun, with order of magnitude lower $\dot{M}_{\rm BH}$.  Given the mass resolution for the (preexisting) stellar and dark matter components ($m_{\star} = 3.3 \times 10^4$\,\Msun~and $m_{\rm DM} = 1.7 \times 10^5$\,\Msun), it is possible that we are not fully capturing the dynamics and gravitational capture of gas by the $M_{\rm BH} = 10^6$\,\Msun~black hole.
Overall, while changing $M_{\rm BH}$ may affect other properties such as gas density and relative velocity,
these results are roughly consistent with the weak $M_{\rm BH}$ dependence in the gravitational torque accretion model of \citet{Hopkins2011_Analytic}, and in stark contrast with the strong $\sim$$M_{\rm BH}^{2}$ dependence in Bondi accretion.

\vspace{1cm}

\section{Discussion}\label{sec:discussion}

We have presented a new technique to model black hole growth at sub-pc resolution in a full cosmological galaxy formation context.  We achieve an unprecedented dynamic range by augmenting the Lagrangian MFM hydrodynamics method in GIZMO \citep{Hopkins2015_Gizmo} with a new dynamic, hyper-Lagrangian refinement technique that increases the resolution dramatically approaching the central black hole.  
The success of this technique relies on key physics and algorithmic implementations in FIRE-2 \citep{Hopkins2018_FIRE2methods} and our explicit sub-pc scale treatment of black holes:

\begin{itemize}[itemsep=-10pt]

\item {\it Adaptive gravitational softenings}:  We solve gravitational forces for identical gas mass distribution as hydrodynamic forces, decreasing the gravitational softenings dynamically as required by the increased resolution in the hyper-refinement region.\\

\item {\it Adaptive star formation criteria}:  Only gas that is locally self-gravitating and Jeans unstable is allowed to form stars, which is crucial in galactic nuclei where more standard density/molecular thresholds can overestimate the SFR owing to high mean densities even for Jeans stable/unbound gas.\\

\item {\it Predictive star formation efficiency ({\rm SFE})}:  Gas satisfying the star formation criteria is not artificially forced to have a low SFE per free-fall time, which is rather an outcome of local self-regulation by stellar feedback \citep{Orr2018}.  This is crucial to predict whether gas can inflow faster than it can turn into stars without imposing artificial constraints in the sub-grid star formation model. \\

\item {\it Resolved multi-phase interstellar medium}:  We do not include a pressure floor nor impose an effective equation of state for ISM gas, which would prevent us from modeling stellar feedback and turbulence directly or predicting the relative efficiencies of local fragmentation and radial inflow.\\

\item {\it Adaptive stellar feedback}:  The adopted prescriptions are explicitly designed and tested to be consistent across mass resolution in the full range $m_{\rm b} \sim 1$--$10^5$\,\Msun~without re-calibration or tuning of free parameters \citep{Hopkins2018_SNeFeedback}. This is crucial to maintain physical consistency and model stellar feedback robustly 
as we transition from standard resolution at $>$1\,kpc to the maximum hyper-refinement resolution level at $<$100\,pc, where individual SNe are well resolved.\\

\item {\it Time-resolved stellar evolution}:  All stellar feedback quantities depend on the actual stellar ages rather than e.g. scaling with SFR, which is crucial in our hyper-refinement simulations with nuclear dynamical times vastly shorter than stellar evolution times.\\

\item {\it Cooling physics for large dynamic range}:  Our cooling module includes dust-gas thermal coupling, cosmic ray-gas coupling, and a treatment for optically-thick cooling which is crucial to capture the thermodynamic state of gas at the densities reached in the nuclear disk ($n_{\rm H} > 10^7$\,cm$^{-3}$).  The same cooling physics has been used for individual molecular cloud-scale simulations \citep{Grudic2018,Guszejnov2020} reaching $10^{-5}$\,pc resolution.\\

\item {\it Resolved gas gravitational capture}:  We avoid many of the uncertainties suffered by sub-grid accretion parameterizations by resolving the gravitational capture of gas by the central black hole at $<$0.1\,pc.\\

\item {\it Resolved black hole dynamics}:  We follow the response of the black hole to the time varying gravitational potential without including artificial drag forces or repositioning algorithms, which is crucial to capture the complex dynamics in galactic nuclei that lead to the development of non-axisymmetries, missalignements, and a variety of other phenomena.      

\end{itemize}

We stress that accurately following such a large dynamic range is not simply a matter of ``brute force'' resolution increase: many reasonable numerical implementations of the physics applicable on galaxy scales $\gtrsim $\,kpc and resolution $\sim$10$^{5}\,M_{\odot}$ would break/diverge or give un-physical results if naively applied to simulations with four orders-of-magnitude smaller-scale resolution.
At our highest resolution, we reach timesteps as low as months (Table~\ref{tbl:sims}), which makes this technique computationally infeasible for cosmological integrations down to $z=0$.  Crucially, we circumvent this limitation by exploiting existing FIRE-2 simulations to identify interesting phases for re-simulation at ultra-high resolution.
As a first application, we have focused on gas transport and black hole growth in a massive galaxy ($M_{\star} \sim 6$--20\,$\times 10^{10}$\,\Msun) before, during, and after its peak of nuclear gas density at $z\sim 2.5\rightarrow1$.

\subsection{Comparison to previous methodologies}

\citet{Levine2008,Levine2010} carried out adaptive mesh refinement (AMR) simulations of a Milky Way-mass progenitor at $z=3$--6, performing one of the few studies of gas transport down to sub-pc scales in a cosmological galaxy formation context.  They included prescriptions for star formation and stellar feedback but neglected their effects when reaching the maximum resolution in the nuclear region, missing also a treatment for optically-thick cooling.
While their extremely gas rich conditions are rather different, the diversity of transient features, high time variability, and strong misalignment across scales in our simulations are qualitatively consistent with their findings.

Smoothed particle hydrodynamics (SPH) simulations with static particle splitting have also been used to study the development of multi-scale non-axisymmetries in idealized systems \citep{Escala2004,Mayer2010,Mayer2015,Hopkins2010_MultiScale,Hopkins2011_Analytic}.
These valuable studies are nonetheless limited by the use of a pressurized ISM via an effective equation of state, lack of low-temperature cooling, constant density threshold for star formation, low sub-grid star formation efficiency, or stellar feedback prescriptions designed for lower resolution.
Some of our key results are in good agreement with earlier idealized models, including the dominant role of stellar torques and the weak dependence of inflow rate on $M_{\rm BH}$ \citep{Hopkins2011_Analytic}, confirmed in more recent simulations with stellar feedback and multiphase cooling \citep{Hopkins2016_NuclearSims}.

Complementary approaches are becoming possible with the implementation of super-Lagrangian refinement techniques in other cosmological simulations codes. 
\citet{Curtis2015,Curtis2016} implemented a dynamic refinement scheme in the moving-mesh code AREPO \citep{Springel2010} to increase the resolution dynamically around the black hole, which has been used to improve black hole feedback implementations in cosmological simulations \citep{Curtis2016_cosmo} and idealized simulations of galaxies \citep{Koudmani2019} and clusters \citep{Bourne2017}.
\citet{Beckmann2019} presented a new refinement scheme to resolve the black hole radius of influence in RAMSES \citep{Teyssier2002} and studied the growth of black hole seeds in isolated cooling halos.  
These novel techniques could also provide sub-pc resolution in a full cosmological context, but exploiting the increased dynamic range while maintaining physical consistency requires including ISM/stellar physics such as those included here, capable of extrapolating to very different resolution and density regimes.

\subsection{Reproducing luminous QSO inflows}

The observed radiative output of QSOs with bolometric luminosities $L_{\rm bol} \sim 10^{46}$--$10^{47}$\,erg\,s$^{-1}$\citep[e.g.][]{Fan2001,Kollmeier2006,Mortlock2011,Trakhtenbrot2011,Banados2018,Zakamska2019} imply black hole accretion rates $\dot{M}_{\rm BH} \sim 2$--20\,\Msunyr~assuming a radiative efficiency $\epsilon_{\rm rad} \equiv L_{\rm bol} / \dot{M}_{\rm BH}\,c^2 \approx 0.1$ \citep[e.g.][]{YuTremaine2002,Marconi2004}, and up to an order of magnitude higher for the most extreme QSOs known \citep[e.g.][]{Wu2015_hyperLqso}.  Reproducing such high inflow rates from galactic scales down to the black hole accretion disk represents a significant challenge for models.  
Cold flows, major mergers, minor mergers, and secular processes are generically identified as AGN triggering mechanisms in cosmological and galaxy scale simulations \citep[e.g.][]{Springel2005_BHmodel,Li2007,Hopkins2010_MultiScale,Bournaud2011,DiMatteo2012,Bellovary2013,Angles-Alcazar2015,Pontzen2017,Steinborn2018,Ricarte2019}, but predicted inflow rates have been often limited by either resolution, sub-grid ISM physics, idealized initial conditions, or uncertainties in black hole accretion parameterization.
For the first time in a full cosmological context, we have shown that gas inflow rates of up to $\sim$10\,\Msunyr~can indeed occur at sub-pc scales in a host halo with $M_{\rm vir} \sim 10^{12.5}$\,\Msun~at $z\sim2$, sufficient to power a luminous QSO.\\

Our simulated \qso~phase corresponds to very particular conditions selected at the peak of nuclear gas density.
The four FIRE-2 massive halos simulated in \citet{Angles-Alcazar2017_BHsOnFIRE} all show a similar phase of extreme nuclear densities, reaching $\Sigma_{\rm gas} > 10^{10.5}$\,\Mkpc~and $\Sigma_{\star} > 10^{12}$\,\Mkpc~on $\sim$10\,pc scales, suggesting that all massive halos undergo at least one phase conducive to a luminous QSO \citep[e.g.][]{Soltan1982,YuTremaine2002,Shankar2009}.
Constraints on QSO lifetimes from black hole demographics \citep[e.g.][]{YuTremaine2002}, clustering \citep[e.g.][]{Martini2001,White2012_clustering}, radiation proximity effects \citep[e.g.][]{Trainor2013,Schmidt2017_QSOlifetime,Eilers2017,Eilers2020}, gas fueling arguments \citep{Martini2003}, and others vary from $\sim$0.1--100\,Myr \citep{Martini2004}, though there is indication that AGN flicker on shorter timescales \citep{Schawinski2015,Sartori2018}.
In the \qso~phase, we measure an inflow rate of $\sim$6\,\Msunyr~at $<$0.1\,pc almost in steady state during the full $\sim$4\,Myr duration of the simulation, with no sign of decline.  The full history of the host galaxy (Figure~\ref{fig:ini}) shows nuclear $\Sigma_{\rm gas} > 10^{10.5}$\,\Mkpc~for $\sim$50\,Myr, suggesting that QSO-like inflow rates could be maintained for up to a few times longer than the $\sim$4\,Myr period simulated here.   
These results are consistent with inferred QSO lifetimes.  We note, however, that we have neglected the effects of AGN feedback throughout this study.

\subsection{Diversity of accretion phases on cosmological timescales}

Hyper-refinement simulations on different regimes (Figure~\ref{fig:mdot_vs_z}) suggest that the \qso~conditions may indeed correspond to a special period in the host galaxy's lifetime.
We find strikingly different conditions a mere $\sim$40\,Myr preceding the \qso~phase.  While the gas and stellar surface density profiles at $>$100\,pc are roughly similar, the \eqso~conditions show extreme variability in sub-pc scale inflow rate.  During the $\sim$3\,Myr duration of the simulation, the accretion rate varies by four orders of magnitude from $\dot{M}_{\rm BH} = 0.001$--10\,\Msunyr, corresponding to $L_{\rm bol} \sim 6\times10^{42-46}$\,erg\,s$^{-1}$.  With a factor of ten lower average accretion rate, $\langle \dot{M}_{\rm BH} \rangle = 0.6$\,\Msunyr, order of magnitude changes in \eqso~accretion can occur in only $\sim$0.01\,Myr, in stark contrast with the steady inflow rate in the \qso~phase.  
This transition likely originates from the significantly deeper stellar potential in the \qso~phase, providing stronger torques, enhanced retention of gas against turbulent motions and stellar feedback, and stabilizing the black hole location in the inner $\sim$10\,pc, compared to the significantly more dynamic \eqso~conditions with a flat stellar surface density profile in the central $\sim$100\,pc.

Another important qualitative change in accretion properties occurs at late times ($z \sim 1$), when a thin, rotationally supported gas disk forms, with lower surface density and star formation activity  \citep[$\Sigma_{\rm gas} \sim 10^{8-9}$\,\Mkpc; SFR$\sim$10\,\Msunyr; see also][]{Stern2020,Wellons2020}.  These \agn~conditions show significantly lower average accretion rate, $\langle \dot{M}_{\rm BH} \rangle = 0.01$\,\Msunyr, in this case characterized by $\sim$2\,Myr-long active phases with $\dot{M}_{\rm BH} = 0.005$--0.1\,\Msunyr~followed by longer periods of inactivity owing to the formation of a hot, low-density cavity in the central $\sim$100\,pc that inhibits black hole growth.  Gas consumption by star formation and stellar feedback are likely responsible for the formation of the cavity \citep{Hopkins2016_NuclearSims,Torrey2017}, and possibly an inner Lindblad-like resonance \citep[e.g.][]{Garcia-Burillo2005}, only penetrated by cold gas clumps and filaments falling toward the black hole when destabilized out of the inner edge of the circumnuclear disk (Figure~\ref{fig:zoom_agn}).  These conditions are likely representative of more frequent, low-luminosity AGN in massive galaxies after their peak in star formation activity, with their central black holes accreting predominantly at low-Eddington ratios in the radiatively inefficient, kinetic regime \citep[e.g.][]{Fabian2012,Heckman2014}.

Interestingly, the \qso~phase occurs at the time when the host galaxy is transitioning from bursty star formation driving prominent galactic winds at early times to a more steady mode of star formation that promotes the formation of extended disks at late times.  This transition has been identified in the FIRE simulations for massive ($M_{\star} > 10^{10}$\,\Msun) galaxies at different redshifts  \citep{Muratov2015,Angles-Alcazar2017_BaryonCycle,Angles-Alcazar2017_BHsOnFIRE,Faucher-Giguere2018,Hayward2017,Ma2017_DiverseGradients,Ma2017_StellarDisk} and coincides with the virialization of the inner halo \citep{Stern2020}.  The early bursty epoch represents yet another qualitatively distinct regime for black hole growth, where sub-grid accretion calculations predict short ($\sim$1\,Myr) accretion episodes that can reach or even exceed the Eddington rate followed by longer periods of inactivity owing to stellar feedback continuously evacuating gas from galactic nuclei (Figure~\ref{fig:mdot_vs_z}, middle panel; \citealt{Angles-Alcazar2017_BHsOnFIRE}; \citealt{Catmabacak2020}; Byrne et al. in prep.).  These early conditions should be investigated in future hyper-refinement simulations.

\subsection{Multi-scale torques and AGN triggering}\label{sec:tordis}

Our simulations show a diverse range of morphologies across scales: from large scale filaments to nested spirals, bars, ring-like features, irregular streams, and clumps (Figures~\ref{fig:resmap}--\ref{fig:zoom_agn}). 
The cascade of non-axisymmetric gravitational instabilities seen on different scales resembles the classic ``bars-within-bars'' scenario \citep{Shlosman1989,Shlosman1990}, generalized to a diverse range of morphologies as found in previous idealized models \citep{Escala2007,Hopkins2010_MultiScale,Hopkins2011_Analytic,Mayer2015} and some cosmological simulations \citep{Levine2008,Levine2010,Prieto2016}.
Non-axisymmetric structures with a variety of morphologies are indeed often observed in galactic nuclei \citep[e.g.][]{Knapen2000,Martini2003,Garcia-Burillo2005,Storchi-Bergmann2007,Haan2009,Stoklasova2009,Combes2014,Combes2019} and connected to kinematic evidence for radial gas flows.

Despite the diversity in physical conditions and accretion properties, a common feature identified for the \eqso, \qso, and \agn~phases is that gravitational torques from the stars dominate gas angular momentum transport down to our $\sim$0.1\,pc resolution limit (Figure~\ref{fig:trqRat}), in agreement with \citet{Hopkins2010_MultiScale}.  
While gas shows strong non-axisymmetries on all scales, gravitational self-torquing can become comparable to the stellar torques only on galactic ($\sim$\,kpc) scales for our most gas-rich \eqso~conditions.  Hydrodynamic torques by pressure gradients are stronger than gas self-torquing but still sub-dominant compared to stellar torques.
Other mechanisms for angular momentum transport may be more important in very gas-rich systems with efficient local fragmentation, including scattering of dense gas clumps and turbulence driven by gravitational instability and stellar feedback \citep[e.g.][]{Levine2008,Bournaud2011,Hopkins2016_NuclearSims,Prieto2016,Prieto2017}.  However, comparing our \eqso~and \qso~conditions suggests that gravitational torques from a dominant asymmetric stellar nuclear structure is the most efficient mechanism to drive sustained large inflow rates.

We have explicitly shown that it is generally local torques from nearby stars that are most relevant for driving gas inflows (Figure~\ref{fig:trqStarRat}), as assumed in models of local angular momentum transport \citep{Hopkins2011_Analytic}.  Our best example is the \agn~phase, where the total torque on gas at a given scale is clearly dominated by stars living in the same radial domain, implying that the instantaneous gas inflow rate at $\sim$\,0.1pc is decoupled from the larger scale stellar distribution.  However, the unique dynamic range of our simulations demonstrates that this is not always the case: the net torque on sub-pc scale gas in the \eqso~phase contains significant contributions from stars in the entire radial range 0.1\,pc--1\,kpc, which cannot be explicitly captured in idealized nuclear scale simulations or local transport models.
In either case, the characteristic time variability of stellar non-axisymmetries varies widely on different scales (Figure~\ref{fig:starmap}) and the net torque on gas can change sign both radially, in a time average sense, and temporally, in a given spatial scale \citep[Figure~\ref{fig:trq}; see also][]{Hopkins2010_MultiScale}, with important implications for observations connecting large scale torques and AGN activity.
The small scale ($<10$\,pc) motion of the black hole relative to the stellar center of mass in the \eqso~phase plays a minor role in the observed non-axisymmetries at $>$100\,pc, but more off-centered black holes in low mass hosts and/or higher redshift \citep[e.g.][]{Menezes2016,Bellovary2019,Pfister2019,Shen2019,Reines2020,MaLinhao2021} could significantly affect the resulting large scale gravitational torques.

The non-trivial dependence of sub-pc inflow rates on multi-scale non-axisymmetries and their time variability can explain why observations provide inconclusive evidence for a connection between large scale bars \citep{Knapen2000,Cisternas2013,Cisternas2015,Goulding2017}, clumpy disks \citep{Trump2014}, or nuclear non-axisymmetries \citep[e.g.][]{Martini2003,Garcia-Burillo2005,Jogee2006} and instantaneous AGN activity \citep{AlexanderHickox2012}.
For instance, our simulated \qso~phase shows time-averaged positive torque and correspondingly net outflow rate on $\sim$100\,pc scales (Figures~\ref{fig:MdotSFR} and~\ref{fig:trq}) while sustaining a large inflow rate of $\sim$10\,\Msunyr~on $\sim10$\,pc scales (of which  $\sim$6\,\Msunyr~reach $<$0.1\,pc) owing to strong time-averaged negative torquing in the inner region.

Observations also appear inconclusive regarding the relative roles of galaxy mergers \citep[e.g.][]{Treister2012,HopkinsKocevski2014,Kocevski2015,Wylezalek2016,Ricci2017,Diaz-Santos2018,Donley2018} and secular processes \citep[e.g.][]{Kocevski2012,Simmons2012,Mechtley2016,Villforth2017,Smethurst2019} driving black hole growth, 
with idealized and cosmological simulations sometimes producing contrasting results \citep[e.g.][]{Springel2005_BHmodel,Li2007,Hopkins2010_MultiScale,Bournaud2011,DiMatteo2012,Angles-Alcazar2015,Pontzen2017,Steinborn2018,Ricarte2019,Sharma2021}.
A massive satellite galaxy ($M_{\star} \sim 10^{10} $\,\Msun) is approaching the central galaxy in its second passage toward finale coalescence at the time of the \eqso~and \qso~phases, located $\sim$20\,kpc and $\sim$15\,kpc away, respectively. 
These accretion phases occur within only $\sim$40\,Myr of each other and produce accretion rates that vary by four orders of magnitude while the stellar distribution on $\sim$kpc scales does not change appreciably, highlighting the difficulty of connecting global morphological features with instantaneous accretion rate.  
The merging galaxy contributes less than $\sim$1\,\%~of the total stellar torque on nuclear scale gas ($<100$\,pc) during the \eqso~and \qso~phases (Figure~\ref{fig:trqStarRat}), suggesting a minor role feeding the central black hole, but the overall impact of the ongoing merger on longer timescales (e.g. increasing the density of the central $\sim$1\,kpc) should be investigated in future work.  The lower nuclear activity in the \agn~conditions is clearly unrelated to mergers and driven by secular processes.

\subsection{The SFR--AGN connection}\label{sec:sfragn}

In addition to the self-gravity of stars and gas, our simulations show that it is crucial to include star formation and stellar evolution in gas transport calculations \citep[e.g.][]{Thompson2005,Nayakshin2007,Hopkins2010_MultiScale,Hopkins2011_Analytic,Forbes2014,Goldbaum2016,Krumholz2018}.  Stellar feedback can produce galaxy-scale perturbations in the form of galactic winds and coherent re-accretion events \citep{Muratov2015,Christensen2016,Angles-Alcazar2017_BaryonCycle,Angles-Alcazar2017_BHsOnFIRE}, 
 but also regulates star formation locally and generally impacts the dynamics and multi-phase structure of the ISM \citep{Torrey2017,Orr2018,Moreno2019,Gurvich2020,Orr2021}.  This has important implications for AGN variability and duty cycle as seen for the \eqso~and \agn~conditions (Figure~\ref{fig:mdot_vs_z}), where the accretion rate becomes significantly suppressed when the black hole is embedded in the hot gas phase.
Stellar mass return can become an important source of fueling at late times in gas poor galaxies \citep[e.g.][]{Ciotti2007}, 
but star formation itself can decrease substantially the gas supply available for black hole growth: in our simulations, star formation can consume as much gas as is provided by inflows in the full range of scales (1\,pc--10\,kpc), with some indication for comparatively lower star formation on pc scales (Figure~\ref{fig:MdotSFR}; see also \citealt{Hopkins2010_MultiScale}).  Recently formed stars ($<$5\,Myr) can contribute $\sim$10--20\%~of the total stellar torque on gas at $<$10\,pc for gas-rich conditions (\eqso~and \qso~phases), further emphasizing the importance of including star formation down to the smallest scales.

Our results also have important implications for the connection between global SFR and instantaneous black hole accretion.  Positive average correlations between global SFR and AGN luminosity have been reported for a number of observations \citep{Netzer2009,Mullaney2012_AGN_MainSequence,Chen2013,Feltre2013,Azadi2015,Delvecchio2015,Dai2018,Masoura2018}, while others suggest that they hold only for bulge-dominated galaxies \citep{Yang2019} or the highest luminosity systems \citep{Lutz2010,Rosario2012,Rovilos2012,Duras2017}. 
Other studies suggest little or no connection between average SFR and black hole accretion \citep{Harrison2012,Mullaney2012_NoSfAgnCorr,Stanley2015} or link luminous AGN activity with a suppression of star formation \citep{Page2012}, possibly only for galaxies above the star formation main sequence \citep{Masoura2018}.
Part of the discrepancy may originate from biases and selection effects \citep[e.g.][]{Trump2015}, but different characteristic timescales for star formation and AGN activity are likely crucial as well \citep{Hickox2014,Angles-Alcazar2015,Volonteri2015,Ricarte2019,Thomas2019}.

Our simulations demonstrate that the instantaneous accretion rate (measured at $<$0.1\,pc) can indeed vary by orders of magnitude for a given host galaxy SFR \citep[see also][]{Hopkins2010_MultiScale,Levine2010,Novak2011,Gabor2013}. 
We find that the average accretion rate roughly follows a linear proportionality $\langle \dot{M}_{\rm BH}\rangle \propto \langle {\rm SFR}\rangle$ for star formation within small radii, but time variability decouples AGN activity and global star formation.  
These results are consistent with observations in the local universe indicating correlations between nuclear star formation and instantaneous black hole accretion in individual systems but no clear connection to global star formation \citep{Davies2007,Diamond-Stanic2012_AgnSF,LaMassa2013,Esquej2014,Esparza-Arredondo2018}.

We find time-averaged accretion to global star formation ratios $\dot{M}_{\rm BH} / {\rm SFR} \sim 1/1000$ for the \eqso~and \agn~conditions, in agreement with expectations from the $M_{\rm BH}/M_{\rm bulge}$ ratios measured in the local universe \citep[e.g.][]{Kormendy2013,McConnell2013} and similar to the typical $\dot{M}_{\rm BH} / {\rm SFR}$ ratios observed for AGN in star forming galaxies \citep[e.g.][]{Mullaney2012_AGN_MainSequence,Chen2013,Delvecchio2015,Sun2015,Dai2018}.  
In our \qso~phase, the central black hole is overgrowing its host galaxy, with $\dot{M}_{\rm BH} / {\rm SFR} \sim 1/50$, also in good agreement with observations of luminous QSOs and AGN-dominated sources \citep[e.g.][]{Netzer2009,Fan2016_HotDogs,Netzer2016,Duras2017}.
This suggests that a common gas supply for star formation and black hole growth regulated by gravitational torques may play a key role driving the black hole--galaxy connection \citep{Escala2007,Angles-Alcazar2013,Chen2013,Trump2013,Angles-Alcazar2015,Cen2015,Angles-Alcazar2017_BHfeedback}.

\subsection{From galaxy scales to the accretion disk}

The inflow rate at 0.1\,pc corresponds roughly to the scale below which the non self-gravitating accretion disk is expected to form according to analytic models \citep[e.g.][]{Goodman2003} and reverberation mapping observations \citep[e.g.][]{Homayouni2019}: $10^3$--$10^4\,R_{\rm s} \approx 0.1$\,pc for the Schwarzschild radius $R_{\rm s}$ of black holes with $M_{\rm BH} = 10^{8-9}$\,\Msun.  
Our results have interesting implications for accretion disk models, providing realistic outer boundary conditions that differ significantly from the idealized, torus-like, equilibrium conditions often employed as initial conditions \citep[e.g.][]{Bu2013}.  We have shown that radial flows are present at $\sim$0.1\,pc and the outer disk feeding rate can vary by orders of magnitude on $\sim$3\,Myr (Figure~\ref{fig:mdot_vs_z}), which would imply drastic changes in the geometric and radiative properties of the accretion disk, from super-Eddington fueling conditions \citep{Jiang2014,Jiang2019} to low-Eddington, hot accretion flows \citep{Quataert2000,Narayan2008}, on timescales comparable to the standard $\alpha$-disk viscous time \citep{ShakuraSunyaev1973}.

Gas on sub-pc scales is preferentially rotationally supported but the specific angular momentum direction can change on timescales $<$0.1\,Myr.  Sub-pc scale disks can thus be significantly misaligned with the galaxy-scale disk, in agreement with observations of megamaser disks \citep{Greene2013,Pjanka2017} and jets \citep{Kinney2000,Gallimore2006}. 
Nuclear scale disks are also often decoupled from the angular momentum of the host galaxy \citep{Martini2003,Davies2014,Garcia-Burillo2016,Pjanka2017,Alonso-Herrero2018,Combes2019}, in agreement with our results \citep[see also][]{Shlosman2002,Levine2010,Hopkins2012_AGNmisaligned,Emsellem2015,Renaud2015,Capelo2017,Starkenburg2019}.
Strikingly, we find persistent, increasingly misaligned orientation across scales in the \qso~conditions, with near spin flip in the pc-scale disk orientation relative to kpc scales (Figure~\ref{fig:Jdir}).
The changing orientation of sub-pc scale disks can have drastic implications for the coupling efficiency of black hole feedback \citep{Hopkins2016_NuclearSims,Torrey2020,Su2021_Jets} and the spin evolution of massive black holes \citep{Volonteri2005_BHspin,King2008,Fanidakis2011,Hopkins2012_AGNmisaligned,Dotti2013,Dubois2014_spin,Fiacconi2018,Bustamante2019}, which itself has strong implications for black hole mergers \citep{Bogdanovic2007,Blecha2016} and the origin of radio jets \citep{Tchekhovskoy2011}.

\subsection{Black hole accretion parameterizations}\label{sec:bhpres}

Most modern cosmological simulations employ sub-grid black hole accretion prescriptions based on the spherical \citet{Bondi1952} model\footnote{A notable exception is the hybrid accretion prescription in the SIMBA simulation \citep{Dave2019_Simba}, where cold gas accretes following the gravitational torque accretion model \citep{Hopkins2011_Analytic,Angles-Alcazar2017_BHfeedback} and hot gas accretes following the Bondi parameterization.} (e.g. Horizon-AGN, Eagle, and IllustrisTNG).
Despite the success of these models at reproducing global galaxy and black hole observables, previous idealized simulations have shown that Bondi accretion can fail to reproduce gas inflow rates by orders of magnitude under a variety of conditions \citep{Hopkins2010_MultiScale,Hopkins2011_Analytic,Hobbs2012,Gaspari2013,Hopkins2016_NuclearSims,Negri2017,Beckmann2018}, limiting the predictive power of cosmological simulations.
Our results indicate that the inflow at sub-pc scales is dominated by cool gas with significant rotational support, and it is primarily driven by stellar gravitational torques.  
This is inconsistent with the assumption in Bondi accretion of spherically symmetric, pressure-supported gas in hydrostatic balance.

If we apply Bondi accretion to the \agn~phase (arguably the most favorable conditions for Bondi, with higher presence of hot, pressure-supported gas compared to the other phases), we obtain $\dot{M}_{\rm Bondi} = 4\pi \, G^{2} \, M_{\rm BH}^{2} \, \rho  /  c_{\rm s}^{3}  > 10^7$\,\Msunyr~for $M_{\rm BH} = 10^{8}$\,\Msun~and the gas density $\rho$ and sound speed $c_{\rm s}$ found during the active phases on pc scales, in striking contrast with the range $\dot{M}_{\rm BH}\sim 0.01$--0.1\,\Msunyr~that we measure in the hyper-refinement simulation (Figure~\ref{fig:mdot_vs_z}).  If we consider ``turbulent Bondi-Hoyle'', replacing $c_{\rm s}^{3} \rightarrow (c_{\rm s}^{2}+\sigma_{\rm z}^{2})^{3/2}$ where $\sigma_{\rm z}$ is the vertical velocity dispersion, we obtain $\dot{M}_{\rm Bondi,turb} > 10^4$\,\Msunyr, still many orders of magnitude higher than the explicit inflow rate \citep[see also][]{Hopkins2016_NuclearSims}.  If we further include the gas azimuthal velocity $v_{\phi}$, replacing\footnote{Including the azimuthal velocity in this way is not well motivated theoretically.} $c_{\rm s}^{3} \rightarrow (c_{\rm s}^{2}+\sigma_{\rm z}^{2}+v_{\phi}^{2})^{3/2}$ similar to the prescription in Eagle \citep{Schaye2015}, the accretion rate is still overestimated by more than two orders of magnitude.  Considering the black hole--gas relative velocity has a minimal impact compared to $\sigma_{\rm z}$ and $v_{\phi}$, while adding a density-dependent ``boost" factor \citep[e.g.][]{Booth2009,Dubois2014} only increases these discrepancies.
In addition, Bondi-based models predict a decrease in gas inflow rate with increasing radial distance by factors $>10^3$ in the range 0.1--100\,pc, in clear contrast with the opposite radial trend shown in Figure~\ref{fig:MdotSFR}, and do not capture the rather different accretion conditions in the \eqso, \qso, and \agn~phases.  
In practice, Bondi-based models in cosmological simulations do not yield such unrealistically high accretion rates owing to self-regulation by AGN feedback and imposing the Eddington limit, but this illustrates the inapplicability of current black hole fueling models.

In contrast, a simple free-fall accretion estimator evaluated at $\sim$100\,pc (with overall normalization such that $M_{\rm BH}/M_{\rm bulge} \sim 0.002$ at $z=1$) reproduces the sub-pc inflow rates to within an order of magnitude (Figure~\ref{fig:mdot_vs_z}).  While the free-fall estimator over/under-predicts the low/high accretion phases\footnote{A trend of overpredicted inflow rate during low-accretion phases (\agn) and underpredicted inflow rate during high-accretion phases (\eqso~and \qso) is expected: the constant normalization factor in the free-fall estimator reproduces the average behavior but fails to capture the full variability.  In the more complex gravitational torque model of \citet{Hopkins2011_Analytic}, the normalization depends on the amplitude of non-axisymmetric perturbations, which is higher in the \eqso~and \qso~phases and lower in the \agn~phase, in the right direction toward reproducing the sub-pc inflow rates.}, it reproduces the correct relative differences between the \eqso, \qso, and \agn~phases, unlike Bondi accretion.
We defer a detailed comparison to accretion parameterizations to future work, but note that a key element of sub-grid accretion prescriptions is the intrinsic dependence on black hole mass, $\dot{M}_{\rm BH} \propto M_{\rm BH}^{p}$ \citep{Angles-Alcazar2015}: strong $M_{\rm BH}$ dependence ($p>1$) requires strong feedback self-regulation, as is the case in Bondi accretion, while it is not a necessary condition for weak $M_{\rm BH}$ dependence ($p<1$).
The sub-pc accretion rates predicted in our simulations indicate a weak $M_{\rm BH}$ dependence for a variety of conditions, 
in good agreement with the idealized simulations of \citet{Hopkins2010_MultiScale,Hopkins2011_Analytic} and their gravitational torque accretion model.  
In addition to being crucial for the origin of black hole--galaxy scaling relations \citep{Angles-Alcazar2013,Angles-Alcazar2015,Angles-Alcazar2017_BHfeedback}, the weak $M_{\rm BH}$ dependence also has important implications for super-Eddington feeding.  
In the absence of AGN feedback, we show that sustained sub-pc fueling rates can reach up to $\times$3 and $\times$10 higher than the Eddington limit for $M_{\rm BH} = 10^{8}$\,\Msun~and $M_{\rm BH} = 10^{7}$\,\Msun, respectively, in the extreme conditions of the \qso~phase.
The Eddington limit is often imposed in cosmological simulations but can be greatly exceeded according to recent accretion disk models \citep{Jiang2014,Jiang2019}, provided that sufficiently high inflow rates can be driven from larger scales as our simulations predict.

\subsection{Caveats and future prospects}

We have neglected AGN feedback in order to study how gas is transported down to the black hole accretion disk in the first place.  Reported accretion rates should thus be regarded as upper limits for how efficiently stellar torques can drive gas inflows across scales.  A variety of idealized, high-resolution simulations have shown that AGN feedback can open a central cavity and thus regulate the growth of the central black hole \citep[e.g.][]{Gabor2014,Curtis2016,Hopkins2016_NuclearSims,Richings2018_MolecularOutflow,Costa2020,Torrey2020}. 
However, the coupling efficiency of AGN feedback in the crucial 0.1--10\,pc regime and thus the extent to which self-regulation occurs remains uncertain under highly non-isotropic gas distributions and extreme inflow conditions such as those seen in the \qso~phase.  Future cosmological hyper-refinement simulations will address the role of self-regulation by incorporating a novel implementation of fast, accretion-driven winds \citep{Torrey2020,Su2021_Jets} and other black hole feedback channels.

On the other hand, we have neglected accretion of stars by the central black hole, which could represent an additional source of growth \citep[e.g.][]{Magorrian1999_TDEs,Madigan2011,Pfister2021}.  Future work should consider the incidence of tidal disruption events under the extreme nuclear stellar densities reached in the \qso~and \agn~condition, though these are probably unphysically large owing to the lack of AGN feedback in the original FIRE-2 simulation employed for re-simulation \citep{Angles-Alcazar2017_BHsOnFIRE,Wellons2020}.  Synthetic observations including dust radiative transfer show that the effective size and stellar mass surface density are in general agreement with observational scaling relations at $z>2$ but that the stars are too compact at lower redshift \citep{Parsotan2021}, comparable only to the most extreme compact galaxies observed \citep[e.g.][]{vanDokkum2015,Diamond-Stanic2021}.  Encouragingly, our results show that increasingly large inflow rates fuel the central black hole at the time that AGN feedback is becoming most needed.

Despite the unprecedented dynamic range with full galaxy formation physics, our simulations are still $\sim$3 orders of magnitude from actually resolving accretion down to the innermost stable circular orbit.
Nonetheless, our hyper-refinement technique opens up the possibility of implementing black hole feedback exactly at the scale where accretion disk simulations predict the mass, momentum, and energy output (and their angular dependence) from winds and radiation \citep[e.g.][]{Ohsuga2011,Jiang2014}, as well as studying their coupling efficiency and impact from sub-pc scales to the CGM.
Additional physics missing in our calculations but readily available in GIZMO include magnetohydrodynamics, conduction, viscosity, and cosmic ray feedback \citep{Su2017_multiphys,Hopkins2020_FIREmultiphys}, whose effects have not been explored yet in a galaxy formation context at the resolution achieved here.

Our simulations have been limited to exploring only three different galactic nuclei conditions corresponding to a massive galaxy ($M_{\star} \sim 6$--20\,$\times 10^{10}$\,\Msun) before, during, and after its peak of nuclear gas density at $z\sim 2.5\rightarrow1$.  The \eqso, \qso, and \agn~phases simulated predict strikingly different black hole accretion properties, which motivates expanding this work to a much broader set of host galaxy conditions.  Future cosmological hyper-refinement simulations should model black hole growth and feedback in the full range of environments, from active dwarf galaxies \citep{Reines2013,Mezcua2017,Bellovary2019,Greene2020_ARAA} and nearby Seyferts with obscuring ``torus'' structures \citep{RamosAlmeidaRicci2017} to the extreme conditions triggering the highest redshift QSOs \citep{Banados2018}.

\acknowledgments 

We thank Ena Choi, Romeel Dav\'e, Robert Feldmann, John Forbes, Shy Genel, Melanie Habouzit, Yan-Fei Jiang, Yuan Li, Xiangcheng Ma, Eve Ostriker, Roxana Pop, Julissa Rojas-Sandoval, Matthew Smith, Tjitske Starkenburg, Kung-Yi Su, Paul Torrey, Rainer Weinberger, and Sarah Wellons for many insightful discussions and suggestions during the development of this work.  
We thank the referee for a detailed and constructive review that helped improve the paper.
We acknowledge outstanding support by the Scientific Computing Core group and the Center for Computational Astrophysics at the Flatiron Institute as part of the SMAUG project, which are supported by the Simons Foundation.
DAA was supported in part by NSF grant AST-2009687.
EQ was supported in part by a Simons Investigator Award from the Simons Foundation and by NSF grant AST-1715070.
Support for PFH was provided by NSF Collaborative Research Grants 1715847 \&\ 1911233, NSF CAREER grant 1455342, NASA grants 80NSSC18K0562, JPL 1589742. 
CAFG was supported by NSF through grants AST-1517491, AST-1715216, and CAREER award AST-1652522; by NASA through grant 17-ATP17-0067; and by a Cottrell Scholar Award and a Scialog Award from the Research Corporation for Science Advancement.
GLB was supported in part by the NSF throughs grants OAC-1835509 and AST-2006176, as well as support from the STScI under NASA contract NAS5-26555.
DK was supported by NSF grant AST-1715101 and the Cottrell Scholar Award from the Research Corporation for Science Advancement.
The simulations were run on Flatiron Institute's research computing facilities ({\it Gordon-Simons}, {\it Popeye}, and {\it Iron} compute clusters), supported by the Simons Foundation, and XSEDE allocation TG-AST160048, supported by NSF grant ACI-1053575.  Additional numerical calculations were run on the Caltech compute cluster ``Wheeler,'' allocations FTA-Hopkins supported by the NSF and TACC, and NASA HEC SMD-16-7592. \\

\appendix

\section{}\label{sec:appendix:tests}

\subsection{Resolution convergence}\label{sec:appendix:res}
 
\begin{figure}
\begin{center}
\includegraphics[width=0.4\textwidth]{\pathL/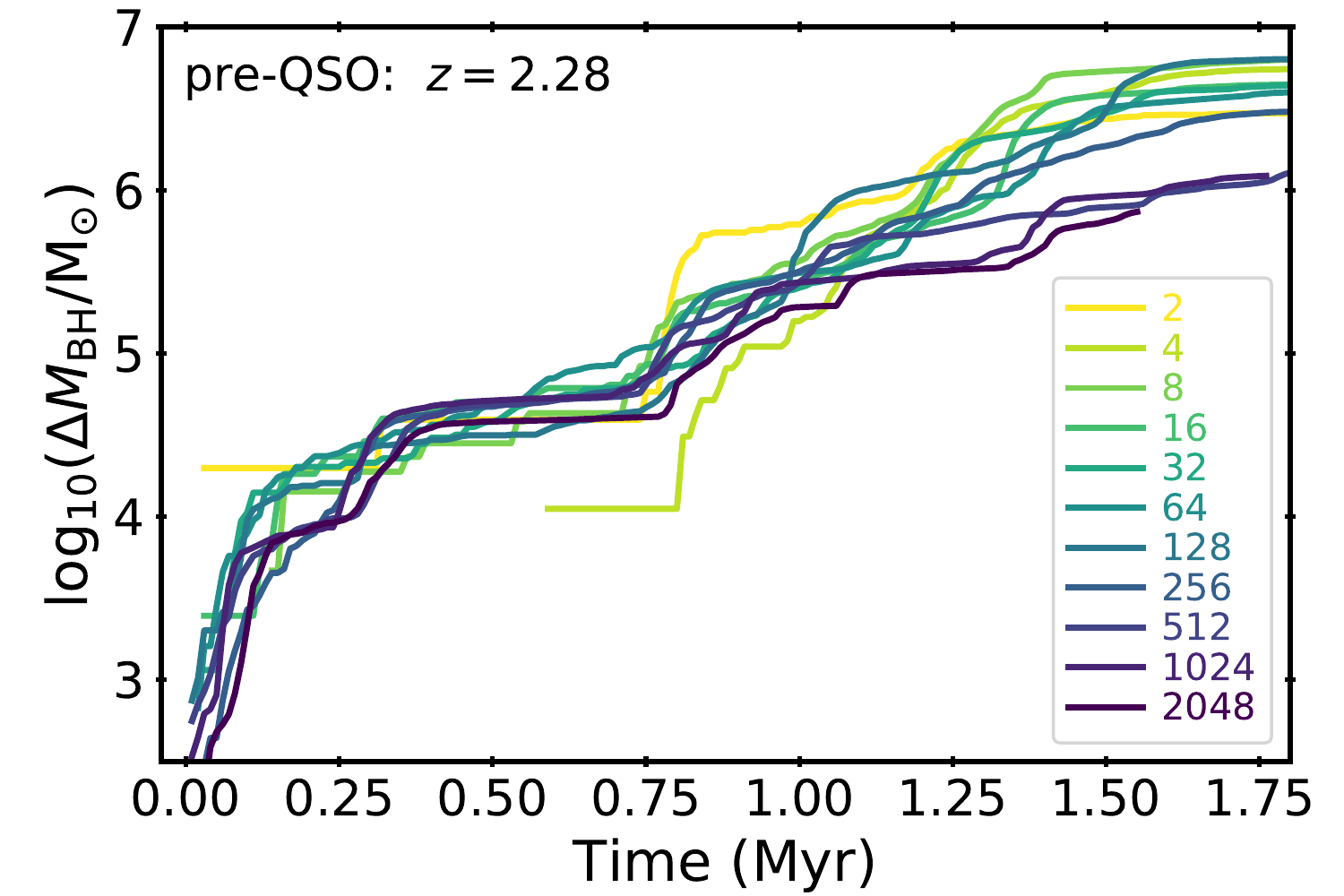}
\includegraphics[width=0.4\textwidth]{\pathL/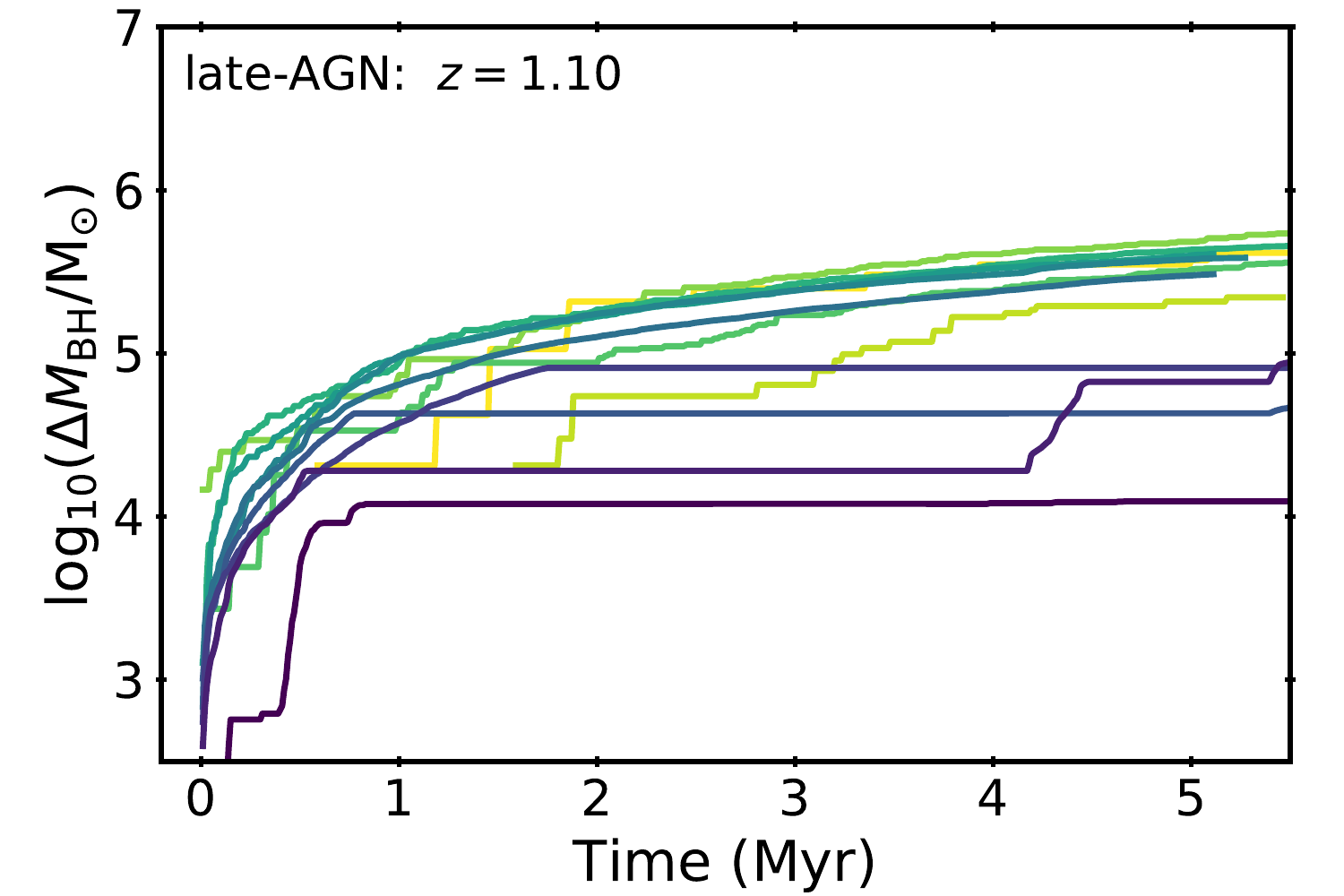}
\end{center}
\vspace{\vsf}
\caption{
Cumulative black hole growth as a function of time for the \eqso~(top) and \agn~(bottom) conditions simulated with increasing mass resolution, from $\times$2 higher resolution than the original cosmological zoom-in simulation (yellow) to $\times$2048 for the highest hyper-refinement level (purple).
Low resolution simulations overpredict the inflow rate at 0.1\,pc by up to an order of magnitude, with indication of numerical convergence for the highest resolution levels.
}
\label{fig:res_test}
\end{figure}

Figure~\ref{fig:res_test} investigates the convergence properties of predicted accretion rates relative to mass resolution for the \eqso~and \agn~conditions.  We show cumulative black hole growth as a function of time in simulations starting from identical initial conditions but implementing dynamic hyper-refinement of gas to different levels, where the color scale indicates the factor by which the mass resolution is increased relative to the initial conditions (see Figure~\ref{fig:ref} for the radial profile of gas mass resolution in each level).  All other simulation parameters, including gravitational softenings and accretion aperture, are kept identical.
There is no clear direct relationship between change in resolution and variation in accretion rate, but we find that lower resolution simulations tend to overpredict black hole growth by up to one order of magnitude relative to our highest resolution simulations.  Encouragingly, there is indication of convergence for the highest resolution levels, when the gas mass resolution in the nuclear region satisfies $m_{\rm g} \lesssim 100$\,\Msun.

It is interesting to note that the average black hole growth rate, measured as explicit gravitational capture of gas at $<$0.1\,pc, varies by less than one order of magnitude in \eqso~simulations with mass resolution changing by $\times$1000.  The weak dependence of \eqso~conditions to mass resolution can be attributed to a number of factors:
\begin{itemize}[itemsep=-10pt]

\item The initial black hole ($M_{\rm BH} = 10^8$\,\Msun) is already very massive compared to individual gas particles even at  the lowest resolution ($m_{\rm g} \sim10^4$\,\Msun), implying that the black hole radius of influence is well resolved and gas particles can indeed be gravitationally bound to the black hole in a resolved sense.\\

\item The total, time-integrated black hole growth ($\sim10^6$\,\Msun~at the highest resolution) is also significantly larger than a single gas particle mass even at the lowest resolution, still allowing for a particle-by-particle representation of accretion over the episode.\\

\item The gas inflow rate is roughly constant within a radial distance $R \lesssim 50$\,pc (Figure~\ref{fig:MdotSFR}) which is large enough to be resolved even at the lowest resolution, and the amount of gas flowing into this central region is roughly independent of resolution over the tested range.
\end{itemize}

\begin{figure}
\begin{center}
\includegraphics[width=0.4\textwidth]{\pathL/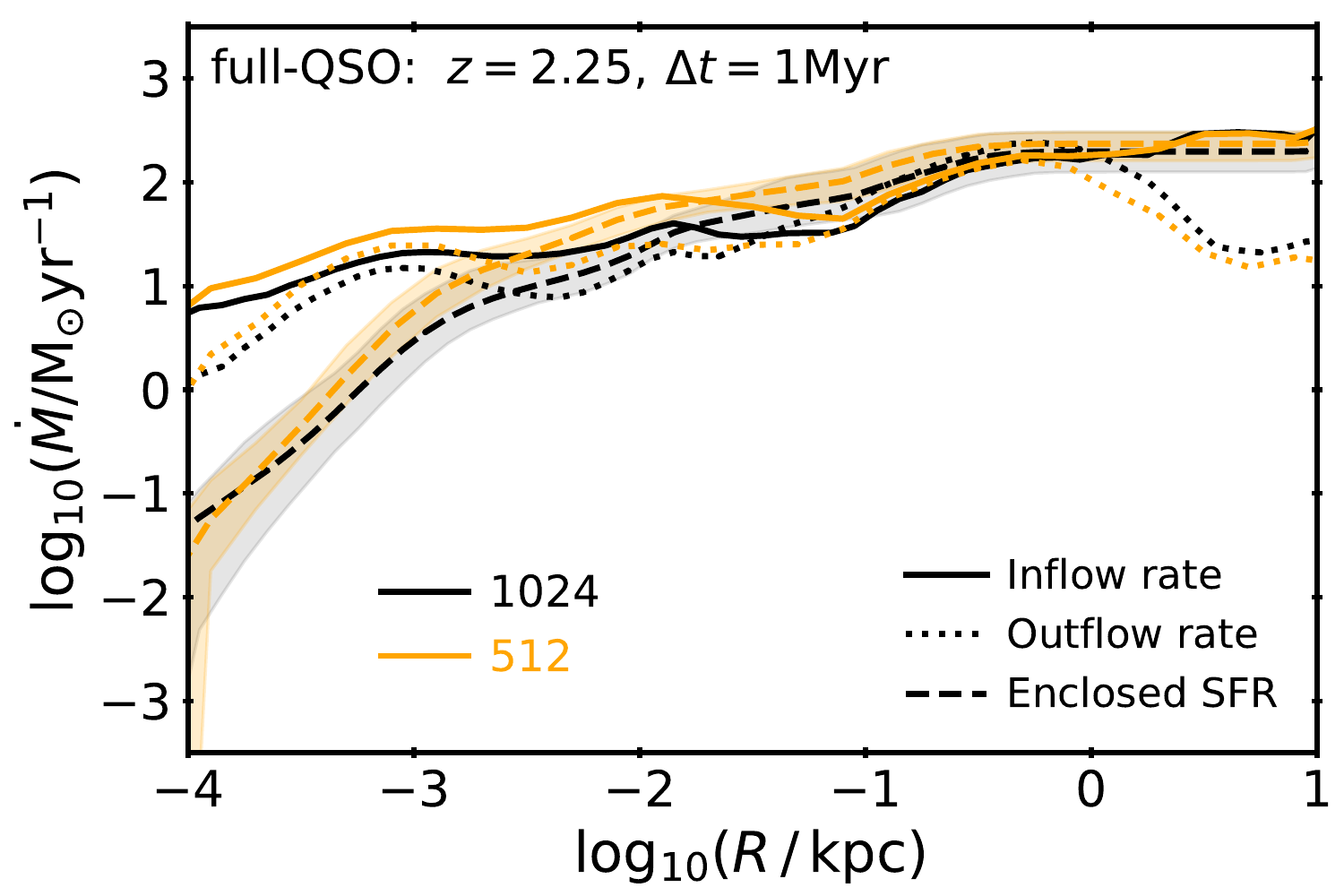}
\end{center}
\vspace{\vsf}
\caption{
Time-averaged gas inflow rate (solid), outflow rate (dotted), and enclosed star formation rate (dashed) as a function of spherical radial distance $R$ from the central black hole for the \qso~conditions simulated with $\times$512 (orange) and $\times$1024 (black) higher resolution than the original cosmological simulation.
Shaded regions indicate the 10--90\% percentile range of SFR achieved in each radial bin.
Both resolution levels show similar radial flows and star formation profiles during the same evolution time $\Delta t$, with sub-pc inflow rates converging to within $\sim$50\%.
}
\label{fig:res_qso}
\end{figure}

The accretion aperture $R_{\rm acc}=0.1$\,pc is not resolved at low resolution, where individual particles represent a much larger volume, but particles can still be accreted when their center of mass happens to be within $R_{\rm acc}$.  The probability of finding a low resolution particle within $R_{\rm acc}$ is low, but a single accretion event contributes the equivalent of many high resolution particles.  Since the radial inflow is roughly independent of $R$ below a scale which is also resolved at our lowest resolution, the time-integrated black hole growth becomes roughly independent of resolution.
These conditions can clearly break down for a number of reasons, including lower mass black holes, lower accretion rates, or strong radial dependence of inflow rates.  
For example, the \agn~conditions show more variation with resolution owing to the lower accretion rate and total integrated black hole growth.  In this case, even the highest resolution levels are not expected to fully converge during the $\Delta t \sim 5$\,Myr period for which simulations with different resolution levels are available, which is not long enough to average over the stochastic fueling conditions and low duty cycle driven by the formation of a hot, low density cavity in the central $\sim$100\,pc (see Figure~\ref{fig:mdot_vs_z}).   

Figure~\ref{fig:res_qso} shows an additional resolution convergence test for the \qso~phase, in this case comparing the full radial profile of gas inflow rate, outflow rate, and enclosed SFR (similar to Figure~\ref{fig:MdotSFR} but separating the inflow/outflow components) for the two simulations with the highest resolution available (corresponding to $m_{\rm g} \sim 30$--60\,\Msun) during the same evolution time $\Delta t \sim 1$\,Myr.  Note that 1\,Myr corresponds to $\sim$10--100 dynamical times at 10--1\,pc for the \qso~phase, reaching a reasonable statistical steady state for the much larger and more steady accretion rates.  We find roughly similar time-averaged radial profiles for both resolution levels over the full dynamic range 0.1\,pc--10\,kpc, with a slight trend for higher rates at $<$10\,pc in the lower resolution simulation.  While these measurements can be affected by stochasticity in black hole dynamics (with the black hole location used as center of reference), the sub-pc inflow rates at our highest resolution levels converge to within $\sim$50\%.
Overall, these tests show that our results are robust to changes in resolution and motivate the development of hybrid accretion models for full cosmological simulations, switching between (1) a continuous sub-grid accretion model when the inflow rate and black hole radius of influence are unresolved to (2) direct gravitational capture of individual gas particles when they are resolved.

\subsection{Gravitational softenings and stellar splitting}\label{sec:appendix:gravtest}

We test the robustness of our results relative to the treatment of gravitational softenings and the mass resolution of the stellar component by comparing our fiducial model described in \S\ref{sec:methods} with simulations implementing:
\begin{itemize}[itemsep=-10pt]
\item {\it Larger stellar gravitational softenings}, which we increase from $\epsilon_{\star} = 0.1$\,pc in the fiducial model to the same stellar softening employed in the original FIRE-2 simulation, $\epsilon_{\star} =7$\,pc.\\

\begin{figure}
\begin{center}
\includegraphics[width=0.40\textwidth]{\pathL/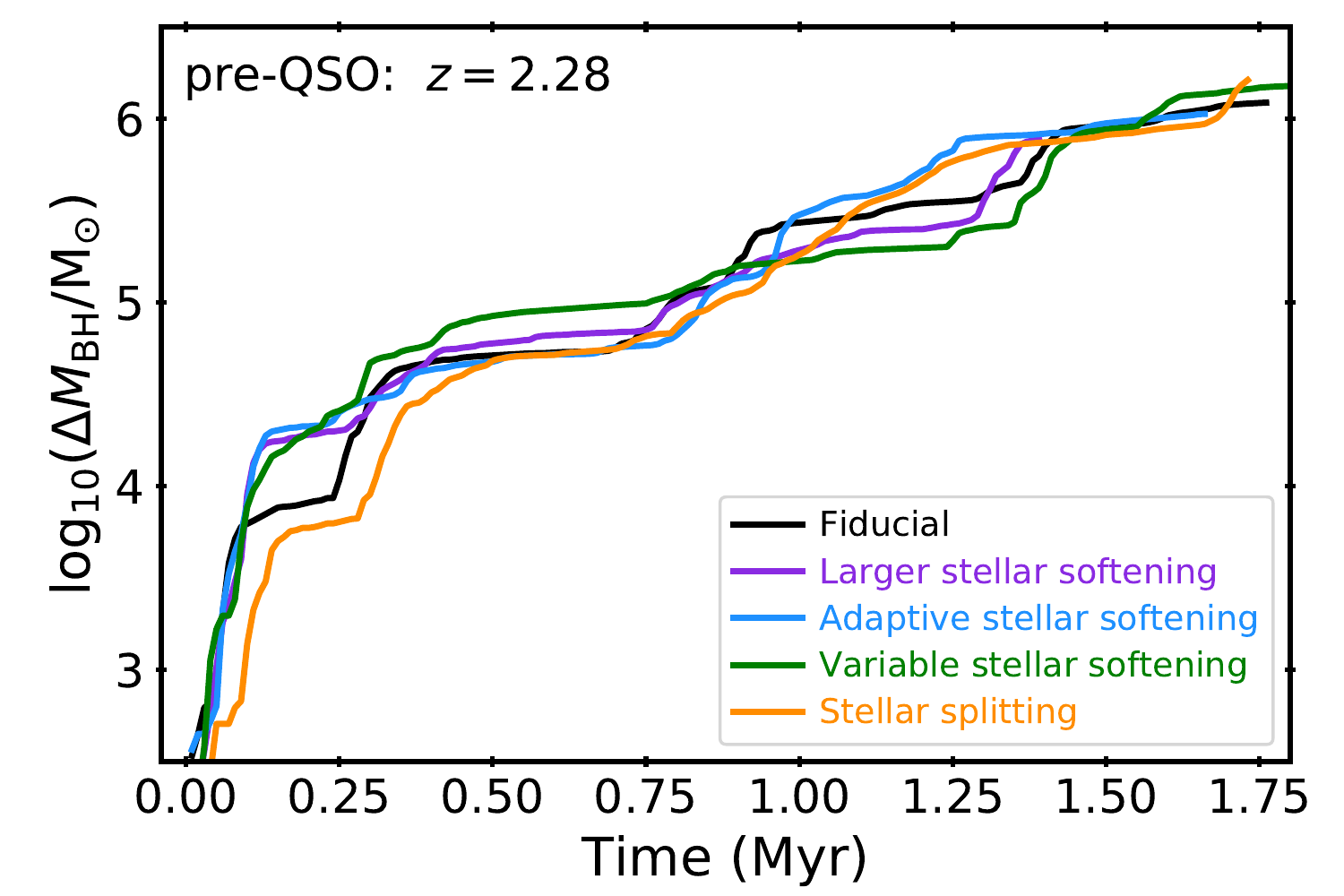}
\end{center}
\vspace{\vsf}
\caption{Cumulative black hole growth as a function of time for the fiducial \eqso~simulation (black solid line) compared to various test simulations with (1) larger stellar gravitational softenings (purple),
(2) adaptive gravitational softenings (blue),
(3) variable stellar gravitational softenings (green), and
(4) splitting of star particles (orange).
See text for details.
Predicted inflow rates are robust to the details of the gravitational softening implementation.
}
\label{fig:GS_test}
\end{figure}

\item {\it Adaptive gravitational softenings for collisionless particles}, which are computed as the mean (kernel-averaged) interparticle spacing within the kernel as for the gas component \citep{Hopkins2015_Gizmo,Hopkins2018_FIRE2methods}.  For star particles, $\epsilon_{\star} \propto \rho_{\rm b}^{-1/3}$, where $\rho_{\rm b}$ is the total baryonic (gas+stars) mass density while for the dark matter the adaptive softening is determined from the dark matter particle neighbor distribution, $\epsilon_{\rm DM} \propto \rho_{\rm DM}^{-1/3}$.  The minimum adaptive softenings for stars and dark matter are $\epsilon_{\star, \rm min} = 0.1$\,pc and $\epsilon_{\rm DM, min} = 57$\,pc, respectively.\\

\item {\it Variable gravitational softenings for star particles}, with $\epsilon_{\star} = 7$\,pc\,$\times (m_{\star}/{\rm 3e4}\,{\rm M}_{\odot})^{1/3}$ (and upper limit $\epsilon_{\star, \rm max} = 7$\,pc) such that star particles represent the same stellar mass density regardless of the particle mass $m_{\star}$.  This assigns the same stellar softening as the original cosmological zoom-in simulation for pre-existing stars at the initial conditions ($\epsilon_{\star} =7$\,pc) while providing higher force resolution for star particles forming out of higher resolution gas in the nuclear region ($\epsilon_{\star, \rm min} \sim 0.6$\,pc for $m_{\star} = 20$\,\Msun).\\

\item {\it Dynamic stellar splitting}, where we increase the mass resolution of the stellar component by splitting star particles progressively according to their distance to the central black hole.  We define a mass resolution profile as in Figure~\ref{fig:ref} with maximum refinement factor $\chi_{\rm ref} = 1/64$, which yields $m_{\star, \rm min} \approx 500$\,\Msun~in the central $\sim$10\,pc.  The splitting procedure for star particles is identical to that of the gas component except that split stars retain the same velocity as the parent star particle.
Note that this is a simplified treatment of stellar splitting, which should involve interpolation in the 6D phase space that describes collisionless systems.  In this case, we use fixed softenings with $\epsilon_{\star} = 0.1$\,pc as in the fiducial model.
\end{itemize}

Figure~\ref{fig:GS_test} shows the cumulative black hole growth predicted for the \eqso~phase in simulations testing different implementation details.  While the instantaneous accretion rate can vary by up to two orders of magnitude between different models, owing to stochastic variability in the very dynamic \eqso~conditions, the cumulative black hole growth follows a similar trend regardless of the exact values of gravitational softenings or numerical implementation details.  
Analysis of gravitational torques (as in Figures \ref{fig:trq}, \ref{fig:trqRat}, and~\ref{fig:trqStarRat}) shows no significant differences between simulations except for a decrease in the contribution of stars within 1\,pc to the total gravitational torque on gas at $<$1\,pc when using larger stellar gravitational softenings ($\epsilon_{\star} =7$\,pc), as expected.  While the integrated black hole growth is not significantly affected by the larger softenings (possibly due to the very strong large scale torques but also stochasticity in the \eqso~conditions), these results suggest that it is more important to achieve high gravitational force resolution than specifically hyper-refining the stellar component, with the overall results robust to numerical details.

\section{}
\subsection{Multi-scale gas in the \eqso~phase}\label{sec:appendix:qsomap}

\begin{figure*}
\begin{center}
\includegraphics[width=1\textwidth]{\pathL/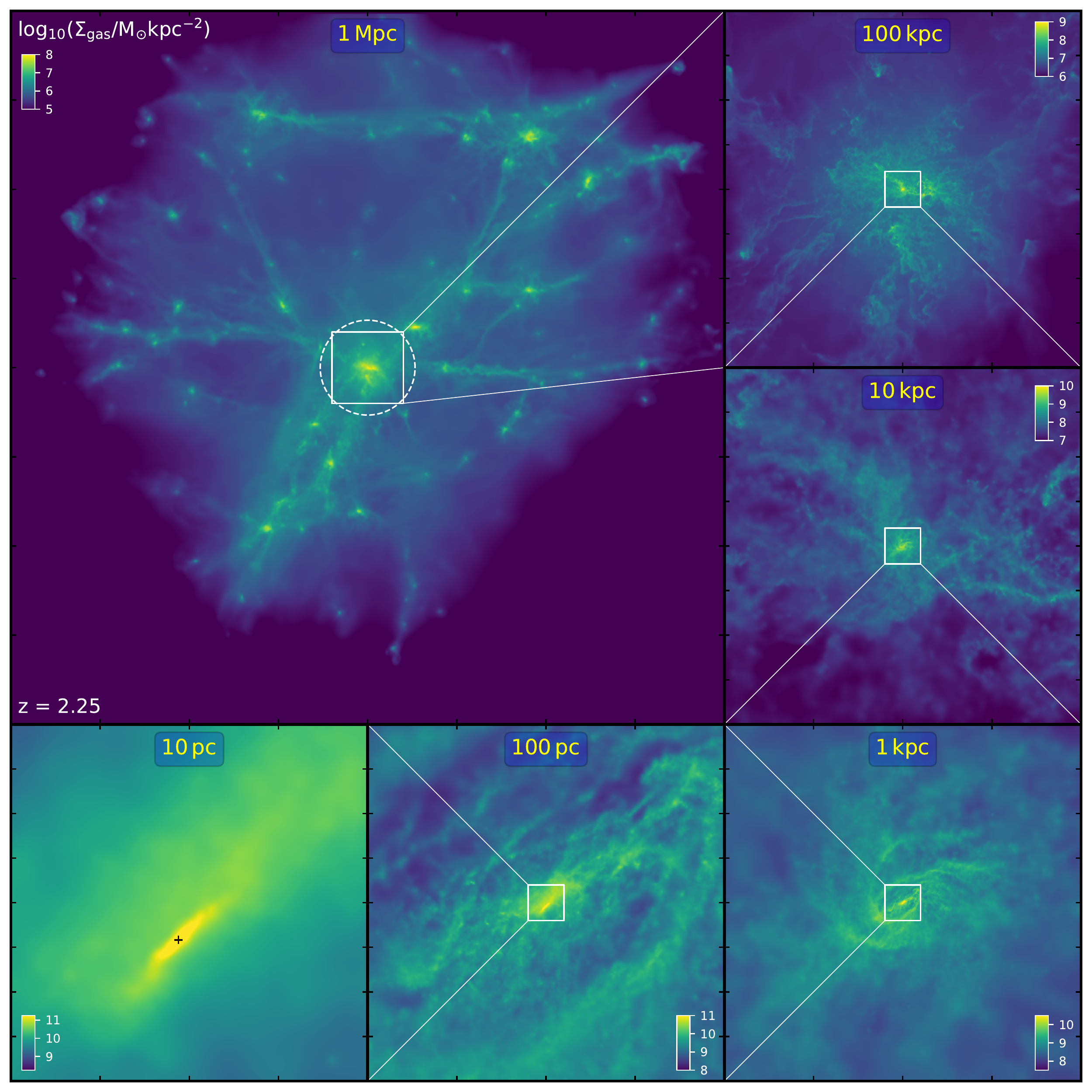}
\end{center}
\vspace{\vsf}
\caption{
Same as Figure~\ref{fig:zoom_qso} for the multi-scale gas mass surface density distribution in the \qso~phase at $z=2.25$, corresponding to $\sim$3.6\,Myr after the start of the hyper-refinement simulation.
The top left panel shows the central 1\,Mpc, with the white dashed line indicating the virial radius of the central halo.  Subsequent panels progressively zoom into the central 10\,pc of the main galaxy.  
All panels are centered on the center of mass of the stellar component within the inner 1\,kpc of the main halo.
The location of the central massive black hole is indicated by a + symbol in the lower left panel.
}
\label{fig:qsomap} 
\end{figure*}

Figure~\ref{fig:qsomap} shows the multi-scale gas mass surface density distribution for the \qso~conditions, for direct comparison to the gas density distributions in the \eqso~and \agn~phases (Figures~\ref{fig:zoom_eqso} and~\ref{fig:zoom_agn}) as well as the temperature distribution for the \qso~conditions (Figure~\ref{fig:zoom_qso}).  Despite the similar high level of turbulence, the \qso~phase shows significantly more centrally concentrated gas density distribution on scales $<$1\,kpc compared to the \eqso~conditions, reaching $\Sigma_{\rm gas} > 10^{11}$\,\Mkpc~in the central 10\,pc and feeding the central black hole in steady state through an extremely dense and cold pc-scale gas disk.

\bibliography{../../references}
\bibliographystyle{aasjournal}

\end{document}